\documentclass[english,twocolumn,prb,amsmath,amssymb,floatfix]{revtex4-2}
\pdfpageattr {/Group << /S /Transparency /I true /CS /DeviceRGB>>}
\usepackage[T1]{fontenc}
\usepackage[latin9]{inputenc}
\usepackage{dcolumn}
\usepackage{bm}
\usepackage{graphicx}
\usepackage{color}
\usepackage{babel}
\usepackage{amsfonts}
\usepackage{mathtools}
\usepackage{slashed}
\usepackage{enumerate}
\usepackage{braket}
\usepackage[multiple]{footmisc}

\usepackage{tikz}
\usetikzlibrary{calc,shapes.arrows,shapes.geometric}
\begin{document}

%###########################
%###########################

\title{Linear response quantum transport through interacting multi-orbital nanostructures}

\author{Emma L. Minarelli}
\affiliation{School of Physics, University College Dublin, Belfield, Dublin 4, Ireland}
\affiliation{Centre for Quantum Engineering, Science, and Technology, University College Dublin, Ireland}
\author{Jonas B. Rigo}
\affiliation{School of Physics, University College Dublin, Belfield, Dublin 4, Ireland}
\affiliation{Centre for Quantum Engineering, Science, and Technology, University College Dublin, Ireland}
\author{Andrew K. Mitchell}
\affiliation{School of Physics, University College Dublin, Belfield, Dublin 4, Ireland}
\affiliation{Centre for Quantum Engineering, Science, and Technology, University College Dublin, Ireland}

%###########################
%###########################

\begin{abstract}
\noindent Nanoelectronics devices, such as quantum dot systems or single-molecule transistors, consist of a quantum nanostructure coupled to a macroscopic external electronic circuit. Thermoelectric transport between source and drain leads is controlled by the quantum dynamics of the lead-coupled nanostructure, through which a current must pass. Strong electron interactions due to quantum confinement on the nanostructure produce nontrivial conductance signatures such as Coulomb blockade and Kondo effects, which become especially pronounced at low temperatures. 
In this work we first provide a modern review of standard quantum transport techniques, focusing on the linear response regime, and highlight the strengths and limitations of each. In the second part, we develop an improved numerical scheme for calculation of the ac linear electrical conductance through generic interacting nanostructures, based on the numerical renormalization group (NRG) method, and explicitly demonstrate its utility in terms of accuracy and efficiency. In the third part we derive low-energy effective models valid in various commonly-encountered situations, and from them we obtain simple analytical expressions for the low-temperature conductance. This indirect route via effective models, although approximate, allows certain limitations of conventional methodologies to be overcome, and provides physical insights into transport mechanisms. Finally, we apply and compare the various techniques, taking the two-terminal triple quantum dot and the serial multi-level double dot devices as nontrivial benchmark systems.
\end{abstract}
\maketitle

%###########################
%###########################

\section{Introduction}\label{sec:intro}

\begin{figure}[b!]
\includegraphics[width=8.7cm]{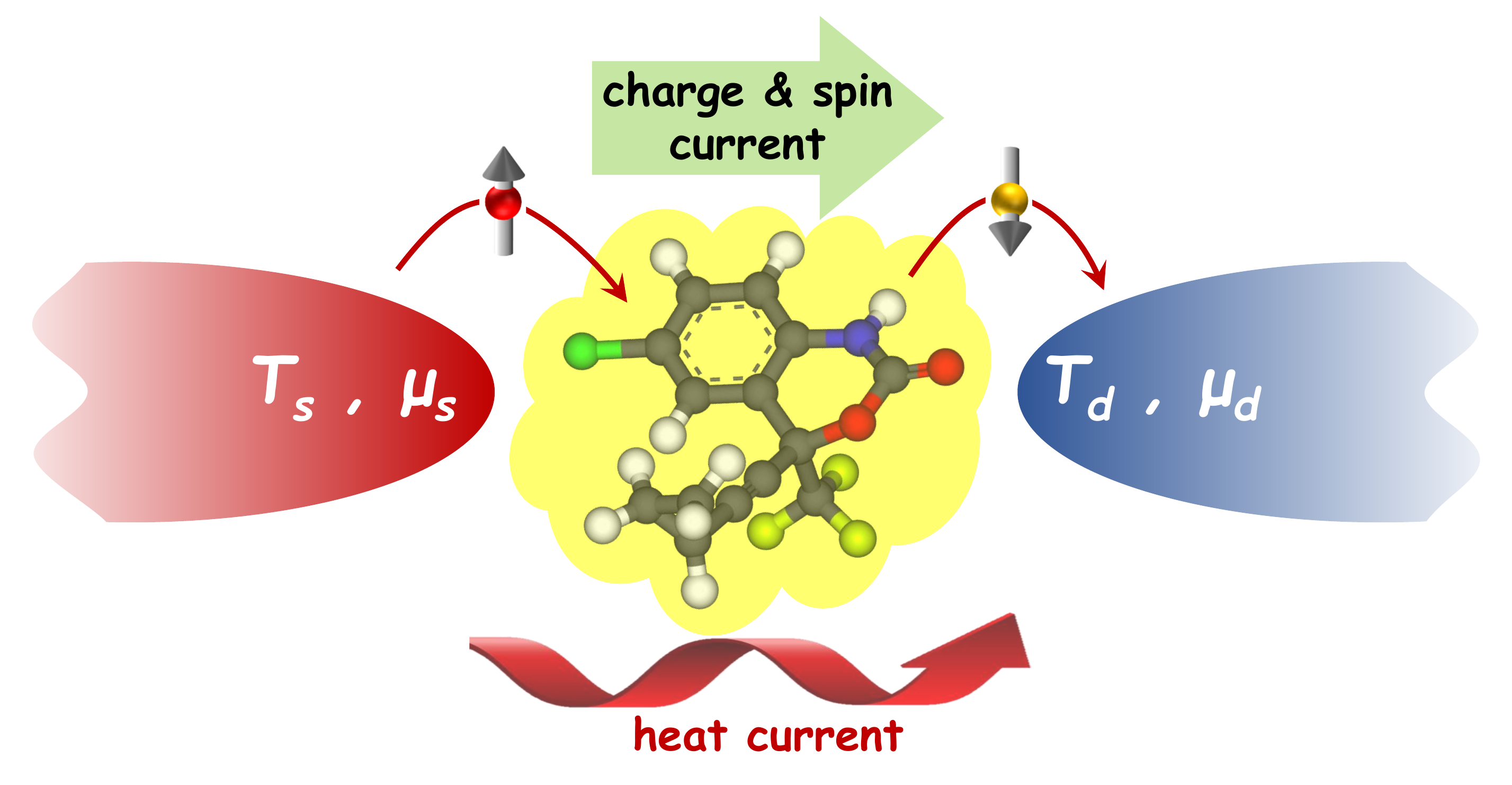}
  \caption{Schematic of the basic quantum transport setup: a voltage bias and/or temperature gradient between source and drain leads results in current flow through the nanostructure.
  }
  \label{fig:schematic}
\end{figure}

The theory of mesoscopic quantum transport\cite{datta2005quantum,nazarov2009quantum,beenakker1991quantum,cohen2020green} has gained importance in recent years due to the development of advanced nanoelectronics devices, such as semiconductor quantum dots\cite{goldhaber1998kondo} or single-molecule transistors.\cite{liang2002kondo} Cryogenic quantum devices can now even be fabricated in commercial CMOS technology.\cite{ruffino2022cryo} These systems typically consist of a nanostructure coupled to source and drain leads, and are tunable \textit{in-situ} by applying gate voltages. The charge and/or heat current that flows between the macroscopic leads due to an applied voltage bias and/or temperature gradient is controlled by the quantum dynamics of the nanostructure. The setup is illustrated in  Fig.~\ref{fig:schematic}. A deep understanding of the underlying theory of quantum transport in such systems is a pre-requisite for the development and design of devices underpinning possible future quantum technologies.

% types of physics

To theoretically capture the diverse range of complex physics that can be realized in quantum nanoelectronics devices, and to accurately predict the resulting transport signatures, three steps must be undertaken. First, an accurate microscopic model of the system must be formulated, taking into account both the interacting, multi-orbital nanostructure and the macroscopic metallic leads -- as well as the coupling between them. Secondly, this model must be solved. Typically, this must be done numerically using sophisticated quantum many-body computational techniques. Often, the bare microscopic model is too complex to be treated directly, and simplified effective models that capture the physics of interest must be derived. Finally, the desired physical observables (in this case quantum transport coefficients) must be obtained from the solution of the model. Standard techniques for the latter require the calculation of dynamical correlation functions for the interacting, lead-coupled nanostructure -- a formidable task in itself. Furthermore, these standard techniques each have strengths and limitations and there is no all-purpose black-box method.

In this paper we focus on quantum transport through interacting multi-orbital nanostructures in the linear response regime, and discuss each of the above three steps in detail.

%%%%%%%%%%%%%%%%

The paper is organized as follows. In Sec.~\ref{sec:model} we introduce and discuss in detail the microscopic models describing nanoelectronics systems. We focus on two-terminal setups, with non-interacting metallic leads connected in an arbitrary way to an interacting, multi-orbital nanostructure (the generalization to structured leads in a multi-terminal setup is discussed in the Appendixes). We discuss two important and commonly-used limiting cases: the proportionate coupling hybridization geometry (Sec.~\ref{sec:PC}) and the single-orbital nanostructure  (Sec.~\ref{sec:AIM}). We emphasize that these limits are typically not satisfied for realistic systems, and our main results do \emph{not} rely on either condition being met. Secs.~\ref{sec:GFs} and \ref{sec:obs} formalize the dynamical quantities and quantum transport coefficients considered in this paper, while Sec.~\ref{sec:nrg} briefly summarizes the NRG method employed for the numerical calculations.

In Sec.~\ref{sec:LR} we give a modern review of various standard approaches to the calculation of thermoelectric quantum transport coefficients. We relate the different techniques in the appropriate limits where equivalences can be made, and emphasize the benefits and limitations of each. We provide generalizations of the conventional formulae where possible. We consider the Landauer approach (Sec.~\ref{sec:landauer}), the renormalized Landauer method (Sec.~\ref{sec:oguri}), the Meir-Wingreen formula (Sec.~\ref{sec:MW}), and the Kubo formula (Sec.~\ref{sec:kubo}) as specific paradigms.

In Sec.~\ref{sec:newkubo} we introduce an ``improved'' version of the Kubo formula for the ac linear electrical conductance for use with NRG, demonstrating via explicit benchmark calculations the benefit in terms of accuracy and efficiency over the conventional Kubo approach. By contrast, in Sec.~\ref{sec:heatkubo} we see that the Kubo formula for heat conductance in NRG breaks down.

In Sec.~\ref{sec:emPC_CB} we derive low-energy effective models valid for nanostructures in the Coulomb-blockade regime (and consider separately the situations arising for  ground states with different spin). These effective models are analyzed in terms of their RG flow, and the ``emergent proportionate coupling'' concept is introduced, in which an effective single-channel description pertains at low temperatures. This allows us to obtain analytic results for the low-temperature conductance, directly in terms of the effective model parameters. Analytical results are supported by numerical calculations. In Sec.~\ref{sec:emPC_MV} we do this instead near the charge degeneracy points separating two Coulomb blockade diamonds. Depending on the spin states of the two degenerate sectors, different effective models result (and low-$T$ transport is in consequence found to be markedly affected).

In Sec.~\ref{sec:aux} we apply the recently-developed `auxiliary field' mapping for interacting quantum impurity-type models to derive a novel approximation for conductance calculations. The approach exploits an equivalent non-interacting formulation of an interacting model, so that the Landauer-B\"uttiker formula can be employed. 

Finally, in Sec.~\ref{sec:apps} we apply the techniques discussed above to two nontrivial quantum dot models (the triangular triple quantum dot system in both one- and two-channel geometries; and the multi-level double quantum dot system in a serial geometry). We compare and contrast the numerical methods and benchmark the analytical approximations. A conclusion and outlook is provided in Sec.~\ref{sec:conc}, while various derivations are given in the Appendixes.

%%%%%%%%%%%%%%

\section{Models of nanodevices}\label{sec:model}
We consider systems described by the generic Hamiltonian $H=H_{\rm leads} + H_{\rm nano} + H_{\rm hyb} + H_{\rm gate} +H_{\rm field}$. Here,
\begin{eqnarray}\label{eq:Hleads}
H_{\rm leads} =  \sum_{\alpha}H_{\rm leads}^{\alpha}=\sum_{\alpha,k,\sigma} \epsilon_{k}^{\phantom{\dagger}} c_{\alpha k \sigma}^{\dagger} c_{\alpha k \sigma}^{\phantom{\dagger}} \;,
\end{eqnarray}
describes source ($\alpha=s$) and drain ($\alpha=d$) leads as non-interacting fermionic reservoirs at thermal equilibrium, with  $f_{\alpha}(\omega)=(1+\exp[(\omega-\mu_{\alpha})/k_{\rm B} T_{\alpha}])^{-1}$
the Fermi-Dirac distribution of lead $\alpha$ held at chemical potential $\mu_{\alpha}$ and temperature $T_{\alpha}$. An electron with momentum $k$ and spin $\sigma=\uparrow$ or $\downarrow$ is created in lead $\alpha$ by $c_{\alpha k \sigma}^{\dagger}$, and $c_{\alpha k \sigma}$ is the corresponding annihilation operator. We take the dispersion $\epsilon_k$ to be independent of lead $\alpha$ and spin $\sigma$. 

The isolated nanostructure is modelled by,
\begin{eqnarray}\label{eq:Hnano}
H_{\rm nano} = \sum_{i,j,\sigma} t_{ij}d^{\dagger}_{i\sigma}d^{\phantom{\dagger}}_{j\sigma} + H_{\rm int} \;,
\end{eqnarray}
where $t_{ij}=t_{ji}^*$ are elements of the single-particle matrix $\textbf{H}_{\rm nano}^{(0)}$, and  
%$H_{\rm int}=\sum_{i} U_{i} \hat{n}_{i\uparrow}\hat{n}_{i\downarrow} + \sum_{i < j}U'_{ij} \hat{n}_{i}\hat{n}_{j}$ describes electron-electron interactions on the nanostructure within an extended Hubbard model approximation. 
$H_{\rm int}$ describes electron-electron interactions on the nanostructure. Here $d^{\dagger}_{i\sigma}$ ($d_{i\sigma}$) creates (annihilates) a spin-$\sigma$ electron in orbital $i$ of the nanostructure, while  $\hat{n}_{i\sigma}=d^{\dagger}_{i\sigma}d^{\phantom{\dagger}}_{i\sigma}$ and $\hat{n}_{i}=\sum_{\sigma}\hat{n}_{i\sigma}$ are number operators. In Sec.~\ref{sec:tqd} we study a triple quantum dot model of this form.

The tunnel-coupling between nanostructure and leads is described by the hybridization Hamiltonian,
$H_{\rm hyb}=\sum_{\alpha,k,i,\sigma} (V_{\alpha,ik} d^{\dagger}_{i\sigma} c_{\alpha k \sigma}^{\phantom{\dagger}}  + \text{H.c.} ) $. 
Hereafter we shall take a common simplification of this expression to the \emph{two-channel} limit in which the tunneling matrix elements can be factorized as $V_{\alpha,ik}=V_{\alpha,i} U_{\alpha,k}$ with $\sum_k | U_{\alpha,k}|^2=1$. In this case, we define real-space lead orbitals localized at the nanostructure position, given by $c_{\alpha\sigma}=\sum_{k} U_{\alpha,k} c_{\alpha k \sigma}^{\phantom{\dagger}}$. The hybridization Hamiltonian then becomes,
\begin{eqnarray}\label{eq:Hhyb}
H_{\rm hyb}=\sum_{\alpha,i,\sigma}\left (V_{\alpha,i} d^{\dagger}_{i\sigma} c_{\alpha  \sigma}^{\phantom{\dagger}}  + \text{H.c.} \right ) \;.
\end{eqnarray}
This two-channel limit is a very good approximation when the dominant nanostructure-lead hybridization is localized, and may be justified by the contacting geometry of the device. 

We can now identify the local retarded Green's function of the \emph{uncoupled} leads at the nanostructure position. It is given by $\mathcal{G}_{\alpha\beta}^0(\omega) \equiv \langle\langle c_{\alpha\sigma}^{\phantom{\dagger}} ;c_{\beta\sigma}^{\dagger} \rangle\rangle^0_{\omega} = \delta_{\alpha\beta}\sum_k |U_{\alpha,k}|^2 (\omega^+-\epsilon_k)^{-1}$, where $\omega^+=\omega+i0^+$. We take for simplicity a flat density of states in equivalent leads,  $\rho_{\alpha}(\omega)=-\tfrac{1}{\pi}\text{Im}~\mathcal{G}_{\alpha\alpha}^0(\omega)=\rho_0\Theta(D-|\omega|)$ independent $\alpha$, inside a band of half-width $D$ and with $\rho_0=1/(2D)$.

The `frontier orbitals' of the nanostructure hybridizing with the leads can also be identified as
\begin{equation}
\label{eq:frotnier}
\bar{d}_{\alpha\sigma}^{\dagger}=\tfrac{1}{V_{\alpha}}\sum_i V_{\alpha,i} d_{i\sigma}^{\dagger} \; ,
\end{equation}
with $V_{\alpha}^2=\sum_{i}|V_{\alpha,i}|^2$, which yields,
$H_{\rm hyb}=\sum_{\alpha,\sigma} (V_{\alpha} \bar{d}^{\dagger}_{\alpha\sigma} c_{\alpha  \sigma}^{\phantom{\dagger}}  + \text{H.c.} )$. 
However, note that in general  $\{\bar{d}_{s\sigma}^{\dagger},\bar{d}^{\phantom{\dagger}}_{d\sigma}\} \ne 0$ due to overlap between the frontier orbitals coupling to source and drain leads, and hence these operators are not proper canonical fermions (for example $\bar{d}_{s\sigma}=\bar{d}_{d\sigma}\equiv d_{\sigma}$ in the case of the single-orbital nanostructure\cite{hewson1997kondo}).

The quantum dynamics of the lead-coupled nanostructure depend on the complex hybridization matrices $\boldsymbol{\Delta}_{\alpha}(\omega)$, whose elements are the individual hybridization functions $\Delta_{\alpha,ij}(\omega)=V_{\alpha,i}(V_{\alpha,j})^* \mathcal{G}_{\alpha\alpha}^0(\omega)$. The low-energy physics of such nanoelectronics devices,\cite{eickhoff2020strongly} including the low-temperature transport properties, are controlled by the low-frequency form of $\boldsymbol{\Delta}_{\alpha}(\omega)$. We therefore introduce the matrix $\boldsymbol{\Gamma}_{\alpha}=-\text{Im}~\boldsymbol{\Delta}_{\alpha}(\omega=0)$, with elements $ \Gamma_{\alpha,ij}=\pi\rho_0 V_{\alpha,i}(V_{\alpha,j})^*$, and we define $\Gamma_{\alpha}=\pi\rho_0 |V_{\alpha}|^2$.

Physical properties of the nanodevice so defined may be tuned in-situ by application of suitable gate voltages. These may provide experimental control \cite{Kastner_Artificialatoms1994,goldhaber1998kondo} over the tunneling matrix elements $V_{\alpha,i}$ entering in the hybridization, or allow to manipulate the nanostructure single-particle level energies through a term $H_{\rm gate} = eV_{\rm g} \sum_i \hat{n}_i$. Transport properties are also strongly affected by the application of a local magnetic field, modelled by the term $H_{\rm field}$.

%%%%%%%%%%%%%%

\subsection{Proportionate Coupling}\label{sec:PC}
A system is said to be under `proportionate coupling' (PC) if $\Gamma_{d,ij}=\lambda \Gamma_{s,ij}$ (or equivalently $V_{d,i} = \sqrt{\lambda}V_{s,i}$), with $\lambda$ some real constant, independent of the nanostructure level indices $i$ and $j$. \cite{meir1992landauer,*MeirWingreen_AMoutEquilib_PRL1993,jauho1994time} Evidently, this is not the generic scenario for a multi-orbital nanostructure with spatially separated leads. 

A system under PC simplifies to an effective \emph{single-channel} description, with 
\begin{eqnarray}\label{eq:Hhyb_PC}
H_{\rm hyb} ~\overset{\text{PC}}{\longrightarrow}~  \sum_{i,\sigma}\left (V_i^{\phantom{\dagger}} d^{\dagger}_{i\sigma} c_{e  \sigma}^{\phantom{\dagger}}  + \text{H.c.} \right )
\end{eqnarray}
where only the `even' combination of source and drain leads couples to the nanostructure, with $c_{e\sigma}=\tfrac{1}{\sqrt{1+\lambda}}(c_{s\sigma}+\sqrt{\lambda}c_{d\sigma})$ and $V_i=V_{s,i}\sqrt{1+\lambda}$. The `odd' combination of leads, defined by $c_{o\sigma}=\tfrac{1}{\sqrt{1+\lambda}}(\sqrt{\lambda} c_{s\sigma} -c_{d\sigma})$, is formally decoupled on the level of the bare Hamiltonian (we contrast with with the `emergent decoupling' of a channel in Secs.~\ref{sec:emPC_CB} and \ref{sec:emPC_MV}). However, note that quantum transport still proceeds in the physical real-space basis of source and drain leads.

In this case, one can also define a single linear combination of nanostructure operators, the `PC frontier orbital' $\bar{\bar{d}}_{\sigma}^{\dagger}$, to which the even channel couples, given by
\begin{equation}
\label{eq:PCforntier}
\bar{d}_{\alpha\sigma}^{\dagger}  ~\overset{\text{PC}}{\longrightarrow}~ \bar{\bar{d}}_{\sigma}^{\dagger}=\tfrac{1}{V}\sum_i V_i d_{i\sigma}^{\dagger} \; , 
\end{equation}
with $V^2=\sum_i |V_i|^2$ (and $\Gamma = \pi\rho_0 V^2$). The hybridization Hamiltonian then takes the same form as that of a single-channel, single-impurity model, 
\begin{equation}
H_{\rm hyb} ~\overset{\text{PC}}{\longrightarrow}~  V\sum_{\sigma} (\bar{\bar{d}}_{\sigma}^{\dagger} c_{e\sigma}^{\phantom{\dagger}} + \rm{H.c.} )    
\end{equation}
However, this is somewhat deceptive, since the other nanostructure operators $d_{i\sigma}^{\dagger}$ are still coupled to $\bar{\bar{d}}_{\sigma}^{\dagger}$ via $H_{\rm nano}$. Therefore one cannot in general determine the dynamics or transport properties of a multi-orbital system from a pure single impurity Anderson model, despite the form of Eq.~\ref{eq:Hhyb_PC}. 

The PC concept generalizes to systems with 2 or more leads with arbitrary densities of states, for which the condition on the hybridization matrices reads $\boldsymbol{\Delta}_{\alpha}(\omega) = \lambda_{\alpha} \boldsymbol{\Delta}(\omega)$, with $\lambda_{\alpha}$ a real scalar and $\boldsymbol{\Delta}(\omega)$ equal for all leads. Note that the PC condition can be satisfied only with equivalent leads, where $\rho_{\alpha}(\omega)\equiv \rho(\omega)$. The case of inequivalent leads is considered in Appendix~\ref{app:inequiv}.

%%%%%%%%%%%%%%

\subsection{Anderson impurity model}\label{sec:AIM}
The simplest nontrivial interacting nanostructure model in terms of which quantum transport can be studied, comprises a single interacting site tunnel-coupled to two leads -- the Anderson impurity model (AIM),\cite{hewson1997kondo}~~\cite{Anderson1961,*Anderson_LocalisedMoments1978}
\begin{equation}\label{eq:aim}
H_{\rm AIM} = H_{\rm leads}+\epsilon_d (\hat{n}_{\uparrow} +
\hat{n}_{\downarrow}) + U_d\hat{n}_{\uparrow}\hat{n}_{\downarrow}+
\sum_{\alpha,\sigma}\left ( V_{\alpha}^{\phantom{\dagger}} d^{\dagger}_{\sigma} c_{\alpha \sigma}^{\phantom{\dagger}} + \rm{H.c.} \right )
\end{equation}
where $\hat{n}_{\sigma}=d_{\sigma}^{\dagger}d_{\sigma}$ and $\alpha=s,d$. The model automatically satisfies the PC condition, and so reduces to a single-channel model,
\begin{equation}\label{eq:aim1ch}
H_{\rm AIM} = H_{\rm leads}^e+\epsilon_d (\hat{n}_{\uparrow} +
\hat{n}_{\downarrow}) + U_d\hat{n}_{\uparrow}\hat{n}_{\downarrow}+
V\sum_{\sigma}\left ( d^{\dagger}_{\sigma} c_{e \sigma}^{\phantom{\dagger}} + \rm{H.c.} \right )
\end{equation}
where $H_{\rm leads}^e$ describes the even lead, coupled to the impurity site at the local orbital $c_{e\sigma}$, and $V^2=V_s^2+V_d^2$, such that $\Gamma=\pi\rho_0 V^2$ is the total hybridization. The focus of this paper is however multi-orbital two-channel systems.

%%%%%%%%%%%%%%

\subsection{Green's functions and t-matrix}\label{sec:GFs}
The retarded Green's functions for the interacting, lead-coupled nanostructure at equilibrium in Zubarev \cite{zubarev1960double} notation are specified by $G_{ij,\sigma}(\omega,T) \equiv \langle\langle d_{i\sigma}^{\phantom{\dagger}} ; d_{j\sigma}^{\dagger} \rangle\rangle_{\omega,T}$. These are the Fourier transforms of the time-domain retarded correlators $G_{ij,\sigma}(t,T)=-i\theta(t)\langle \{ d_{i\sigma}^{\phantom{\dagger}} , d_{j\sigma}^{\dagger}(t) \} \rangle_T$. The corresponding non-interacting Green's functions, $G_{ij,\sigma}^{(0)}(\omega)$, are defined for $H_{\rm int}=0$ and are temperature independent. These are related through the Dyson equation by the interaction self-energy,
\begin{eqnarray}\label{eq:dyson}
[\boldsymbol{G}_{\sigma}(\omega,T)]^{-1} = [\boldsymbol{G}_{\sigma}^{(0)}(\omega)]^{-1} - \boldsymbol{\Sigma}_{\sigma}(\omega,T) \;, 
\end{eqnarray}
where $[\boldsymbol{G}_{\sigma}^{(0)}(\omega)]^{-1} = (\omega^+-eV_g +B\sigma)\textbf{I}-\textbf{H}_{\rm nano}^{(0)}-\sum_{\alpha}\boldsymbol{\Delta}_{\alpha}(\omega)$ with $\sigma=\pm\tfrac{1}{2}$ for $\uparrow/\downarrow$.

The local t-matrix equation for the system is \cite{hewson1997kondo},
\begin{eqnarray}\label{eq:tmatrix}
\mathcal{G}_{\alpha\beta,\sigma}(\omega,T) = \mathcal{G}_{\alpha\beta}^0(\omega) +\mathcal{G}_{\alpha\alpha}^0(\omega) T_{\alpha\beta,\sigma}(\omega,T) \mathcal{G}_{\beta\beta}^0(\omega) \;,\;\;\;
\end{eqnarray}
with the t-matrix itself defined as $T_{\alpha\beta,\sigma}(\omega,T)= \sum_{i,j} (V_{\alpha,i})^*V_{\beta,j}G_{ij,\sigma}(\omega,T)
\equiv V_{\alpha}^*V_{\beta} 
 \overline{G}_{\alpha\beta,\sigma}(\omega,T)$, where the latter is given in terms of the frontier orbital Green's function $\overline{G}_{\alpha\beta,\sigma}=\langle\langle \bar{d}_{\alpha\sigma}^{\phantom{\dagger}} ; \bar{d}_{\beta\sigma}^{\dagger} \rangle\rangle$. 
 
 In proportionate coupling, this simplifies further to $T_{\alpha\beta,\sigma}(\omega,T)=V_{\alpha}^*V_{\beta}\overline{\overline{G}}_{\sigma}(\omega,T)$ with $\overline{\overline{G}}_{\sigma}=\langle\langle \bar{\bar{d}}_{\sigma}^{\phantom{\dagger}} ; \bar{\bar{d}}_{\sigma}^{\dagger} \rangle\rangle$. We may also write a t-matrix equation in the even/odd channel basis, 
 $\mathcal{G}_{ee,\sigma}(\omega,T) = \mathcal{G}_{ee}^0(\omega) +[\mathcal{G}_{ee}^0(\omega)]^2 T_{ee,\sigma}(\omega,T)$, where $T_{ee,\sigma}(\omega,T)=V^2 \overline{\overline{G}}_{\sigma}(\omega,T)$ and $T_{oo,\sigma}(\omega,T)= 0$ in PC.

%%%%%%%%%%%%%%

\subsection{Observables}\label{sec:obs}
We consider the electrical and heat conductances due to a voltage bias $\Delta V_b$ and/or temperature gradient $\Delta T$ between leads.

The electrical current 
$I_C^{\alpha} = -eI_N^{\alpha}$ into lead $\alpha=s,d$ is given in terms of the particle current $I_N^{\alpha} =\langle \dot{N}^{\alpha} \rangle$. Here, $\dot{\Omega}=\tfrac{d}{dt}\hat{\Omega}$ denotes the time derivative of an operator, and  $\hat{N}^{\alpha}=\sum_{k,\sigma} c_{\alpha k \sigma}^{\dagger} c_{\alpha k \sigma}^{\phantom{\dagger}}$ is the total number operator for lead $\alpha$. 

Similarly, the energy current is defined as $I_E^{\alpha}=\langle \dot{H}_{\rm leads}^{\alpha}\rangle$, where the Hamiltonian for lead $\alpha$ is given in Eq.~\ref{eq:Hleads}. From the first law of thermodynamics, $dE=dQ+\mu dN$, the heat current operator follows as $I^{\alpha}_{Q}=I^{\alpha}_E-\mu_{\alpha} I^{\alpha}_{N}$, where $\mu_{\alpha}$ is the chemical potential of lead $\alpha$. Note that all thermal expectation values $\langle \dots \rangle$ must be calculated in the full lead-coupled system. 

Due to current conservation in the two-channel setup, $I^d=-I^s\equiv I$, and so we drop the lead index $\alpha$ hereafter.

The bias voltage is related to a chemical potential difference between leads, $e\Delta V_b=\mu_s-\mu_d$. For $\Delta V_b >0$, electrical current flows from source to drain. In particular, an ac bias can be incorporated on the level of the Hamiltonian by adding a time-dependent source term, $H\to H + H_{\rm bias}(t)$ with,
\begin{eqnarray}\label{eq:Hbias}
H_{\rm bias}(t) =-e\Delta V_b\cos(\omega t) \hat{N}_s \;
\end{eqnarray}
where $\omega$ is the ac driving frequency. The dc limit is obtained as $\omega \to 0$.

In contrast, the effect of a temperature gradient $\Delta T= T_s-T_d$ cannot be incorporated by adding a source term to the Hamiltonian. The temperature of the leads enters through their respective distribution functions $f_{\alpha}(\omega)$. 

In this paper we will focus on the linear response regime, $\Delta V_b\to 0$ and $\Delta T \to 0$, in which case the nanostructure remains close to equilibrium, and quantum transport coefficients can be calculated from equilibrium correlation functions. We expand\cite{luttinger} the electrical and heat currents to first order in $\Delta V_{b}$ and $\Delta T$ to obtain, 
\begin{equation}
\begin{pmatrix}I_C\\I_Q\end{pmatrix}=\begin{pmatrix}\chi_{CC} & \chi_{CQ}\\ \chi_{QC} & \chi_{QQ}\end{pmatrix}\begin{pmatrix}\Delta V_b\\ \Delta T/T\end{pmatrix} \;,\label{eq:offdiag}
\end{equation}
where the susceptibilities $\chi$ are related to the desired transport coefficients. 
With $I_C = G_C \Delta V_b$ defined at $\Delta T = 0$, and $I_Q = -\kappa\Delta T$ defined at
$I_C = 0$, we identify the electrical conductance $G_C=\chi_{CC}$ and the heat conductance $\kappa =(\chi_{CQ}\chi_{QC}/\chi_{CC}-\chi_{QQ} )/T$. We also define the heat conductance at $\Delta V_b=0$ as $K_Q=-\chi_{QQ}/T$. 
Similarly, $\Delta V_b=-S\Delta T$ measured at $I_C=0$ defines the thermopower (Seebeck coefficient) $S=\chi_{CQ}/T\chi_{CC}$. In the following we denote the charge and heat conductance quanta as $G_0=e^2/h$ and $K_0=1/h$.

The Lorenz ratio $L=\kappa / TG_C$ plays an important role in understanding the nature of thermoelectric transport in nanoelectronic devices. In particular, quantum transport in the Fermi liquid regime typically satisfies the Wiedemann-Franz (WF) law,\cite{franz1853ueber,costi2010thermoelectric} in which the Lorenz ratio as $T\to 0$ takes the special value $L_0=\pi^2 k_{\rm B}^2/3e^2$. This is expected when the carriers of both charge and heat are the same fermionic quasiparticles (or simply the bare electrons themselves). 
Although violations of the WF law are often associated with non-Fermi liquid physics in bulk systems,\cite{mahajan2013non} in the context of mesoscopic quantum devices, violations are known in Fermi liquid systems\cite{vavilov2005failure,kubala2008violation,bergfield2009thermoelectric} as well as in non-Fermi liquids.\cite{buccheri2022violation} Counter-intuitively the WF law can also be \textit{satisfied} in certain non-Fermi liquid systems.\cite{van2020wiedemann,*van2020electric,karki2020quantum} This highlights the richer thermoelectric capabilities of complex nanostructures. 

Another important quantity is the Johnson-Nyquist noise, which describes the equilibrium fluctuations of the current in time.\cite{johnson1928thermal,*nyquist1928thermal} Its frequency-resolved spectrum is defined as,
\begin{equation}\label{eq:noise}
C(\omega)=e^2\int_{-\infty}^{\infty} dt~e^{i\omega t}\left [ \langle \dot{N}_d(0)\dot{N}_d(t)\rangle - \langle \dot{N}_d\rangle^2 \right ] \;.
\end{equation}
The equilibrium current fluctuations\cite{KuboFDT1966} are given by $\Delta I_C\equiv e^2\tfrac{d}{dt}\langle\hat{N}_d-\langle \hat{N}_d\rangle\rangle^2 = 2 C(0)$.

Finally, we note that the average entropy production rate $\langle \hat{\sigma}\rangle$ can also be obtained from the particle and energy currents from the second law of thermodynamics,\cite{seifert2012stochastic}
\begin{equation}\label{eq:entprod}
\langle \hat{\sigma}\rangle = \left(\frac{\mu_s}{k_{\rm B} T_s} - \frac{\mu_d}{k_{\rm B} T_d} \right) I_N - \left( \frac{1}{k_{\rm B} T_s}-\frac{1}{k_{\rm B} T_d}\right) I_E \;\;
\end{equation}
The signal-to-noise ratio of the currents in nanoelectronic devices is bounded by the entropy production rate, according to the thermodynamic uncertainty relations (TURs).\cite{guarnieri2019thermodynamics}

%%%%%%%%%%%%%%

\subsection{NRG}\label{sec:nrg}

The gold-standard method of choice for solving equilibrium generalized quantum impurity models of the type discussed above is often considered to be Wilson's Numerical Renormalization Group (NRG).\cite{wilson1975renormalization,*bulla2008numerical}  Thermodynamical quantities and can be accurately obtained within NRG at essentially any energy scale or temperature, down to $T=0$. The numerical results for quantum transport coefficients presented in this work are obtained from dynamical correlation functions calculated using the full-density-matrix NRG approach.\cite{weichselbaum2007sum} 
An advantage of this method is that real-frequency dynamics with excellent resolution at low energies and temperatures can be obtained for strongly interacting models. A disadvantage is that the impurity part of the Hamiltonian must be diagonalized exactly, which limits the ability of NRG to treat more complex nanostructures; and two-lead calculations are computationally expensive. This motivates the development of new techniques, simplified effective models, and reusable approximate results for these models. We undertake this in Secs.~\ref{sec:newkubo}-\ref{sec:emPC_MV}.

However, we emphasize that our results are not specific to NRG and that any suitable quantum impurity solver can be used.

%###########################
%###########################

\section{Exact techniques for\\linear response}\label{sec:LR}
In this section we review some commonly used techniques for the calculation of quantum transport. We focus on the linear response regime for 2-lead models of the type introduced in Sec.~\ref{sec:intro}, and highlight the limitations of each. Here we assume that source and drain leads are metallic with a constant density of states $\rho_0$. The generalization to equivalent leads with arbitrary density of states $\rho(\omega)$ is given in Appendix~\ref{app:dos}. The multi-terminal case is discussed briefly in Appendix~\ref{app:LB}, while the situation for inequivalent leads is considered in Appendix~\ref{app:inequiv}.

%%%%%%%%%%%%%%

\subsection{Landauer}\label{sec:landauer}
For non-interacting systems (Eq.~\ref{eq:Hnano} with $H_{\rm int}=0$), the Landauer formula for the dc currents may be used \cite{Landauer1957,*Landauer1970,*LandauerButtiker1985,*Buttiker1986},
\begin{subequations}\label{eq:land_I}
    \begin{align}
I_C(T) &= \frac{e}{h} \int d\omega~[f_d(\omega)-f_s(\omega)]\mathcal{T}_{\rm L}(\omega) \;, \label{eq:IC_land}\\
I_Q(T) &= \frac{1}{h} \int d\omega~(\mu_d-\omega)[f_d(\omega)-f_s(\omega)]\mathcal{T}_{\rm L}(\omega)\;. \label{eq:IQ_land}
\end{align}
\end{subequations}
 Within linear response, the transmission function $\mathcal{T}_{\rm L}(\omega)$ assumes its equilibrium form, and the Fermi functions $f_{\alpha}(\omega)$ can be expanded to first order in $\Delta V_b$ and $\Delta T$. This yields the susceptibilities $\chi_{CC}=e^2 \Phi_0$, $\chi_{CQ}=\chi_{QC}=e\Phi_1$ and $\chi_{QQ}=\Phi_2$, in terms of the transport integrals, $\Phi_n(T) = h^{-1} \int d\omega~\omega^n [-\partial_{\omega} f_{\rm eq}(\omega)  ]\mathcal{T}_{\rm L}(\omega) $, where 
$f_{\rm eq}(\omega)=(1+\exp[\omega/k_{\rm B} T])^{-1}$ is the equilibrium Fermi function (and its derivative is $-\partial_{\omega} f_{\rm eq}(\omega) = \text{sech}[\omega/(2 T)]^2/(4 T)$). 
In particular, the electrical and heat conductances are,
\begin{subequations}\label{eq:land}
\begin{align}
G_C(T) &= \frac{e^2}{h} \int d\omega~ [-\partial_{\omega} f_{\rm eq}(\omega)  ]\mathcal{T}_{\rm L}(\omega) \;, \label{eq:G_land}\\
K_Q(T) &= \frac{1}{hT} \int d\omega~ \omega^2 [-\partial_{\omega} f_{\rm eq}(\omega)  ]\mathcal{T}_{\rm L}(\omega) \;, \label{eq:K_land}
\end{align}
\end{subequations}
such that $G_C(T\to 0)=(e^2/h) \mathcal{T}_{\rm L}(0)$ and $K_Q(T\to 0) = (\pi^2 k_{\rm B}^2T/3h) \mathcal{T}_{\rm L}(0)$ provided the transmission function remains finite as $\omega \to 0$. Note also that $\Phi_1(T)=0$ under particle-hole symmetry $\mathcal{T}_{\rm L}(\omega)=\mathcal{T}_{\rm L}(-\omega)$, such that $\kappa=K_Q$. Under these rather typical conditions, $\kappa/TG_C=L_0$ and the Wiedemann-Franz law is automatically satisfied, independently of microscopic details. In the marginal case $\mathcal{T}_{\rm L}(\omega)\sim 1/\ln^2|\omega|$ characteristic of spin-flip scattering, the Lorenz ratio picks up logarithmic temperature corrections to the Wiedemann-Franz limit, $L-L_0\sim a/\ln|bT|$. The Lorenz ratio is typically strongly enhanced at a transmission node; for example $L=9\zeta(3) k_{\rm B}^2/(\ln(4)e^2)$ for $\mathcal{T}_{\rm L}(\omega)\sim |\omega|$ and $L=7\pi^2 k_{\rm B}^2/(5e^2)$ for $\mathcal{T}_{\rm L}(\omega)\sim \omega^2$. 

The transmission function can be expressed in the Caroli \cite{Caroli1970} form in terms of Green's functions for the lead-coupled (but non-interacting) nanostructure and the hybridization matrices,\cite{FisherLeeRelation1981,*Beenakker_RandomMatrix1997}
\begin{subequations}\label{eq:land_T}
\begin{align}
\mathcal{T}_{\rm L}(\omega)&=4\sum_{\sigma} {\rm Tr}[\boldsymbol{G}_{\sigma}^{(0)}(\omega)\boldsymbol{\Gamma}^{d}\boldsymbol{G}_{\sigma}^{(0)}(\omega)^* \boldsymbol{\Gamma}^{s}] \label{eq:caroli}\\
&\equiv 4\Gamma_d\Gamma_s\sum_{\sigma} \overline{G}^{(0)}_{sd,\sigma}(\omega)\overline{G}^{(0)}_{ds,\sigma}(\omega)^* \;, \label{eq:caroli2}
\end{align}
\end{subequations}
where the simplified form of Eq.~\ref{eq:caroli2} follows from the definition of the non-interacting frontier orbital Green's functions  $\overline{G}_{\alpha\beta,\sigma}^{(0)}=\langle\langle \bar{d}_{\alpha\sigma}^{\phantom{\dagger}} ; \bar{d}_{\beta\sigma}^{\dagger} \rangle\rangle^{(0)}$ and the corresponding hybridizations $\Gamma_{\alpha}=\pi\rho_0 |V_{\alpha}|^2$. Note that the transmission function in the non-interacting limit $H_{\rm int}=0$ is seen from this expression to be temperature-independent (since non-interacting retarded Green's functions are temperature-independent). Furthermore, in the absence of a magnetic field (whence we have time-reversal symmetry), the expression simplifies further to $\mathcal{T}_{\rm L}(\omega)=8\Gamma_d\Gamma_s |\overline{G}_{ds,\uparrow}^{(0)}(\omega)|^2$. This has a physically intuitive interpretation: combined with Eqs.~\ref{eq:land}, the expressions relate the conductance to the probability amplitude for electrons to propagate across the nanostructure.

Finally, we note that under proportionate coupling (see Sec.~\ref{sec:PC}), Eq.~\ref{eq:caroli2} reduces to,
\begin{eqnarray}\label{eq:TlandauerPC}
\mathcal{T}_{\rm L}(\omega) ~\overset{\text{PC}}{\longrightarrow}~ 4 \Gamma_d\Gamma_s\sum_{\sigma} |\overline{\overline{G}}_{\sigma}^{(0)}(\omega)|^2 
\end{eqnarray}
where $\overline{\overline{G}}_{\sigma}^{(0)} =\langle\langle \bar{\bar{d}}_{\sigma}^{\phantom{\dagger}} ; \bar{\bar{d}}_{\sigma}^{\dagger} \rangle\rangle^{(0)}$ is the local non-interacting nanostructure Green's function. 
In the special case of a non-interacting single-orbital nanostructure (resonant level model),\cite{hewson1997kondo} $|\overline{\overline{G}}_{\sigma}^{(0)}(\omega)|^2 \simeq [(\omega-\epsilon_d)^2 + (\Gamma_s+\Gamma_d)^2]^{-1}$ takes the form of a Lorentzian of width $\Gamma_s+\Gamma_d$, centered on the level position $\epsilon_d$.

These results can be generalized to two or more inequivalent leads with arbitrary densities of states (Appendices~\ref{app:LB} and \ref{app:inequiv}). The key constraint for the standard Landauer approach, however, is the non-interacting limit.

%%%%%%%%%%%%%%%%

\subsection{Renormalized Landauer}\label{sec:oguri}
As first discussed by Oguri,\cite{Oguri_FermiLiquidTheoryConductance1997,*Oguri_TransmissionProb_2001,*Oguri_QuasiParticlePRB2001} all the results of the previous section can be immediately applied to interacting systems in the Fermi liquid regime at low temperatures $T\to 0$. One simply replaces \cite{HewsonOguri_RenormalizedImp_2004} the non-interacting nanostructure Green's functions by their interacting counterparts, $G_{ij,\sigma}^{(0)} \to G_{ij,\sigma}$. The Green's functions are evaluated at the Fermi energy $\omega=0$ because the $T\to 0$ limit of Eqs.~\ref{eq:land} picks out the zero frequency value of the transmission function, $\mathcal{T}_{\rm L}(0)$. 

For a Fermi liquid, the imaginary part of the interaction self-energy vanishes at low temperatures and energies \cite{nozieres1974fermi},
\begin{eqnarray}\label{eq:FLcondition}
\text{Im}~\boldsymbol{\Sigma}_{\sigma}(\omega = 0, T= 0) ~\overset{\text{FL}}{\longrightarrow}~  0
\end{eqnarray}
while the real part can be viewed as renormalizing the single particle nanostructure Hamiltonian,
\begin{eqnarray}
\textbf{H}_{\rm nano,\sigma} = \textbf{H}_{\rm nano}^{(0)} + \text{Re}~\boldsymbol{\Sigma}_{\sigma}(0,0) \;.
\end{eqnarray}
This follows from the form of the Dyson equation, Eq.~\ref{eq:dyson}. At $T=0$ and $\omega=0$, the Green's functions of the interacting system are the same as those of a system with $H_{\rm int}=0$ but with a renormalized single-particle Hamiltonian $\textbf{H}_{\rm nano,\sigma}$.\cite{HewsonOguri_RenormalizedImp_2004}  Quantum transport properties of the renormalized non-interacting system can then be obtained within the Landauer formalism. We emphasize that the form of the renormalized non-interacting system need not be found explicitly: only the Green's functions $G_{ij,\sigma}(0,0)$ of the interacting system are required.

The electrical conductance through a generic nanostructure in the Fermi liquid regime as $T \to 0$ therefore follows as,
\begin{eqnarray}\label{eq:G_oguri}
G_C(0)=\left(\frac{e^2}{h}\right)
4\Gamma_d\Gamma_s\sum_{\sigma} \overline{G}_{sd,\sigma}(0,0)\overline{G}_{ds,\sigma}(0,0)^* \;,\qquad
\end{eqnarray}
and in the case of proportionate coupling,
\begin{eqnarray}\label{eq:G_oguri_pc}
G_C(0) ~\overset{\text{PC}}{\longrightarrow}~ \left(\frac{e^2}{h}\right)
4\Gamma_d\Gamma_s\sum_{\sigma} |\overline{\overline{G}}_{\sigma}(0,0)|^2 \;.
\end{eqnarray}

The above equations can be reformulated in terms of the scattering t-matrix. Using current conservation, one may write $\mathcal{T}_{\rm L}(0,0)=(2/\pi\rho_0)^2\sum_{\sigma}\mathcal{G}_{sd,\sigma}(0,0)\mathcal{G}_{ds,\sigma}(0,0)^*$, and thence from Eq.~\ref{eq:tmatrix} we find,
\begin{eqnarray}\label{eq:oguri_tm}
G_C(0)=\left(\frac{e^2}{h}\right)
(2\pi\rho_0)^2\sum_{\sigma}T_{sd,\sigma}(0,0)T_{ds,\sigma}^*(0,0) \;,\qquad 
\end{eqnarray}
or in the case of time-reversal symmetry ($B=0$), $G_C(0)=\tfrac{2e^2}{h} \times 4 |\pi\rho_0 T_{ds,\uparrow}(0,0)|^2$. These expressions allow the conductance to be obtained from standard diagrammatic techniques for the scattering t-matrix \cite{Yamada1286AMPertTMat,*Yamada316ImSelfDiag}, or from Fermi liquid theory. \cite{nozieres1974fermi,nozieres1980kondo}

Similar expressions can be obtained for the heat conductance. Alternatively, since the approach is only valid for Fermi liquids in the low-temperature limit, the heat conductance may be obtained from the electrical conductance via the Wiedemann-Franz law, $\kappa / T G_C =L_0$ (assuming $\mathcal{T}_{\rm L}(0,0)\ne 0$ as above).

We note that the renormalized Landauer formulation only holds for systems described by Fermi liquid theory at low temperatures and energies. Genuine non-Fermi liquid states\cite{logan2014common} (characterized by non-vanishing $\text{Im}~\Sigma(0,0)$ and fractional residual entropy), arising due to a frustration of Kondo screening for example, can also be engineered in quantum dot devices, and alternative approaches must then be used. However, we emphasize here that the Fermi liquid scenario is very much the standard one. Indeed, even in cases where the nanostructure hosts free local moments as $T\to 0$ (so-called `singular Fermi liquids'), arising due to ferromagnetic interactions or Kondo underscreening for example, Eq.~\ref{eq:FLcondition} still holds asymptotically.\cite{VarmaSingularNFL2002}

Rather, the key limitation of the renormalized Landauer approach is the $T\to 0$ limit. This may be particularly problematical when comparing with results of experiments performed at base temperatures larger than emergent energy scales (such as the Kondo temperature). For example, the general considerations of Sec.~\ref{sec:landauer} suggest that the thermoelectric Lorenz ratio for standard quantum dot devices should saturate to $L_0=\pi^2 k_{\rm B}^2T/3h$ as $T \to 0$. However, experiments often appear to show non-Fermi liquid signatures \cite{seaman1991evidence,*ralph19942} because the low-temperature limit is not reached in practice, and electron-electron interactions give a finite $\text{Im}~\Sigma(\omega,T)$.

%%%%%%%%%%%%%%

\subsection{Meir-Wingreen}\label{sec:MW}
A general framework for calculating electrical conductance through generic interacting nanostructures was devised by Meir and Wingreen (MW).\cite{meir1992landauer,*MeirWingreen_AMoutEquilib_PRL1993,jauho1994time} The approach was subsequently extended to thermoelectric transport by Costi and Zlati{\'c}.\cite{costi2010thermoelectric} The dc electrical and energy currents into the drain lead are given by,
\begin{subequations}\label{eq:MW_I}
\begin{align}
I_C(T) &= \frac{e}{h}\int d\omega~ X(\omega,T) \;,\label{eq:MW_IC} \\
I_Q(T) &= \frac{1}{h}\int d\omega~(\mu_d-\omega)X(\omega,T) \;,\label{eq:MW_IQ} 
\end{align}
\end{subequations}
where the kernel 
\begin{equation}
X(\omega,T) =-\text{Im}~\text{Tr}\sum_{\sigma}\boldsymbol{\Gamma}^d \left ( \boldsymbol{G}_{\sigma}^<(\omega,T) + 2 f_d(\omega)\boldsymbol{G}_{\sigma}(\omega,T) \right) \;,\label{eq:MW_X}
\end{equation}
now involves the nanostructure lesser Green's functions, $\boldsymbol{G}_{\sigma}^<(\omega,T)$. The above equations hold in the general non-equilibrium situation at finite bias $\Delta V_b$ and temperature gradient $\Delta T$. However, in this case the full non-equilibrium retarded and lesser Green's functions must be obtained for the interacting problem. Since this is typically deeply nontrivial,\cite{Dias_noPC2017} we again focus on the linear response regime where the nanostructure remains close to equilibrium. But even in this limit, the explicit first-order variation of the Green's functions with respect to $\Delta V_b$ and $\Delta T$ is required to calculate the conductances $G_C$ and $K_Q$. In particular, $\partial_{\Delta V_b}\boldsymbol{G}_{\sigma}^< \vert_{\Delta V_b=0}$ and $\partial_{\Delta T}\boldsymbol{G}_{\sigma}^< \vert_{\Delta T=0}$ are not accessible using standard equilibrium techniques.

An important simplification arises in the case of proportionate coupling (Sec.~\ref{sec:PC}), since then the lesser Green's functions in Eq.~\ref{eq:MW_X} can be eliminated to yield the Landauer form,
\begin{subequations}\label{eq:MW_PC}
\begin{align}
X(\omega,T) &~\overset{\text{PC}}{\longrightarrow}~ \left [ f_d(\omega) - f_s(\omega)\right ] \mathcal{T}_{\rm MW}(\omega,T) \label{eq:MW_PC_X}\\
\mathcal{T}_{\rm MW}(\omega,T) &= \frac{4\Gamma_d\Gamma_s}{\Gamma_d+\Gamma_s} \sum_{\sigma}\left [-\text{Im}~\overline{\overline{G}}_{\sigma}(\omega,T)\right]
\label{eq:MW_PC_T}
\end{align}
\end{subequations}

Importantly, the Landaurer form of Eq.~\ref{eq:MW_PC_X} in proportionate coupling implies that the conductances in linear response can be obtained using Eqs.~\ref{eq:G_land} and \ref{eq:K_land}, but with $\mathcal{T}_{\rm L}(\omega) \to \mathcal{T}_{\rm MW}(\omega,T)$ given now by Eqs.~\ref{eq:MW_PC_T}. This is because the leading linear variations in $\Delta V_b$ and $\Delta T$ of the currents $I_C$ and $I_Q$ are now carried solely by the lead Fermi functions, and not the nanostructure transmission function (first order variations in $\Delta V_b$ and $\Delta T$ of $\mathcal{T}_{\rm MW}(\omega,T)$ appear only to quadratic order in the currents). Linear response conductances in proportionate coupling can therefore be obtained for interacting nanostructures at finite temperatures using only \emph{equilibrium} retarded nanostructure Green's functions,
\begin{subequations}\label{eq:MW_cond}
\begin{align}
G_C(T) &~\overset{\text{PC}}{\longrightarrow}~ \frac{e^2}{h} \int d\omega~ [-\partial_{\omega} f_{\rm eq}(\omega)  ]\mathcal{T}_{\rm MW}(\omega,T) \;, \label{eq:MW_G}\\
K_Q(T) &~\overset{\text{PC}}{\longrightarrow}~ \frac{1}{hT} \int d\omega~ \omega^2 [-\partial_{\omega} f_{\rm eq}(\omega)  ]\mathcal{T}_{\rm MW}(\omega,T) \;, \label{eq:MW_K}
\end{align}
\end{subequations}
with the full energy- and temperature-dependent transmission $\mathcal{T}_{\rm MW}(\omega,T)$ here given by Eq.~\ref{eq:MW_PC_T}. Analogous expressions can be obtained for the cross susceptibility $\chi_{CQ}=\chi_{QC}$ in Eq.~\ref{eq:offdiag}.

Further insight is gained by expressing the transmission function in terms of the t-matrix,
\begin{eqnarray}
\mathcal{T}_{\rm MW}(\omega,T) = \frac{4\Gamma_d\Gamma_s}{(\Gamma_d+\Gamma_s)^2}\sum_{\sigma} t_{ee,\sigma}(\omega,T) \label{eq:MW_PC_TM}
\end{eqnarray}
where $t_{ee,\sigma}(\omega,T)=-\pi 
\rho_0\text{Im}T_{ee,\sigma}(\omega,T)\equiv (\Gamma_d+\Gamma_s)[- \text{Im}~\overline{\overline{G}}_{\sigma}(\omega,T)]$ is the spectrum of the even channel t-matrix.
The dimensionless geometric factor $4\Gamma_d\Gamma_s/(\Gamma_d+\Gamma_s)^2$ in Eq.~\ref{eq:MW_PC_TM} describes the relative coupling to source and drain leads, and takes its maximal value of 1 in the symmetric limit  $\Gamma_s=\Gamma_d$. The spectrum of the t-matrix can take values $0\le t_{ee,\sigma} \le 1$ only. The lower bound arises because $t_{ee,\sigma}$ is proportional to the local nanostructure density of states $A_{\sigma}=-\tfrac{1}{\pi}\text{Im}~\overline{\overline{G}}_{\sigma}\ge 0$. The upper bound can be seen from the t-matrix equation (and similarly noting that $\mathcal{A}_{ee,\sigma}=-\tfrac{1}{\pi}\text{Im}~\mathcal{G}_{ee,\sigma} \ge 0$). The transmission function itself therefore satisfies $0 \le \mathcal{T}_{\rm MW} \le 2$, independently of $\omega$ and $T$. The low temperature and low energy behavior of the t-matrix can be obtained from the Friedel sum rule for Fermi liquid systems, $t_{ee,\sigma}(0,0) = \sin^2(\tfrac{\pi}{2} n_{\rm imp})$ \cite{langreth1966friedel}. Here $n_{\rm imp}$ is the thermodynamic `excess charge' of the system due to the nanostructure at $T=0$. Note however that in a local moment phase (a `singular Fermi liquid' with a residual unscreened spin on the nanostructure), this expression becomes $t_{ee,\sigma}(0,0) = \cos^2(\tfrac{\pi}{2} n_{\rm imp})$ \cite{logan2014common}. For genuine non-Fermi liquid situations, the Friedel sum rule cannot be used.\cite{logan2014common}

The electrical conductance in the low temperature limit follows as,
\begin{eqnarray}\label{eq:MW_G_T0}
G_C(0) ~\overset{\text{PC}}{\longrightarrow}~ \left (\frac{e^2}{h}\right ) \frac{4\Gamma_d\Gamma_s}{(\Gamma_d+\Gamma_s)^2}\sum_{\sigma} t_{ee,\sigma}(0,0) \;,
\end{eqnarray}
which thereby can take a maximum value of $2e^2/h$.
We note that Eq.~\ref{eq:MW_G_T0} is a generalization of Eq.~\ref{eq:G_oguri_pc}. The expressions coincide for a Fermi liquid system since $t_{ee,\sigma}(
\omega,T)= (\Gamma_d+\Gamma_s)[-\text{Im}~\overline{\overline{G}}_{\sigma}(\omega,T)]$ with  $\text{Im}~\overline{\overline{G}}_{\sigma}(0,0)=[\text{Im}~\overline{\overline{\Sigma}}_{\sigma}(0,0)-\Gamma_d-\Gamma_s]\times |\overline{\overline{G}}_{\sigma}(0,0)|^2$, and where the local self-energy $\text{Im}~\overline{\overline{\Sigma}}_{\sigma}(0,0)=0$ vanishes in a Fermi liquid by definition\cite{nozieres1974fermi}.

The above formulation in terms of the t-matrix is particularly important because it allows the low-$T$ transport to be described using simplified low-energy effective models, whose t-matrix $t_{ee,\sigma}(\omega,T)$ still provides a faithful representation of the electronic scattering in the leads induced by the nanostructure. In many cases, an effective Kondo model may be used, which significantly reduces the nanostructure complexity. This is discussed further in Sec.~\ref{sec:emPC_CB} and demonstrated explicitly for the TQD in Sec.~\ref{sec:tqd}.

Finally, we emphasize that quantum transport calculations using the MW approach require the accurate determination of nanostructure Green's functions. These follow immediately from the Dyson equation, Eq.~\ref{eq:dyson}, once the self-energies are known. In NRG the self-energy matrix can be obtained via,
\begin{equation}\label{eq:UFG}
    \boldsymbol{\Sigma}_{\sigma}(\omega,T)=[\boldsymbol{G}_{\sigma}(\omega,T)]^{-1}\boldsymbol{F}_{\sigma}(\omega,T) \;,
\end{equation}
with elements  $F_{ij,\sigma}=\langle\langle  d_{i\sigma}^{\phantom{\dagger}} ; [H_{\rm int} , d_{j\sigma}^{\dagger}] \rangle\rangle$. This generalizes the result for the AIM in Ref.~\onlinecite{bulla1998numerical}, and is found in practice to yield highly accurate results within NRG. Conductance calculated in this way for interacting PC systems reproduces exact analytical results when known,\cite{sela2011exact,*mitchell2012universal} and has been shown to agree with experimental results for a wide range of nanoelectronics devices (see e.g.~Refs.~\onlinecite{anders2008zero,roch2009observation,keller2014emergent,keller2015universal}).

For interacting systems in linear response, the main limitation of the MW formulation in terms of equilibrium retarded Green's functions is the PC requirement \cite{Dias_noPC2017}: this places constraints on the form of the hybridization for nanostructures with multiple degrees of freedom, and requires equivalent source and drain leads.

%%%

\subsubsection{Ng Ansatz}
For systems not in PC, the currents can be obtained from Eq.~\ref{eq:MW_I}, using the general form of Eq.~\ref{eq:MW_X}. This requires a knowledge of the non-equilibrium lesser Green's function matrix,  $\boldsymbol{G}^{<}_{\sigma}(\omega,T)$. From the Keldysh equation \cite{Keldysh1965,*LangrethLinearRespNEq1976} $\boldsymbol{G}^{<}_{\sigma}=\boldsymbol{G}_{\sigma}^{\phantom{<}}\boldsymbol{\Sigma}_{\sigma}^{<}\boldsymbol{G}_{\sigma}^*$ this is related to the usual retarded nanostructure Green's functions $\boldsymbol{G}_{\sigma}$ through the lesser self-energy $\boldsymbol{\Sigma}_{\sigma}^{<}$. Here $\boldsymbol{\Sigma}_{\sigma}^{<} = 
\boldsymbol{\Sigma}_{\sigma;0}^{<}+ \boldsymbol{\Sigma}_{\sigma;{\rm int}}^{<}$ where $\boldsymbol{\Sigma}_{\sigma;0}^{<} = 2i (f_s\boldsymbol{\Gamma}^s + f_d\boldsymbol{\Gamma}^d)$ involves the non-interacting hybridization , while $\boldsymbol{\Sigma}_{\sigma;{\rm int}}^{<}$ accounts for electronic interactions. Similarly, $\boldsymbol{\Sigma}_{\sigma}^{>} = \boldsymbol{\Sigma}_{\sigma;0}^{>}+ \boldsymbol{\Sigma}_{\sigma;{\rm int}}^{>}$ with $\boldsymbol{\Sigma}_{\sigma;0}^{>} = 2i ([f_s-1]\boldsymbol{\Gamma}^s + [f_d-1]\boldsymbol{\Gamma}^d)$. Unfortunately, $\boldsymbol{\Sigma}_{\sigma;{\rm int}}^{</>}$ are notoriously difficult to calculate out of equilibrium.

A simple Ansatz was proposed by Ng in Ref.~\onlinecite{ng1996ac,*Dong2002} for the non-equilibrium single-impurity Anderson model to simplify such calculations. Generalizing to the multi-orbital situation,\cite{sergueev2002spin,zhang2002spin,ferretti2005first} we suppose that the lesser and greater self-energies take a factorized form,
\begin{equation}\label{eq:Sigma_lg}
\begin{rcases*}
 \boldsymbol{\Sigma}_{\sigma}^{<}= \boldsymbol{\Sigma}_{\sigma;0}^{<}~ \boldsymbol{\Lambda}_{\sigma} \\
 \boldsymbol{\Sigma}_{\sigma}^{>}= \boldsymbol{\Sigma}_{\sigma;0}^{>}~ \boldsymbol{\Lambda}_{\sigma} \qquad
\end{rcases*} ~~ {\rm Ng~Ansatz}
\end{equation}
where $\boldsymbol{\Lambda}_{\sigma}\equiv \boldsymbol{\Lambda}_{\sigma}(\omega,T)$ is a common function to be determined. From the identity $\boldsymbol{\Sigma}_{\sigma}^{>}-\boldsymbol{\Sigma}_{\sigma}^{<}=2i{\rm Im} \boldsymbol{\Sigma}_{\sigma}$, where $\boldsymbol{\Sigma}_{\sigma}$ is the usual retarded interaction self-energy defined in Eq.~\ref{eq:dyson}, it then follows that,
\begin{equation}\label{eq:ng_lambda}
    \boldsymbol{\Lambda}_{\sigma} = \boldsymbol{{\rm I}}-[\boldsymbol{\Gamma}^s+ \boldsymbol{\Gamma}^d + 2\delta \boldsymbol{{\rm I}}]^{-1} [{\rm Im}\boldsymbol{\Sigma}_{\sigma}] \;,
\end{equation}
with $\delta\to 0^+$ for regularization. These expressions imply a specific form for the non-equilibrium distribution function matrix $\boldsymbol{f}^{NE}_{\rm int}$ which connects the lesser and retarded interaction self-energies through 
$\boldsymbol{\Sigma}^{<}_{\sigma;{\rm int}}=-2i \boldsymbol{f}_{\rm int}^{\rm NE} ~{\rm Im} \boldsymbol{\Sigma}_{\sigma}$. Specifically, the Ng Ansatz may be stated as $\boldsymbol{f}^{\rm NE}_{\rm int} \to  [f_s\boldsymbol{\Gamma}^s + f_d\boldsymbol{\Gamma}^d] [\boldsymbol{\Gamma}^s+ \boldsymbol{\Gamma}^d + 2\delta \boldsymbol{{\rm I}}]^{-1}$, which is the well-known pseudo-equilibrium distribution\cite{jauho1994time} obtained for \textit{non-interacting} systems out of equilibrium. Of course, the true distribution function is in general modified for interacting systems out of equilibrium.

Using Eqs.~\ref{eq:Sigma_lg}, \ref{eq:ng_lambda} it can be shown that the linear response electrical and heat conductances can be obtained from Eq.~\ref{eq:MW_cond}, with an \emph{effective} transmission function of Landauer-like form,
\begin{equation}\label{eq:ng_T}
    \mathcal{T}_{\rm Ng}(\omega,T)  = 4\sum_{\sigma} {\rm Tr}[\boldsymbol{G}_{\sigma}(\omega,T)\boldsymbol{\Gamma}^{d}\boldsymbol{\Lambda}_{\sigma}(\omega,T)\boldsymbol{G}_{\sigma}(\omega,T)^* \boldsymbol{\Gamma}^{s}] \;,
\end{equation}
where the retarded nanostructure Green's functions $\boldsymbol{G}_{\sigma}(\omega,T)$ are evaluated at equilibrium in the presence of interactions and coupling to the leads. Note that Eq.~\ref{eq:ng_T} is as such the analogue of Eq.~\ref{eq:caroli}.

By construction, the above Ansatz is exact in the non-interacting limit, as well as in equilibrium (no voltage bias or temperature gradient), and automatically satisfies the continuity equation implying current conservation. Indeed, since ${\rm Im}\boldsymbol{\Sigma}_{\sigma}(0,0)=0$ in a Fermi liquid system,  $\boldsymbol{\Lambda}_{\sigma}(0,0)=\boldsymbol{{\rm I}}$ from Eq.~\ref{eq:ng_lambda}, such that Eqs.~\ref{eq:MW_cond}, \ref{eq:ng_T} reduce correctly to the Oguri result, Eq.~\ref{eq:G_oguri}.

In the case of a single-orbital nanostructure, the Ng Ansatz yields $f_{\rm int}^{\rm NE} \to f^{\rm NE}_0 \equiv [f_s \Gamma_s + f_d \Gamma_d]/(\Gamma_s+\Gamma_d)$. To check the validity of the approximation, we computed the non-equilibrium lesser and retarded self-energies of the Anderson impurity model in perturbation theory to second order in the interaction $U$. We found\cite{note_EM_thesis} a non-zero correction to the above expression for $f_{\rm int}^{\rm NE}$ proportional to the voltage bias $\Delta V_b$, which implies a finite correction to the linear-response conductance obtained via Eq.~\ref{eq:ng_T} already at order $U^2$. This correction is given in Appendix~\ref{app:Ng}.

Therefore despite the simplicity and utility of the Ng Ansatz approach, it is clearly a somewhat uncontrolled approximation for interacting systems, and cannot be relied upon to yield accurate results in the general multi-orbital case. For further discussion and rigorous extensions, see Ref.~\onlinecite{ness2010generalization,Aligia2014}. In Sec.~\ref{sec:tqd} we illustrate the use of the Ng Ansatz approach for the TQD, comparing to exact results from the Kubo formula.\cite{Kubo1956,*Kubo1957,*IzumidaSakai1997} In Secs.~\ref{sec:emPC_CB}, \ref{sec:emPC_MV} we devise alternative routes to the low-$T$ conductance of non-PC systems.

%%%%%%%%%%%%%%

\subsection{Kubo}\label{sec:kubo}

The first-order correction to a thermodynamical observable from its equilibrium value due to a source term in the Hamiltonian can be computed using standard perturbation theory.\cite{hewson1997kondo} In the context of linear-response transport coefficients, the result is known as the Kubo formula.\cite{Kubo1956,*Kubo1957,*IzumidaSakai1997} The derivation for electrical conductance due to a bias voltage is straightforward, because a difference in chemical potential between source and drain leads can be incorporated as a perturbation on the level of the Hamiltonian (even though strictly speaking the chemical potential is a statistical property of leads treated as infinite thermal reservoirs at equilibrium).

We treat the ac bias $H_{\rm bias}$ in Eq.~\ref{eq:Hbias} as the perturbation, switched on adiabatically in the infinitely distant past, and calculate the resulting electrical current $I_C(\omega,T)=-e\langle \dot{N}^d\rangle$ into the drain lead at time $t=0$, to first order in the bias voltage $\Delta V_b$. Here, $\omega$ is the frequency of the ac bias voltage. The result is,
\begin{equation}\label{eq:kubo_IC}
I_C(\omega,T) = -ie^2 \Delta V_b {\rm Tr}\int_0^{\infty}e^{-\eta t'} [\hat{N}^s,\hat{\rho}_{\rm eq}(T)] \dot{N}^d(t')\cos(\omega t')~dt' \;,
\end{equation}
where $\hat{\rho}_{\rm eq}(T)$ is the equilibrium density matrix operator evaluated at temperature $T$, $\dot{N}^d(t')=e^{i\hat{H}t'} \dot{N}^d e^{-i\hat{H}t'}$ with $\hat{H}$ the equilibrium Hamiltonian of the full system, and $\eta\to 0^+$ is required for convergence of the integral.

With some manipulation, this expression can be recast in terms of the current-current correlation function $K(\omega,T)=\langle\langle \dot{N}^s ; \dot{N}^d \rangle\rangle_{\omega,T}$ which is itself the Fourier transform to the frequency domain of the retarded function $K(t,T)=-i\theta(t)\langle [\dot{N}^s,\dot{N}^d(t) ] \rangle_T$. The dynamical conductance follows as, 
\begin{equation}\label{eq:kubo}
    G_C(\omega,T) = \left (\frac{e^2}{h}\right ) \frac{-2\pi  ~{\rm Im}~K(\omega,T)}{ \omega} \;.
\end{equation}
In the dc limit $\omega \to 0$ we obtain the steady state conductance,
\begin{equation}\label{eq:kubo_dc}
    G_C(T) \equiv \lim_{\omega \to 0}~ G_C(\omega,T) \;.
\end{equation}

With $\dot{N}^{\alpha}=i [\hat{H},\hat{N}^{\alpha}]$, one can express $K(\omega,T)$ in Eq.~\ref{eq:kubo} in terms of correlation functions involving the nanostructure frontier orbitals and the local lead orbitals,
\begin{equation}\label{eq:K}
K(\omega,T) = V_s V_d\sum_{\sigma,\sigma'}\langle\langle  \bar{d}_{s\sigma}^{\dagger} c_{s\sigma}^{\phantom{\dagger}} - c_{s\sigma}^{\dagger}\bar{d}_{s\sigma}^{\phantom{\dagger}} ~;~ \bar{d}_{d\sigma'}^{\dagger} c_{d\sigma'}^{\phantom{\dagger}} - c_{d\sigma'}^{\dagger}\bar{d}_{d\sigma'}^{\phantom{\dagger}} \rangle\rangle_{\omega,T}
\end{equation}

Eqs.~\ref{eq:kubo}--\ref{eq:K} have the advantage in linear response that they do not require the special proportionate coupling geometry. They are also applicable in the general dynamical regime of an ac driving bias voltage. NRG results for the linear conductance of non-PC systems via the Kubo formula have been shown to reproduce exact results in the rare cases where analytic solutions are known,\cite{mitchell2016universality} and quantitative agreement with experimental results have been demonstrated.\cite{iftikhar2018tunable,han2021extracting,pouse2021exotic}

We now remark on the equivalence of Eqs.~\ref{eq:kubo}--\ref{eq:K} to the MW result Eqs.~\ref{eq:MW_IC} and \ref{eq:MW_PC} in the PC limit. In this case, we may write Eq.~\ref{eq:K} in terms of the even/odd lead basis and the PC frontier orbital $\bar{\bar{d}}_{\sigma}$. The even lead orbital can then be eliminated by exploiting current conservation $\dot{N}^s=-\dot{N}^d$. This yields,
\begin{eqnarray*}
K(\omega,T)\overset{\text{PC}}{\longrightarrow}\tfrac{V_d^2V_s^2}{V_d^2+V_s^2} \sum_{\sigma,\sigma'} 
\langle\langle  \bar{\bar{d}}_{\sigma}^{\dagger} c_{o\sigma}^{\phantom{\dagger}} {\rm -} c_{o\sigma}^{\dagger}\bar{\bar{d}}_{\sigma}^{\phantom{\dagger}} ; c_{o\sigma'}^{\dagger}\bar{\bar{d}}_{\sigma'}^{\phantom{\dagger}} {\rm -} \bar{\bar{d}}_{\sigma'}^{\dagger} c_{o\sigma'}^{\phantom{\dagger}}  \rangle\rangle_{\omega,T}
\end{eqnarray*} 
Provided the original source and drain leads are equivalent, the transformed odd lead is then formally decoupled from the rest of the system (cf. emergent decoupling in Secs.~\ref{sec:emPC_CB} and \ref{sec:emPC_MV}). This means that exact eigenstates of the full system are product states of the form  $|\Psi\rangle=|\psi\rangle_{n-e}\otimes |\phi\rangle_o$, where $|\psi\rangle_{n-e}$ has support in the nanostructure-even-lead Fock space, while $|\phi\rangle_o$ lives in the decoupled odd-lead Fock space. This allows a factorization into single-particle Green's functions. After some manipulation, one obtains
\begin{equation}\label{eq:Kubo_PC}
    \begin{split}
G_C(\omega,T) \overset{\text{PC}}{\longrightarrow} \left(\frac{e^2}{h}\right) \frac{4\Gamma_d\Gamma_s}{\Gamma_d+\Gamma_s}\sum_{\sigma} \int_{-\infty}^{\infty}d\omega' ~{\rm Im}~ \overline{\overline{G}}_{\sigma}(\omega',T)& \\
\times \left [\frac{f_{\rm eq}(\omega'+\omega)- f_{\rm eq}(\omega'-\omega) }{2\omega}\right ]&
    \end{split}
    \end{equation}
where the integral over Fermi functions arises here because of the original trace over decoupled odd lead states. Eq.~\ref{eq:Kubo_PC} is as such the ac generalization of Eq.~\ref{eq:MW_G}, and reduces to it in the dc limit $\omega \to 0$, wherein the expression in the square brackets becomes $\partial_{\omega'} f_{\rm eq}(\omega')$. In the low-temperature limit $T\to 0$, we have
\begin{equation}\label{eq:Kubo_PC_T0}
G_C(\omega,0) \overset{\text{PC}}{\longrightarrow} \left(\frac{e^2}{h}\right) \frac{4\Gamma_d\Gamma_s}{\Gamma_d+\Gamma_s}\sum_{\sigma} \left [\frac{-\int_{-\omega}^{\omega}d\omega' ~{\rm Im}~ \overline{\overline{G}}_{\sigma}(\omega',0)}{2\omega}\right ]
\end{equation}
which is determined by the average of the nanostructure spectral function over a window set by the ac frequency. Ref.~\onlinecite{sindel2005frequency} proposed to use this as a means of measuring the nanostructure spectral function from dynamical conductance measurements.\\

%%%

Finally, we turn to the Kubo formula for heat transport.\cite{luttinger} This is more subtle, since there is no source term in which to do perturbation theory, that can be added to the Hamiltonian that mimics a temperature gradient. An indirect solution utilizing the Tolman-Ehrenfest effect\cite{TolmanEhrenfest} was proposed by Luttinger.\cite{luttinger} The energy current couples to both a gravitational potential gradient as well as a temperature gradient (``heat has weight''). A gravitational potential difference between leads can be incorporated into the Hamiltonian as a source term, leading to an expression for the linear-response energy current analogous to Eq.~\ref{eq:kubo_IC}. Importantly, Luttinger showed\cite{luttinger} that the linear response heat conductance due to a gravitational potential difference with zero temperature difference, is identical to that due to a temperature difference with zero gravitational potential difference. Exploiting this correspondence, one may work the previous steps of this section to obtain,
\begin{equation}\label{eq:kubo_heat}
    K_Q(T)=\left ( \frac{1}{hT}\right) \lim_{\omega \to 0}  \frac{-2\pi~{\rm Im}~K'(\omega,T)}{\omega} \;,
\end{equation}
with $K'(\omega,T)=\langle\langle \dot{H}_{\rm leads}^s ; \dot{H}_{\rm leads}^d \rangle\rangle$. Similarly, the cross susceptibility $\chi_{CQ}=\chi_{QC}$ appearing in Eq.~\ref{eq:offdiag}, which controls thermopower, is given by $\chi_{CQ}(T)=\tfrac{e}{h}\lim_{\omega \to 0} [\tfrac{2\pi}{\omega}{\rm Im}~K''(\omega,T)]$, with $K''(\omega,T)=\langle\langle \dot{N}^s ; \dot{H}_{\rm leads}^d \rangle\rangle$.

Finally, note that in linear response, the Kubo formula provides a connection between the ac conductance and the spectrum of the Johnson-Nyquist noise Eq.~\ref{eq:noise}, via the fluctuation-dissipation theorem,\cite{sindel2005frequency,KuboFDT1966}
\begin{equation}
C(\omega,T)=-2 e^2 f_{\rm BE}(\omega) [{\rm Im}K(\omega,T)] = 2\hbar \omega f_{\rm BE}(\omega) G_C(\omega,T) \;,
\end{equation}
where $f_{\rm BE}(\omega)=(\exp[\omega/k_{\rm B}T]-1)^{-1}$ is the Bose-Einstein distribution.

%###########################
%###########################

\section{``Improved'' charge transport Kubo formula for NRG}\label{sec:newkubo}
A well-defined and finite conductance in the dc limit $\omega \to 0$ implies from Eqs.~\ref{eq:kubo} and \ref{eq:kubo_dc} that the current-current correlator ${\rm Im}~K(\omega,T)\sim \omega$ at low energies. The physics of lead-coupled interacting nanostructures is often controlled by emergent low energy scales such as the Kondo temperature $T_{\rm K}$. The RG structure of such problems means that the limiting $\omega \to 0$ behavior can in practice be extracted for finite $|\omega| \ll T, T_{\rm K}$. However in this regime ${\rm Im}~K(\omega,T)$ is itself very small. Any numerical estimation of $K(\omega,T)$ must therefore be very accurate over a sufficient range of $|\omega| \ll T, T_{\rm K}$ so that the gradient can be identified and the conductance deduced. 

Even with highly accurate methods such as NRG,\cite{wilson1975renormalization,*bulla2008numerical,weichselbaum2007sum,peters2006numerical} this can be a challenge. Good results are only obtained when the number of kept states in NRG is pushed very high. A further issue with the NRG implementation is that, even with the full density matrix approach,\cite{weichselbaum2007sum,peters2006numerical} dynamical information is not reliably obtained for $|\omega|\ll T$. This can make it difficult to extract the dc conductance numerically using Eqs.~\ref{eq:kubo} and \ref{eq:kubo_dc}.

Finally, we note that the explicit form of $K(\omega,T)$ in Eq.~\ref{eq:K} depends on the nanostructure frontier orbitals and the local lead orbitals, and is model-specific. 

Below we propose a conceptually simple but practically useful reformulation that substantially overcomes these limitations and is found to greatly improve accuracy of the Kubo formula for charge transport when implemented in NRG. 

We utilize the Zubarev equations of motion\cite{zubarev1960double} for retarded Green's functions of bosonic operators $\hat{A}$ and $\hat{B}$,
\begin{eqnarray*}
\omega^+\langle \langle \hat{A} ; \hat{B} \rangle \rangle -\langle [\hat{A},\hat{B}] \rangle  =  \langle\langle \hat{A} ;[\hat{H},\hat{B}] \rangle \rangle  = \langle\langle [\hat{A},\hat{H}] ;\hat{B} \rangle \rangle
\end{eqnarray*}
With $\dot{\Omega}\equiv \tfrac{d}{dt}\hat{\Omega}=i[\hat{H},\hat{\Omega}]$, and choosing $\hat{A}=\dot{N}^s$ and $\hat{B}=\hat{N}^d$, we may therefore write, $\langle\langle \dot{N}^s ; \dot{N}^d \rangle\rangle = i \omega^+\langle\langle \dot{N}^s ; \hat{N}^d \rangle\rangle +\langle [[\hat{H},\hat{N}^s] , \hat{N}^d] \rangle $. Since the expectation value 
$\langle [[\hat{H},\hat{N}^s] , \hat{N}^d] \rangle $ is real, we obtain the current-current correlator appearing in Eq.~\ref{eq:kubo} as,
\begin{equation}\label{eq:newK2}
    {\rm Im}~K(\omega,T) = \omega {\rm Re}~\langle\langle \dot{N}^s ; \hat{N}^d \rangle\rangle_{\omega,T} \;.
\end{equation}
Conveniently, the factor of $\omega$ in Eq.~\ref{eq:newK2} cancels with that in the denominator of Eq.~\ref{eq:kubo}, leading to greater stability in the numerical evaluation of the dc conductance via the Kubo formula. In fact, one can apply the equations of motion a second time to obtain,
\begin{equation}\label{eq:newK}
    {\rm Im}~K(\omega,T) = -\omega^2 {\rm Im}~\langle\langle \hat{N}^s ; \hat{N}^d \rangle\rangle_{\omega,T} \;,
\end{equation}
which involves the total number operators for the leads only. Within NRG, this is a convenient reformulation from a technical point of view because the Kubo formula is now independent of the details of the nanostructure or hybridization Hamiltonians: once the NRG code is set up to calculate the correlator $\langle\langle \hat{N}^s ; \hat{N}^d \rangle\rangle_{\omega,T}$, the same code can be used to calculate the conductance of any system without modification. Importantly, we find that Eq.~\ref{eq:newK} yields much more accurate results in practice, and that a greatly reduced number of states $M_K$ need to be kept in the NRG calculations to obtain converged results than when using the standard Kubo formula.  This is in part because nontrivial dynamical contributions to the conductance arise at every step of the NRG within this formulation, since $\hat{N}^{\alpha}$ has weight on all Wilson chain orbitals. 

%%%%%%%%%%%

\subsection{Calculation of $\langle\langle \hat{N}^s ; \hat{N}^d \rangle\rangle_{\omega,T}$}\label{sec:newK}

\noindent The Lehmann representation of generic spectral functions of the type $\mathcal{A^{BC}}(\omega,T) = \int \frac{dt}{2\pi} e^{i\omega t}\langle \hat{\mathcal{B}}(t)\hat{\mathcal{C}}\rangle_T$ is given by,
\begin{equation}\label{eq:lehmann}
    \mathcal{A^{BC}}(\omega,T) = \sum_{a,b}\langle b| \hat{\mathcal{C}}|a\rangle \frac{e^{-E_a/k_{\rm B} T}}{Z} \langle a| \hat{\mathcal{B}}|b\rangle \delta(\omega-E_b+E_a) \;,
\end{equation}
where $\{|a\rangle\}$ constitutes the \emph{complete} set of eigenstates of the Hamiltonian satisfying the Schr\"odinger equation $H |a\rangle = E_a |a\rangle$, and $Z$ is the partition function. 
In principle, the desired retarded correlator $\tilde{K}(\omega,T)=\langle\langle \hat{N}^s ; \hat{N}^d \rangle\rangle_{\omega,T}$ can be obtained from Eq.~\ref{eq:lehmann} as 
$\frac{1}{\pi}{\rm Im}~\tilde{K}(\omega,T)= \mathcal{A}^{N^dN^s}(\omega,T) - \mathcal{A}^{N^sN^d}(-\omega,T)$.

However, calculating such an object within NRG is somewhat unconventional because the operators $\hat{N}^s$ and $\hat{N}^d$ are not local impurity operators, but rather live on the Wilson chains for the leads.

Within NRG, $H_{\rm leads}$ in Eq.~\ref{eq:Hleads} is discretized logarithmically and mapped to Wilson chains of the form,\cite{wilson1975renormalization,*bulla2008numerical}
\begin{equation}\label{eq:Hleads_WC}
    H_{\rm leads} \to H_{\rm leads}^{\rm WC} = \sum_{\alpha,\sigma}\sum_{n=0}^{\infty} \left ( t_n^{\phantom{\dagger}} f_{\alpha n \sigma}^{\dagger}f_{\alpha (n+1) \sigma} +{\rm H.c.} \right ) \;,
\end{equation}
where $t_n \sim D \Lambda^{-n/2}$ at large $n$, $D$ is the bare conduction electron bandwidth as before, and $\Lambda>1$ is the discretization parameter. We have assumed for simplicity leads with particle-hole symmetry, although this is immaterial. The mapping is defined such that the nanostructure hybridizes only with the Wilson `zero' orbitals $f_{\alpha 0 \sigma}$,
\begin{eqnarray}\label{eq:Hhyb_WC}
H_{\rm hyb} \to H_{\rm hyb}^{\rm WC}= \sum_{\alpha,i,\sigma}\left (V_{\alpha,i} d^{\dagger}_{i\sigma} f_{\alpha 0  \sigma}^{\phantom{\dagger}}  + \text{H.c.} \right ) \;.
\end{eqnarray}
The lead number operators are $\hat{N}^{\alpha}=\sum_{n} \hat{n}^f_{\alpha n}$ where $\hat{n}^f_{\alpha n}= \sum_{\sigma} f_{\alpha n \sigma}^{\dagger} f_{\alpha n \sigma}^{\phantom{\dagger}}$.

In NRG, one starts from the `impurity' Hamiltonian $H_{\rm nano}$, and then defines a sequence of Hamiltonians $H_{n}$, obtained iteratively by addition of a successive Wilson chain orbitals,\cite{wilson1975renormalization,*bulla2008numerical,foot_rescale} 
\begin{eqnarray}
H_0 &=& H_{\rm nano} + H_{\rm hyb}^{\rm WC} \;,\\
H_n &=& H_{n-1} + t_{n-1} \sum_{\alpha,\sigma} \left (f_{\alpha (n-1) \sigma}^{\dagger} f_{\alpha n \sigma}^{\phantom{\dagger}} +{\rm H.c.} \right ) \;,\qquad\label{eq:HN}
\end{eqnarray}
with Eq.~\ref{eq:HN} defined for $n>0$. The exact (discretized) Hamiltonian $H^{\rm disc}=H_{\rm nano} + H_{\rm leads}^{\rm WC} + H_{\rm hyb}^{\rm WC}$ is recovered as $H^{\rm disc} = \lim_{n\to \infty} H_{n}$. The traditional rescaling of $H_n$ is omitted here for pedagogical simplicity.

Clearly the Fock space dimension of $H_n$ grows exponentially with $n$, and the exact solution of even the discretized model is intractable. Instead, at each step, $H_n$ is diagonalized and only the low-energy manifold of eigenstates is retained for the next step.
This constitutes an RG transformation;\cite{wilson1975renormalization,*bulla2008numerical} the physics on successively lower energy scales is revealed on increasing $n$.

The iterative scheme is initialized in some convenient basis $\{|\phi_a\rangle_0\}$ spanning $H_0$ by forming the corresponding Hamiltonian matrix $\boldsymbol{H_0}$, with elements $[\boldsymbol{H_0}]_{ab}={_0}\langle \phi_a | H_0 | \phi_b \rangle_0$. The unitary matrix $\boldsymbol{U_0}$ is constructed such that $\boldsymbol{U_0}^{\dagger}\boldsymbol{H_0}\boldsymbol{U_0} = \boldsymbol{D_0}$ is diagonal. Eigenstates $|\psi_k\rangle_0=\sum_a [\boldsymbol{U_0}]_{ak}|\phi_a\rangle_0$ are used to construct $\boldsymbol{H_1}$, with matrix elements 
${_1}\langle \phi_{j;i} | \hat{H}_1 | \phi_{j';i'} \rangle_1$ defined in terms of 
basis states  $|\phi_{j;i}\rangle_1=|\psi_j\rangle_0\otimes |\gamma_i\rangle_1$ spanning the Fock space of $H_1$. Here $\{|\gamma_i\rangle_n\}$ are the 16 states defined on the $n^{\rm th}$ Wilson shell corresponding to $f_{\alpha n \sigma}$. We now diagonalize the matrix $\boldsymbol{H_1}$ to find $\boldsymbol{U_1}$.

Truncation of the $M$-dimensional Fock space of $H_n$ is accomplished in NRG by discarding high energy states at that iteration.\cite{wilson1975renormalization,*bulla2008numerical} $M_K$ eigenstates of $H_n$ are retained up to an energy cutoff $E_{\rm cut}$. The $M\times M$ matrix $\boldsymbol{U_n}$ is therefore reshaped to an $M\times M_K$ matrix $\boldsymbol{\tilde{U}_n}$. This implies that $\boldsymbol{\tilde{U}_n}^{\dagger}\boldsymbol{H_n}\boldsymbol{\tilde{U}_n} = \boldsymbol{\tilde{D}_n}$ is a diagonal matrix of dimension $M_K\times M_K$ containing only the lowest eigenvalues of $H_n$. The restricted set of $M_K$ retained eigenvectors at iteration $n$ are obtained as 
$|\psi_{k}\rangle_n=\sum_{j,i} [\boldsymbol{\tilde{U}_n}]_{(j;i)k}|\phi_{j;i}\rangle_n$. 
Only these retained states at iteration $n$ are used to construct the Hamiltonian matrix at iteration $n+1$. 
The approximate Hamiltonian in the basis of $|\phi_{j;i}\rangle_{n+1}=|\psi_j\rangle_n\otimes | \gamma_i\rangle_{n+1}$ is denoted  $\boldsymbol{\tilde{H}_{n+1}}$. The truncation at each step means that the Fock space dimension is roughly independent of $n$.

Crucially, the structure of the Wilson chain implies that discarded states at a given iteration are unimportant for constructing the low-energy states at a later iteration.\cite{wilson1975renormalization,*bulla2008numerical} Furthermore, useful information can be extracted at each step\cite{wilson1975renormalization,*bulla2008numerical} since $\boldsymbol{\tilde{H}_n}$ may be regarded as a renormalized version of the full Hamiltonian at an effective temperature $T_n \sim D\Lambda^{-n/2}$. 

With this formalism established, we return to the calculation of the correlator $\tilde{K}(\omega,T)$. The exact result for the discretized model could in principle be obtained from Eq.~\ref{eq:lehmann} given the complete set of exact eigenstates of $H^{\rm disc}$. In NRG however, the iterative diagonalization and truncation procedure means that only the approximate (renormalized) eigenstates of $\boldsymbol{\tilde{H}_n}$ at each step are known. The Anders-Schiller (AS) basis\cite{anders2006spin,*anders2005real} comprises the \emph{discarded} states at each step and is a complete, albeit approximate, basis with which to compute spectral functions via the Lehmann sum, Eq.~\ref{eq:lehmann}. Refs.~\onlinecite{weichselbaum2007sum,peters2006numerical} reformulated the problem in terms of the full density matrix (FDM) established on the AS basis, wherein the spectrum $\mathcal{A^{BC}}(\omega,T)=\sum_n w_n \mathcal{A}^{\mathcal{BC}}_n(\omega,T)$ consists of a weighted sum of contributions from each NRG iteration. This requires matrix representations of the operators $\hat{\mathcal{B}}$ and $\hat{\mathcal{C}}$ at iteration $n$ in the eigenbasis of $\boldsymbol{\tilde{H}_n}$. For the correlator $\tilde{K}(\omega,T)$ we require\cite{foot_N} specifically $\boldsymbol{\tilde{N}^{\alpha}_n} = \sum_{m=0}^n \boldsymbol{\tilde{N}^{\alpha,m}_n}$, where
$[\boldsymbol{\tilde{N}^{\alpha,m}_n}]_{kk'} = {_n}\langle \psi_k |\hat{n}^f_{\alpha m} | \psi_{k'}\rangle_n$.  
For $n=0$, $\boldsymbol{\tilde{N}^{\alpha}_0} \equiv \boldsymbol{\tilde{N}^{\alpha,0}_0}$ can be explicitly evaluated from the known exact eigenstates of $H_0$. 
For $n>0$ we split $\boldsymbol{\tilde{N}^{\alpha}_n} =  \boldsymbol{\tilde{N}^{\alpha,<}_n} + \boldsymbol{\tilde{N}^{\alpha,n}_n}$ into two contributions. The on-shell term can be directly evaluated at each step, $[\boldsymbol{\tilde{N}^{\alpha,n}_n}]_{kk'}=\sum_{j,i} [\boldsymbol{\tilde{U}_n}]^*_{(j;i)k}[\boldsymbol{\tilde{U}_n}]_{(j;i)k'}\times{_n}\langle\gamma_i|\hat{n}^f_{\alpha n} |\gamma_i\rangle_n$ since the matrix elements involved are trivial; while 
$[\boldsymbol{\tilde{N}^{\alpha,m}_n}]_{kk'}=\sum_{j,j',i} [\boldsymbol{\tilde{U}_n}]^*_{(j;i)k}[\boldsymbol{\tilde{U}_n}]_{(j';i)k'}\times [\boldsymbol{\tilde{N}^{\alpha,m}_{n-1}}]_{jj'}$ for $0\le m<n$. The latter property implies the recursion relation
$[\boldsymbol{\tilde{N}^{\alpha,<}_n}]_{kk'}=\sum_{j,j',i} [\boldsymbol{\tilde{U}_n}]^*_{(j;i)k}[\boldsymbol{\tilde{U}_n}]_{(j';i)k'}\times [\boldsymbol{\tilde{N}^{\alpha}_{n-1}}]_{jj'}$. The required operator matrices $\boldsymbol{\tilde{N}^{\alpha}_n}$ can then all be obtained iteratively, starting from $\boldsymbol{\tilde{N}^{\alpha}_0}$.

With this, the correlator $\tilde{K}(\omega,T)$ can be obtained using standard FDM-NRG as described in Refs.~\onlinecite{weichselbaum2007sum}.

We emphasize that the technical implementation of the above is rather straightforward within the usual NRG framework, and is compatible with abelian and non-abelian quantum numbers in general symmetry settings,\cite{weichselbaum2012non} and with differentiable programming within $\partial$NRG.\cite{rigo2021automatic} The number operator matrix elements can also be calculated on the generalized Wilson chain used in the interleaved NRG (iNRG) method.\cite{mitchell2014generalized,*stadler2016interleaved}

%!!!!!!!!!!!!!

\subsection{Numerical Results for the AIM}
\begin{figure}[t!]
\includegraphics[width=8.7cm]{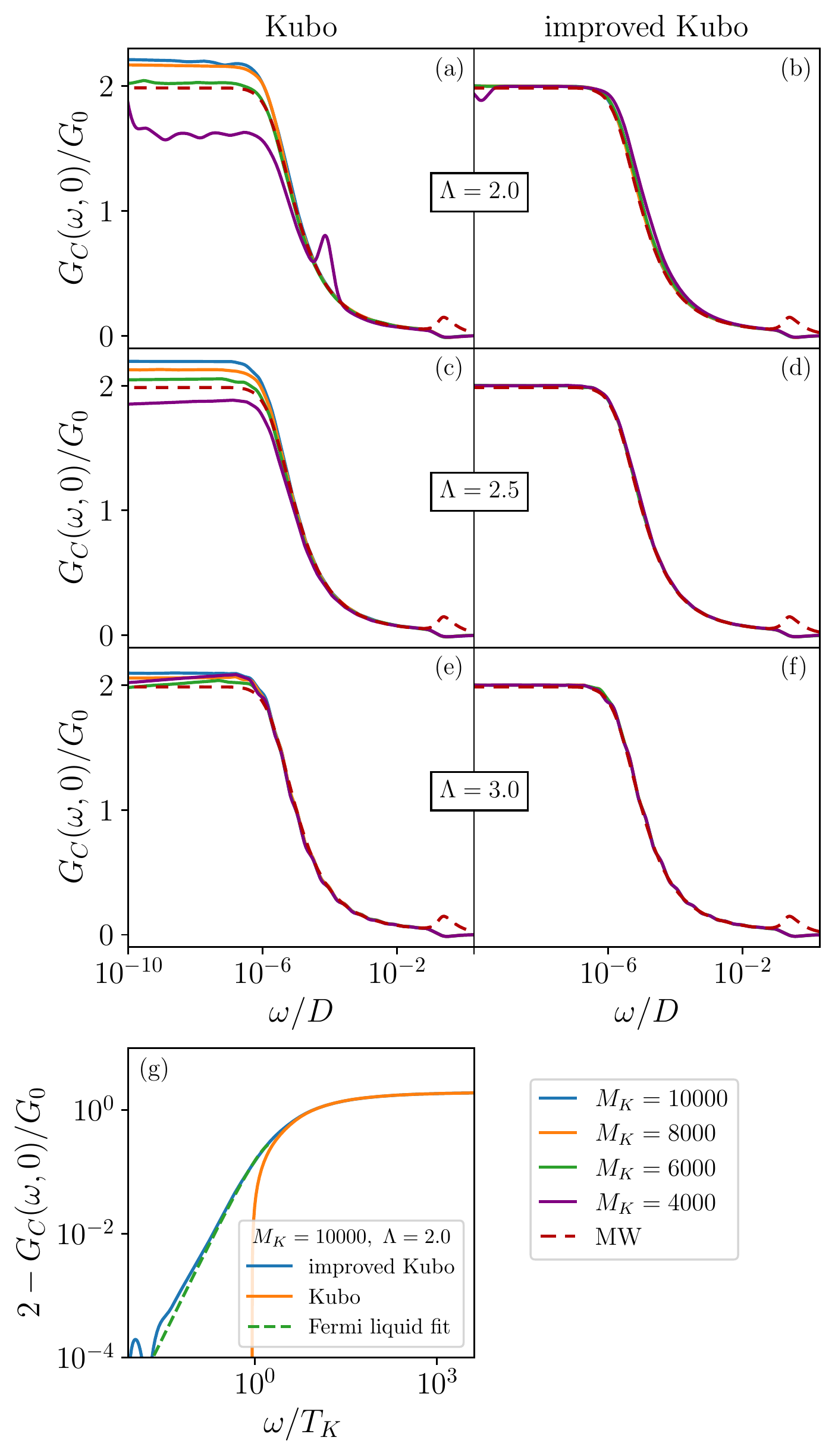}
  \caption{(a)-(f) Comparison of NRG results for the Kubo (left) and ``improved'' Kubo (right) formulae for the ac conductance $G_C(\omega,0)$ at $T=0$ for the two-lead AIM, with different numbers $M_K$ of NRG kept states as given in the legend, and different NRG discretization parameters $\Lambda$. Results  benchmarked against the MW formula, evaluated within the equivalent single-channel AIM (red dashed line). (g) Low-frequency behaviour, comparing Kubo and improved-Kubo with a Fermi liquid fit $2-G_C(\omega,0)/G_0= c_2(\omega/T_{\rm K})^2+c_3|\omega/T_{\rm K}|^3$. AIM parameters: $U_d=0.4D$, $\epsilon_d=-\tfrac{1}{2}U_d$, $V_s=V_d=0.07D$. }
  \label{fig:KvsimpK}
\end{figure}

As a proof-of-principle demonstration, we calculate the ac electrical conductance $G_C(\omega,T)$ vs driving frequency $\omega$ at $T=0$ with NRG for the two-lead AIM (Eq.~\ref{eq:aim}) using the Kubo formula Eq.~\ref{eq:kubo}. In Figs.~\ref{fig:KvsimpK}(a-f) we compare the standard implementation of the Kubo formula Eq.~\ref{eq:K} (left panels) with the ``improved'' Kubo formula Eq.~\ref{eq:newK} (right panels), for different numbers $M_K$ of NRG kept states as specified in the legend. (a-b) are calculated for NRG discretization parameter $\Lambda=2$; (c,d) for $\Lambda=2.5$; and (e,f) for $\Lambda=3$. As a reference, we provide the MW result (dashed line) calculated within NRG within the equivalent single-channel AIM (Eq.~\ref{eq:aim1ch}) obtained via Eq.~\ref{eq:MW_G}, which should be considered the numerically-exact result for this system. 

%%%%%%%%%%%%%%%%
%%%%%%%%%%%%%%%%

Fig.~\ref{fig:KvsimpK}(a) shows that the standard Kubo formula yields rather poor results, even at large $M_K=10000$ for $\Lambda=2$. The situation improves with increasing $\Lambda$ [see panels (c,e)] due to the tradeoff\cite{weichselbaum2011discarded} between $M_K$ and $\Lambda$. The best performance, with an error of $\sim 5\%$, was obtained with $\Lambda=3$ and $M_K=10000$ [blue line, panel (e)], although even here the results are not fully converged with respect to $M_K$. By contrast, the improved Kubo results [panels (b,d,f)] are highly stable, and essentially fully converged even for remarkably low $M_K=4000$ for all $\Lambda$ considered. This shows that accurate results can be obtained within NRG using the improved Kubo formulation at relatively low computational cost. 

Fig.~\ref{fig:KvsimpK}(g) shows the low-frequency behavior, comparing Kubo and improved Kubo with a Fermi liquid fit. The improved Kubo formula is found to capture the expected asymptotic form very accurately. Indeed, for $\Lambda=2$ and $M_K=10000$, the improved Kubo method satisfies the exact dc limit result $G_C(0,0)/G_0=2$ to within $0.01\%$.

Although in this benchmark comparison, accurate results can be alternatively obtained at low computational cost from the MW formula using the equivalent single-channel model, we emphasize that for general non-PC systems this is not possible; and genuine two-channel NRG calculations are notoriously demanding.

Note that the Hubbard satellite feature at $\omega\sim U_d$ seen in the MW result is not captured by either Kubo approach. The lack of resolution at high energies appears to be due to the logarithmic discretization of the bath Hamiltonian in NRG. All other features are well-described; and the improved-Kubo approach confers a significant accuracy and efficiency gain.

%###########################
%###########################

\section{Failure of heat transport\\Kubo formula in NRG}\label{sec:heatkubo}

Although the Kubo formula for charge transport as implemented in NRG can yield highly accurate results, the same is unfortunately \textit{not} true for heat transport. Perhaps surprisingly, we find that the Kubo formula for heat transport, Eq.~\ref{eq:kubo_heat}, does not yield even qualitatively correct results when implemented in NRG. 

In NRG the required current-current correlation function $K'(\omega,T)$ takes the form,
\begin{subequations}\label{eq:Kuboheat_corr}
\begin{align}
\langle\langle \dot{H}_{\rm leads}^s ; \dot{H}_{\rm leads}^d \rangle\rangle &= 
-V_sV_dt_0^2 \sum_{\sigma,\sigma'}\langle\langle  \hat{\mathcal{O}}_{s\sigma} ; \hat{\mathcal{O}}_{d\sigma'} \rangle\rangle \;,\\
 {\rm with,}~~~\hat{\mathcal{O}}_{\alpha\nu}& = \bar{d}^{\dagger}_{\alpha\nu}f_{\alpha 1 \nu} - f^{\dagger}_{\alpha 1 \nu} \bar{d}_{\alpha \nu} \;,
\end{align}
\end{subequations}
where $f_{\alpha 1 \nu}$ is the $n=1$ Wilson orbital defined in Eq.~\ref{eq:Hleads_WC}. Although the $n=0$ Wilson orbital $f_{\alpha 0 \nu}$ may be interpreted as a discretized version of the local lead orbital $c_{\alpha \nu}$, the $n=1$ Wilson orbital $f_{\alpha 1 \nu}$ has no direct physical meaning in the bare model.

%%%%%%%%%%%%%%%%
\begin{figure}[t]
\includegraphics[width=8.4cm]{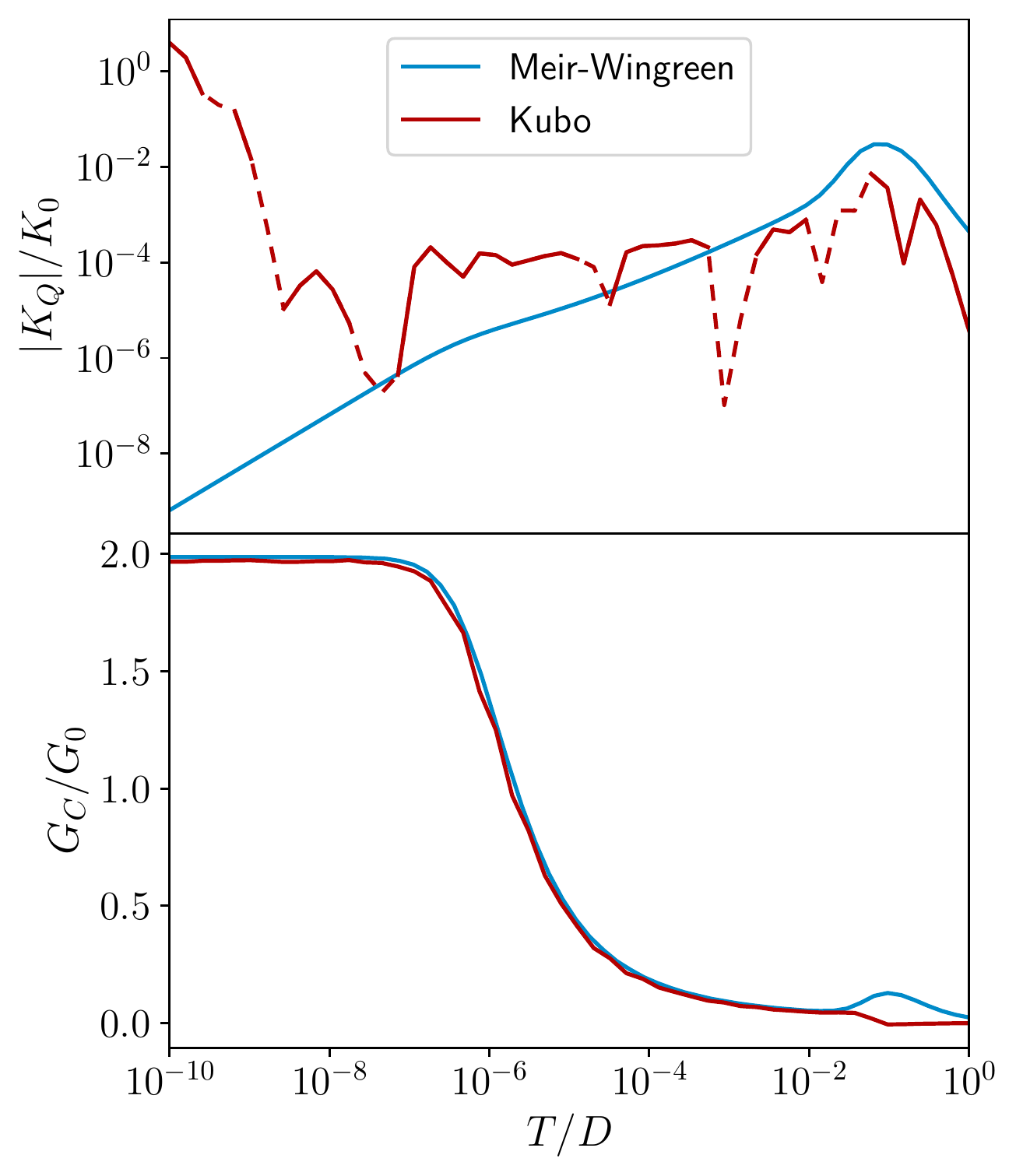}
  \caption{(a) Heat conductance $|K_Q(T)|$ vs $T$ for the two-lead AIM, calculated in NRG via the Kubo formula (Eq.~\ref{eq:kubo_heat}; red line), compared with the MW result (Eq.~\ref{eq:MW_K}; blue line). Solid lines for $K_Q>0$, dashed for $K_Q<0$. Parameters used: $U_d=0.4D$, $\epsilon_d=-U_d/2$, $V_s=V_d=0.07D$, $\Lambda=3$,  $M_K=11000$ (2-channel Kubo) or $4000$ (1-channel MW). (b) Same as panel (a) except for dc charge conductance $G_C(T)$, obtained with NRG via the standard Kubo formula (Eq.~\ref{eq:kubo}; red line) or MW (Eq.~\ref{eq:MW_G}; blue line).  }
  \label{fig:heatkubo}
\end{figure}
%%%%%%%%%%%%%%%%

We use Eq.~\ref{eq:Kuboheat_corr} to calculate $K_Q(T)$ via Eq.~\ref{eq:kubo_heat} for the two-lead AIM using NRG, and present numerical results in Fig.~\ref{fig:heatkubo}(a) as the red line. For comparison, we show the MW result computed via Eq.~\ref{eq:MW_K} for the effective one-channel system as the blue line. The latter should be regarded as the numerically-exact result. We see that $K_Q(T)$ from the Kubo formula fails to capture the correct physics, with apparently noisy data, erratic sign-changes, and diverging behavior at low-$T$ rather than vanishing as $\sim T$. However, other physical quantities computed in the same NRG run, such as the electrical conductance $G_c(T)$ obtained by the Kubo formula Eq.~\ref{eq:kubo}, correctly recover the expected results, see Fig.~\ref{fig:heatkubo}(b). 

The high-quality two-channel NRG calculations here are fully converged (Fig.~\ref{fig:heatkubo}(a) is not improved by increasing the number of kept states $M_K$). We have also implemented an ``improved'' version of the heat Kubo formula along the lines of Sec.~\ref{sec:newkubo}, but the underlying failure of NRG demonstrated above is not resolved by this. 

We note that the heat Kubo formula for the exactly-solvable $U_d=0$ resonant level model coupled to Wilson chains also fails to capture the correct behavior, including the low-$T$ asymptote $K_Q(T)\sim T$. This indicates that the problem lies with the representation of the lead Hamiltonian as a discretized Wilson chain, rather than with the iterative diagonalization procedure in NRG. We therefore believe the failure of NRG in this context to be because Wilson chains are not true thermal reservoirs,\cite{esposito2015nature} as argued in Ref.~\onlinecite{rosch2012wilson} -- arbitrary amounts of energy $\Delta E$ cannot be dissipated by Wilson chains without changing their temperature. We hypothesize that this issue may afflict thermal transport calculations using the Kubo formula for \emph{any} discretized system.

We emphasize that the above breakdown is not a problem with the heat Kubo formula itself, but rather a failure of NRG to capture the proper behavior of the required current-current correlation function defined on the discretized Wilson chain, Eq.~\ref{eq:Kuboheat_corr}. Note that for systems in the PC geometry, the MW formula Eq.~\ref{eq:MW_K} can be used instead. This relates the heat conductance in linear response to nanostructure equilibrium retarded Green's functions, which are very accurately calculated within NRG.\cite{weichselbaum2007sum} This approach within NRG yields the correct behavior of $K_{Q}(T)$, as demonstrated in Ref.~\onlinecite{costi2010thermoelectric}. The problem of how to obtain $K_{Q}(T)$ via NRG for non-PC systems therefore remains an open question.

%###########################
%###########################

\section{Emergent Proportionate Coupling: Coulomb Blockade}\label{sec:emPC_CB}
In most realistic systems (with the notable exception of single-dot devices) the orbital complexity and spatial structure of the nanostructure prevents a PC description. However, in certain situations the low-energy effective model for the system may be in PC even though the bare model is not. This is particularly useful for the MW formulation of quantum transport involving only retarded single-particle nanostructure Green's functions. We explore these scenarios below for the case of systems in the Coulomb blockade (CB) regime.

Deep in the CB regime, the nanostructure has a well-defined number of electrons, with charge fluctuations suppressed at low temperatures $k_{\rm B} T \ll E_C$ by the effective nanostructure charging energy $E_C$ (the microscopic origin of which is the Coulomb repulsion described by $H_{\rm int}$ in Eq.~\ref{eq:Hnano}). Incoherent transport, involving sequential tunneling events between the nanostructure and leads, is therefore also suppressed.\cite{NazarovCoTun1990,sasaki2000kondo,DeFranceschi_ElectronCotunneling2001Exp} This is the dominant transport mechanism for nanostructures with a net spin $S=0$. By contrast, Kondo-enhanced spin-flip scattering in spinful nanostructures can give a substantial boost to low-temperature coherent transport, with the unitarity limit of a perfect single-electron transistor attainable experimentally.\cite{van2000kondo}

In both cases, the structure of the low-energy effective model can be obtained perturbatively to second order in $H_{\rm hyb}$, by projecting onto the ground state manifold of $N$-electron states for the isolated nanostructure $|N;{\rm gs},j\rangle$, and eliminating virtual excitations to nanostructure states $|N\pm 1; {\rm ex},k\rangle$ with $N\pm 1$ electrons. Using Brillouin-Wigner perturbation theory, we may write,
\begin{equation}\label{eq:BWPT}
    H_{\rm eff}^N \simeq H_{\rm leads} + \hat{1}_{\rm gs}^N H_{\rm nano} \hat{1}_{\rm gs}^N + \hat{1}_{\rm gs}^N H_{\rm hyb}(E_{\rm gs}^N-H_{\rm nano})^{-1} H_{\rm hyb} \hat{1}_{\rm gs}^N \;,
\end{equation}
where $\hat{1}_{\rm gs}^N=\sum_j |N;{\rm gs},j\rangle \langle N;{\rm gs},j|$ and $H_{\rm nano} |N;{\rm gs},j\rangle = E_{{\rm gs}}^N |N;{\rm gs},j\rangle$. By construction, $H_{\rm hyb} |N;{\rm gs},j\rangle$ generates (a superposition of) excited eigenstates of $H_{\rm nano}$ satisfying $H_{\rm nano} |N\pm 1; {\rm ex},k\rangle = E_{{\rm ex},k}^{N\pm 1} |N\pm 1; {\rm ex},k\rangle$.

Parameters of the effective model are then related to matrix elements of the type $\langle N+1;{\rm ex},k | \bar{d}_{\alpha\sigma}^{\dagger} | N;{\rm gs},j\rangle$ and $\langle N-1;{\rm ex},k | \bar{d}_{\alpha\sigma}^{\phantom{\dagger}} | N;{\rm gs},j\rangle$, which may be computed straightforwardly for the isolated nanostructure for a given model.

We note that although the structure of the effective model for generalized quantum impurity systems can be obtained from Eq.~\ref{eq:BWPT}, the perturbative estimation of its coupling constants is accurate only in the limit of strong nanostructure interactions and weak hybridization. Recently in Ref.~\onlinecite{rigo2020machine}, machine learning techniques have been employed to determine the numerical value of the effective model parameters non-perturbatively. We illustrate the quantitative accuracy of such a model machine learning approach to describe the low-temperature physics of nanostructures by explicit calculations in Sec.~\ref{sec:tqd}.

In the following we focus on the generic behavior of the underlying effective models. Ultimately, the accuracy of predictions using these models depends on the accurate determination of the effective parameters.

%%%%%%%%%%

\subsection{$S=0$}\label{sec:S0}
For isolated nanostructures with an even number of electrons $N$, the ground state is often (but not always) a unique spin singlet state with total $S=0$. Applying Eq.~\ref{eq:BWPT} yields an effective model in the form of a renormalized tunnel junction, 
\begin{equation}\label{eq:Heff_S0}
H_{\rm eff}=H_{\rm leads} +\sum_{\alpha,\beta}\sum_{\sigma}  W_{\alpha\beta}^{\sigma} c_{\alpha\sigma}^{\dagger} c_{\beta\sigma}^{\phantom{\dagger}}  \;,
\end{equation}
where $\alpha,\beta \in s,d$ and $W_{\alpha\beta}^{\sigma}=(W_{\beta\alpha}^{\sigma})^*$ are effective potential scattering parameters obtained after integrating out the nanostructure. Note that in the presence of a magnetic field on the nanostructure, the excited states $|N\pm 1; {\rm ex},k\rangle$ and excitation energies $(E_{\rm gs}^N-E_{{\rm ex},k}^{N\pm 1})$ depend on spin $\sigma$, which in general endow $W_{\alpha\beta}^{\sigma}$ with a spin dependence, which we retain here for generality. 

Eq.~\ref{eq:Heff_S0} is in some sense trivially in PC. Since the effective model is non-interacting, we may use the Landauer formula for conductance, Eqs.~\ref{eq:land}, \ref{eq:caroli2}. In the low-temperature limit $T\to 0$, the electrical conductance is given by $G_C(0)=(e^2/h) 4 \tilde{\Gamma}_s\tilde{\Gamma}_d \sum_{\sigma} |\mathcal{G}_{sd,\sigma}(0)|^2$. Here $\mathcal{G}_{sd,\sigma}(\omega)\equiv \langle\langle c_{s\sigma}^{\phantom{\dagger}} ; c_{d\sigma}^{\dagger}\rangle\rangle$ is the full Green's function connecting local orbitals in source and drain leads in the presence of the $W_{\alpha\beta}^{\sigma}$ terms in Eq.~\ref{eq:Heff_S0}. It can be expressed in terms of the free lead Green's functions of the isolated $H_{\rm leads}$, which we write as $\mathcal{G}_{\alpha\beta,\sigma}^0(\omega)=\delta_{\alpha\beta}/[\omega^+-\Delta(\omega)]$, with $\Delta(\omega)$ so defined. The effective hybridizations are then $\tilde{\Gamma}_{\alpha}=-{\rm Im}\Delta(0) = 1/(\pi \rho_0)$, equal for both leads. The required full Green's function can be obtained from standard equations of motion techniques as,
\begin{equation}
\mathcal{G}_{sd,\sigma}(\omega)=\frac{W_{sd}^{\sigma}}{[\omega^+-W_{ss}^{\sigma}-\Delta(\omega)][\omega^+-W_{dd}^{\sigma}-\Delta(\omega)]-|W_{sd}^{\sigma}|^2}
\end{equation}

In the zero-field, particle-hole symmetric case where $W_{ss}^{\sigma}=W_{dd}^{\sigma}=0$ and $W_{sd}^{\sigma}\equiv W_{sd}$, the standard result for the $T=0$ transmission of a tunnel-junction is recovered,
\begin{equation}\label{eq:Gqpc}
    G_C(0)=\left ( \frac{2e^2}{h}\right ) \frac{4 |\tilde{W}_{sd}|^2}{(1+|\tilde{W}_{sd}|^2)^2} \;.
\end{equation}
where we have defined the dimensionless quantities  $\tilde{W}^{\sigma}_{\alpha\beta}=\pi \rho_0 W_{\alpha\beta}^{\sigma}$. The maximum conductance $G_{C}=2e^2/h$ is attained when $|\tilde{W}_{sd}|^2=1$, which corresponds to intermediate values of $W_{sd}$.

In the general case, one obtains,
\begin{equation}\label{eq:cond_qpc}
    G_C(0)=\left ( \frac{e^2}{h}\right )\sum_{\sigma} \frac{4 |\tilde{W}_{sd}^{\sigma}|^2}{(1+|\tilde{W}_{sd}^{\sigma}|^2-\tilde{W}_{ss}^{\sigma}\tilde{W}_{dd}^{\sigma})^2+(\tilde{W}_{ss}^{\sigma}+\tilde{W}_{dd}^{\sigma})^2}
\end{equation}
The conductance is therefore seen to depend strongly on particle-hole asymmetry, generated and tuned in practice by application of gate voltages. In particular, the maximum conductance $G_C < 2e^2/h$ for finite $W_{ss}$ and/or $W_{dd}$ is always reduced with respect to its particle-hole symmetric counterpart.

%%%%%%%%%%

\subsection{$S=\tfrac{1}{2}$}\label{sec:Shalf}
The more interesting and subtle case arises when the isolated nanostructure has an odd number of electrons $N$ and hosts a net spin $S=\tfrac{1}{2}$.  Eq.~\ref{eq:BWPT} then becomes a generalized Schrieffer-Wolff transformation.\cite{SchriefferWolff1966,bravyi2011schrieffer} In the absence of other ground state degeneracies, the effective model takes the form of a generalized two-channel Kondo (2CK) model,\cite{Kondo1964}
\begin{equation}\label{eq:H2ck}
H_{\rm eff} = H_{\rm leads} +\sum_{\alpha,\beta}\sum_{\sigma}  W_{\alpha\beta}^{\sigma} c_{\alpha\sigma}^{\dagger} c_{\beta\sigma}^{\phantom{\dagger}} + \sum_{\alpha,\beta} J_{\alpha\beta}\hat{\boldsymbol{S}} \cdot \hat{\boldsymbol{s}}_{\alpha\beta}+ \tilde{B}\hat{S}^z \;,
\end{equation}
where $\hat{\boldsymbol{S}}$ is a spin-$\tfrac{1}{2}$ operator for the nanostructure, $\hat{S}^z$ is its $z$-component, and $\hat{\boldsymbol{s}}_{\alpha\beta}=\tfrac{1}{2}\sum_{\nu,\nu'} c_{\alpha\nu}^{\dagger} \boldsymbol{\sigma}_{\nu\nu'}c_{\beta\nu'}^{\phantom{\dagger}}$ is the spin density of the leads ($\boldsymbol{\sigma}$ is a vector of Pauli matrices). The effective model parameters $J_{\alpha\beta}$ (exchange coupling) and $W_{\alpha\beta}^{\sigma}$ (potential scattering) can be straightforwardly calculated from matrix elements of the isolated nanostructure states, as described in Ref.~\onlinecite{mitchell2017kondo}. Hermiticity requires that $W_{\alpha\beta}^{\sigma}=(W_{\alpha\beta}^{\sigma})^*$ and $J_{\alpha\beta}=(J_{\beta\alpha})^*$. In the absence of a magnetic field or spin-orbit coupling terms in the bare system, the effective model has full SU(2) symmetry and $W_{\alpha\beta}^{\sigma}\equiv W_{\alpha\beta}$ independent of spin, and $\tilde{B}=0$. We consider this case explicitly below. In the following we use the notation $J_{\pm}=J_{ss}\pm J_{dd}$ and $W_{\pm}=W_{ss}\pm W_{dd}$.

The dynamics of the effective model, discussed further below, are characterized by the spectrum of the scattering t-matrix, 
$t_{\alpha\alpha,\sigma}(\omega,T)=-\pi \rho_0 {\rm Im} T_{\alpha\alpha,\sigma}(\omega,T)$, where $T_{\alpha\beta,\sigma}(\omega,T)$ is defined by the t-matrix equation Eq.~\ref{eq:tmatrix}. 
For Eq.~\ref{eq:H2ck}, the t-matrix can be expressed as  $T_{\alpha\beta,\sigma}(\omega,T) = W_{\alpha\beta} + \langle \langle \hat{a}_{\alpha\sigma}^{\phantom{\dagger}} ; \hat{a}_{\beta\sigma}^{\dagger}\rangle\rangle $ where,
\begin{equation}
\hat{a}_{\alpha\uparrow} = \sum_{\gamma} \left [W_{\alpha\gamma} c_{\gamma\uparrow} + \tfrac{1}{2} J_{\alpha\gamma}(c_{\gamma\uparrow}\hat{S}^z + c_{\gamma\downarrow}\hat{S}^{-}) \right ] \;,
\end{equation}
and similarly for $\hat{a}_{\alpha\downarrow}$. The retarded correlator $\langle \langle \hat{a}_{\alpha\sigma}^{\phantom{\dagger}} ; \hat{a}_{\beta\sigma}^{\dagger}\rangle\rangle $ can be computed directly in NRG. However, an `improved' version of the t-matrix can be obtained in a similar fashion to the self-energy method for the retarded nanostructure Green's functions, Eq.~\ref{eq:UFG}. We define a 2x2 matrix Dyson equation for the full (impurity coupled) lead Green's functions, $[\boldsymbol{\mathcal{G}}_{\sigma}(\omega,T)]^{-1}=[\boldsymbol{\mathcal{G}}_{\sigma}^0(\omega)]^{-1} - \boldsymbol{\tilde{\Sigma}}_{\sigma}(\omega,T)$, where the self-energy matrix $\boldsymbol{\tilde{\Sigma}}_{\sigma}(\omega,T)$ incorporates all the effects of the nanostructure. Rearranging Eq.~\ref{eq:tmatrix} gives,
\begin{eqnarray}\label{eq:SE_tm}
\boldsymbol{\tilde{\Sigma}}_{\sigma}(\omega,T) = [\boldsymbol{\mathcal{G}}_{\sigma}(\omega,T)]^{-1}\boldsymbol{\mathcal{G}}_{\sigma}^0(\omega) \boldsymbol{T}_{\sigma}(\omega,T) \;.
\end{eqnarray}
In practice we therefore compute both $\boldsymbol{\mathcal{G}}_{\sigma}(\omega,T)$ and $\boldsymbol{T}_{\sigma}(\omega,T)$ in NRG to obtain $\boldsymbol{\tilde{\Sigma}}_{\sigma}(\omega,T)$ by Eq.~\ref{eq:SE_tm}. This gives an improved estimation of $\boldsymbol{\mathcal{G}}_{\sigma}(\omega,T)$ from the Dyson equation, and hence an improved version of $\boldsymbol{T}_{\sigma}(\omega,T)$ through the t-matrix equation. 

In terms of quantum transport, the Kubo formula can be used to obtain the dc linear response electrical conductance (Eqs.~\ref{eq:kubo}, \ref{eq:kubo_dc}) for the effective model. The retarded current-current correlator $K(\omega,T)=\langle\langle \dot{N}^s ; \dot{N}^d \rangle\rangle \equiv \langle\langle \hat{\Omega} ; \hat{\Omega} \rangle\rangle$ can be expressed in terms of the composite operator,
\begin{eqnarray}\label{eq:KuboKondo}
\hat{\Omega}=-i\left ( J_{sd}\hat{\boldsymbol{S}} \cdot \hat{\boldsymbol{s}}_{sd}+\sum_{\sigma}  W_{sd}^{\phantom{\dagger}}  c_{s\sigma}^{\dagger} c_{d\sigma}^{\phantom{\dagger}} 
\right ) - {\rm H.c.}~~~
\end{eqnarray}
where we have used $\dot{N}^{\alpha}=i[\hat{H},\hat{N}^{\alpha}]$ and current conservation $\dot{N}^{s}=-\dot{N}^{d}$. The form of the `improved' Kubo formula (Sec.~\ref{sec:newkubo}) is model-independent and so Eq.~\ref{eq:newK} remains the same. The corresponding Kubo formula for heat conductance (Eq.~\ref{eq:kubo_heat}) follows similarly.

To gain further insight into the expected behaviour and to make progress analytically, we consider various limiting cases below.

%%%%

\subsubsection{No potential scattering: $W_{\alpha\beta}=0$}\label{sec:W_0}
For $W_{\alpha\beta}=0$, it proves useful to diagonalize the matrix of exchange couplings, $\boldsymbol{U}^{\dagger} \boldsymbol{J} \boldsymbol{U}$, by a canonical rotation of the lead basis $\boldsymbol{c_{\sigma}}=\boldsymbol{U}\boldsymbol{\tilde{c}_{\sigma}}$, where $\boldsymbol{c_{\sigma}}=(c_{s\sigma},c_{d\sigma})^{\rm T}$ and $\boldsymbol{\tilde{c}_{\sigma}}=(c_{e\sigma},c_{o\sigma})^{\rm T}$. In the new `even/odd' basis, the exchange couplings are,
\begin{subequations}\label{eq:JeeJoo}
\begin{align}
J_{ee} &=\tfrac{1}{2} (J_{+}+\sqrt{J_{-}^2+4|J_{sd}|^2}) \label{eq:Jee}\\
J_{oo} &=\tfrac{1}{2}(J_{+}-\sqrt{J_{-}^2+4|J_{sd}|^2}) \label{eq:Joo}
\end{align}
\end{subequations}
and where $J_{eo}=J_{oe}=0$ by construction. 

A special case arises when $J_{oo}=0$, since then the odd lead is decoupled and the effective model reduces precisely to a single-channel Kondo model with a single spin-$\tfrac{1}{2}$ impurity,
\begin{equation}\label{eq:H1ck}
    H_{\rm eff} = H_{\rm leads}^e  + J_{ee} \hat{\boldsymbol{S}} \cdot \hat{\boldsymbol{s}}_{ee} \;.
\end{equation}
From Eq.~\ref{eq:Joo}, this occurs specifically when $|J_{sd}|^2=J_{ss}J_{dd}$. In fact, this condition is always automatically satisfied when the bare model is in PC. Indeed, this is physically natural since Eq.~\ref{eq:H1ck} is the regular Schrieffer-Wolff transformation\cite{SchriefferWolff1966} of the single-channel model using Eq.~\ref{eq:Hhyb_PC} obtained under PC.

The conductance then follows from Eq.~\ref{eq:MW_cond} with 
\begin{eqnarray}
\mathcal{T}_{\rm MW}(\omega,T) \overset{{\rm PC}}{\longrightarrow} \frac{4J_{ss}J_{dd}}{(J_{ss}+J_{dd})^2}\sum_{\sigma} t_{ee,\sigma}(\omega,T) \;.
\end{eqnarray}
With no potential scattering and no magnetic field as considered here, $t_{ee,\sigma}(0,0)=1$ by the Friedel sum rule,\cite{hewson1997kondo} and so $G_C(0)=(2e^2/h)\times \tfrac{4J_{ss}J_{dd}}{(J_{ss}+J_{dd})^2}$ is controlled by a pure geometric factor describing the nanostructure-lead coupling. Indeed for $\omega,T \ll T_K$, with $T_K\sim D \exp[-1/\rho_0 J_{ee}]$ the Kondo temperature for the effective model Eq.~\ref{eq:H1ck}, we have $t_{ee,\sigma}(\omega,T)\simeq 1$ and so the conductance approximately saturates its low-temperature bound, $G_C(T\ll T_K) \simeq G_C(0)$. 

%%%%%%%%%%%%%%%%
\begin{figure}[t!]
\includegraphics[width=8cm]{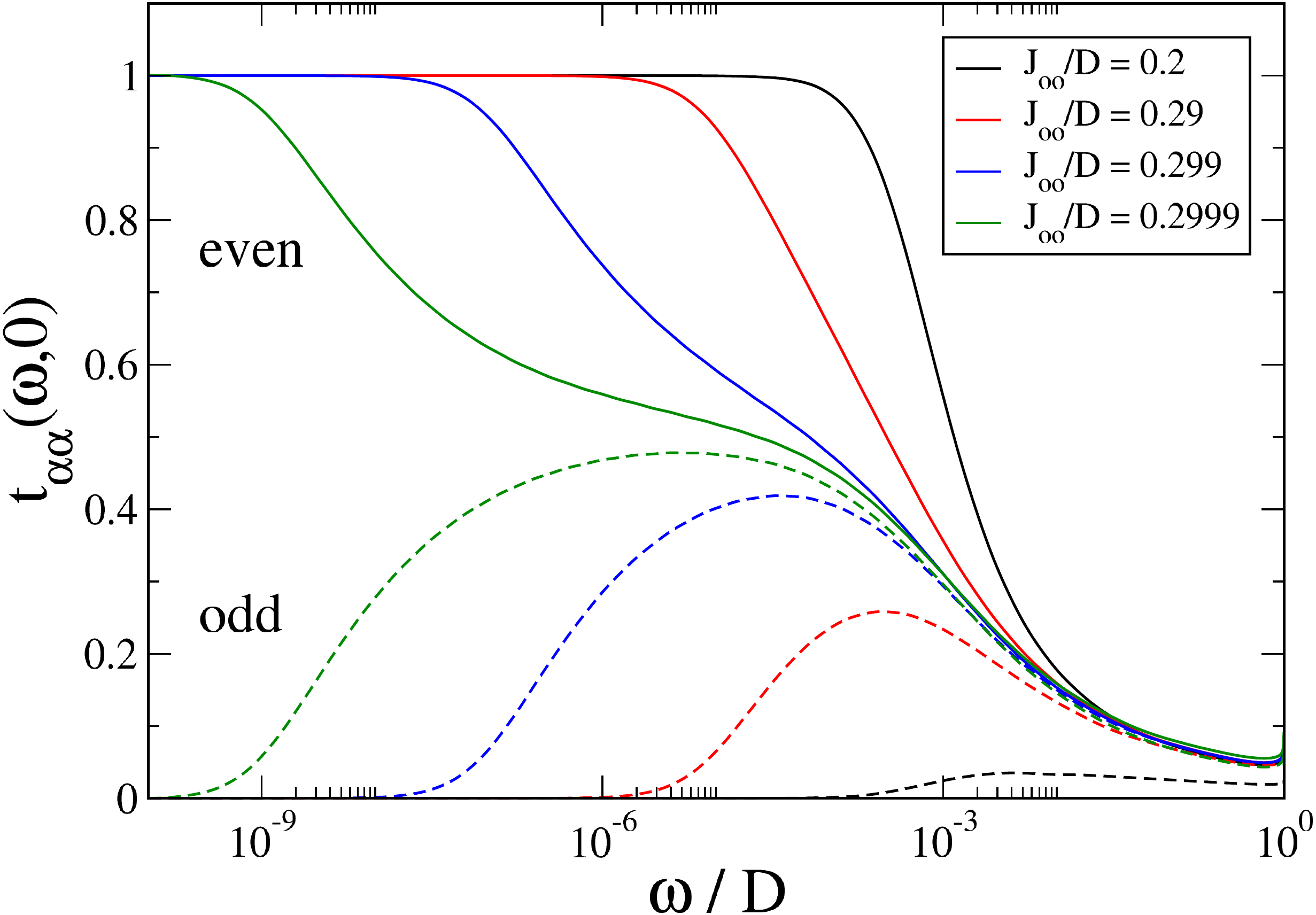}
  \caption{NRG results for the $T=0$ t-matrix spectrum  $t_{\alpha\alpha}(\omega,0)$ of the effective 2CK model, Eq.~\ref{eq:H2ck_eo}. Even (odd) channel spectra shown as the solid (dashed) lines, for fixed $J_{ee}/D=0.3$ and $\boldsymbol{W}=\boldsymbol{0}$, with different $J_{oo}$. In all cases, $t_{ee}(0,0)=1$ and $t_{oo}(0,0)=0$, despite incipient frustration of Kondo screening for small $\delta=J_{ee}-J_{oo}$.
  }
  \label{fig:2ck_tm_W0}
\end{figure}
%%%%%%%%%%%%%%%%

Of course, the above scenario is special, and for most realistic systems the exact PC condition $|J_{sd}|^2=J_{ss}J_{dd}$ is not expected to be satisfied. In such cases, $J_{oo}\ne 0$ and the odd channel remains formally coupled to the effective nanostructure spin-$\tfrac{1}{2}$. In the even/odd basis with $W_{\alpha\beta}=0$, Eq.~\ref{eq:H2ck} reduces to,
\begin{equation}\label{eq:H2ck_eo}
    H_{\rm eff} = H_{\rm leads}  + J_{ee} \hat{\boldsymbol{S}} \cdot \hat{\boldsymbol{s}}_{ee} + J_{oo} \hat{\boldsymbol{S}} \cdot \hat{\boldsymbol{s}}_{oo} \;.
\end{equation}
The physics of this channel-anisotropic 2CK model are rich but well-known, featuring a competition of Kondo screening between the even and odd channels.\cite{coleman1995simple2CK} The model supports a non-Fermi liquid quantum critical point for $J_{ee}=J_{oo}$, but from Eq.~\ref{eq:JeeJoo} we see that this only arises in these kinds of systems when both $J_{ss}=J_{dd}$ and $J_{sd}=0$. Although one may be able to tune to the $sd$-symmetric condition $J_{ee}=J_{oo}$, to realize $J_{sd}=0$ requires the suppression of through-nanostructure exchange-cotunneling processes. In principle, this could arise through gate-tunable many-body quantum interference effects\cite{mitchell2017kondo} that conspire to produce an exact tunneling node, but a real system exhibiting 2CK criticality driven by such effects has not yet been reported. On the other hand, for complex multiorbital nanostructures, electron propagation across the entire structure embodied by the exchange cotunneling $J_{sd}$ is typically small in magnitude (albeit finite) compared with the local terms $J_{ss}$ and $J_{dd}$. Therefore it may still be possible to access the quasi-frustrated quantum critical physics of the 2CK model at small $J_{sd}$ for intermediate temperatures or energies.\cite{mitchell2011two,*vzitko2006kondo} In the following we simply regard $J_{ee}$ and $J_{oo}$ as free independent parameters of the effective model, however obtained, and consider the generic behaviour.

For concreteness we now assume antiferromagnetic couplings $J_{ee}>J_{oo}>0$. Perturbative scaling (poor man's scaling) starting from weak coupling, indicates that both $J_{ee}$ and $J_{oo}$ get renormalized upwards on reducing the energy or temperature scale. The scaling invariant for the RG flow of each is its respective Kondo temperature, $T_{\rm K}^{\alpha} \sim D \exp[-1/\rho_0 J_{\alpha\alpha}]$. However, the low-energy physics is non-perturbative, and the 2CK strong coupling fixed point for a spin-$\tfrac{1}{2}$ impurity is unstable.\cite{nozieres1980kondo} Since $J_{ee}> J_{oo}$ we have $T_{\rm K}^e > T_{\rm K}^o$ and the even channel ultimately flows under RG to strong coupling while the odd channel flows back to weak coupling. 

This physical picture is confirmed by numerically-exact NRG calculations for the effective 2CK model Eq.~\ref{eq:H2ck_eo}, in Fig.~\ref{fig:2ck_tm_W0}. The RG flow is vividly demonstrated in the $T=0$ spectrum of the t-matrix for the even channel (solid lines) and odd channel (dashed lines), plotted for systems with fixed $J_{ee}$ but decreasing $\delta=J_{ee}-J_{oo}>0$ (and $W_{\alpha\beta}=0$). We highlight three features\cite{sela2011exact}: (i) in all cases, the t-matrix spectrum of the even channel is always exactly pinned to 1 at zero frequency and zero temperature, $t_{ee}(0,0)=1$; (ii) by contrast, the odd channel t-matrix vanishes, $t_{oo}(0,0)=0$; (iii) relief of incipient frustration is characterized by a new scale $T^{\rm FL}$. 

Importantly, we see an emergent decoupling of the odd lead at low energies $|\omega|\ll T^{\rm FL}$ in all cases, where $T^{\rm FL}=\min(T_{\rm K}^e,T^*)$ is a dynamically-generated Fermi liquid scale. Here $T^* \sim \delta^2/ T_{\rm K}^e$ is an energy scale embodying the frustration of Kondo screening from the two channels. For large $\delta > T_{\rm K}^e $ we have $T^* > T_{\rm K}^e$ and the RG flow of the odd channel is cut off on the scale of $T_{\rm K}^e$ as the even channel flows to strong coupling. On the other hand, for small $\delta < T_{\rm K}^e $ we have $T^* < T_{\rm K}^e $ and the system is described by the frustrated (non-Fermi liquid) 2CK fixed point over an intermediate temperature range $T^* < T < T_{\rm K}^e $. The even channel flows to strong coupling and the odd channel flows to weak coupling only for $T \ll T^*$ in this case. Without potential scattering, it is then guaranteed that $t_{ee}(0,0)=1$ and $t_{oo}(0,0)=0$, as dictated by phase-shift arguments: due to the decoupling at $\omega=T=0$, we may write $t_{\alpha\alpha}(0,0)=\sin^2(\delta_{\alpha})$. The Kondo effect in the even channel confers a phase shift $\delta_e=\pi/2$ whereas in the decoupled odd channel $\delta_o=0$. 
We note that although the non-Fermi liquid 2CK regime has been accessed experimentally in quantum dot devices,\cite{seaman1991evidence,*ralph19942} this is the result of fine-tuning and ingenious quantum engineering. The more standard situation is for the odd channel to decouple at low temperatures, as described above.

Since such systems flow under RG to an effective single-channel description involving only the even lead combination, we have an \textit{emergent} PC condition. This greatly simplifies the calculation and interpretation of the low-temperature quantum transport. 
On the lowest temperature and energy scales $\omega,T \ll T^{\rm FL} $, Eq.~\ref{eq:oguri_tm} yields the electrical conductance. We may now relate the inter-channel t-matrix $T_{sd,\sigma}$ to the even channel t-matrix $T_{ee,\sigma}$ using the rotation $\boldsymbol{U}$ diagonalizing the exchange matrix, 
$T_{ds,\sigma}(0,0)=\sum_{(\alpha,\beta)\in (e,o)}U_{d\alpha}U_{s\beta}^* T_{\alpha\beta,\sigma}(0,0)$. Note that  $T_{eo,\sigma}(\omega,T)=T_{oe,\sigma}(\omega,T)=0$ since $J_{eo}=W_{eo}=0$, while for $W_{\alpha\beta}=0$ we have $T_{oo}(0,0)=0$ and $|\pi \rho_0T_{ee,\sigma}(0,0)|^2=[t_{ee,\sigma}(0,0)]^2=1$ due to the emergent decoupling. Therefore the $T=0$ conductance follows as,
\begin{equation}\label{eq:G0_emPC_W0}
G_{C}(0)=\left ( \frac{2e^2}{h}\right ) \frac{4 |J_{sd}|^2}{4|J_{sd}|^2 + J_{-}^2}  \;.
\end{equation}
This result also applies accurately for all $T\ll T^{\rm FL}$.
The low-temperature conductance can therefore be determined directly from the effective exchange couplings $J_{ss}$, $J_{dd}$, and $J_{sd}$. Eq.~\ref{eq:G0_emPC_W0} is as such the generalization of Eq.~\ref{eq:MW_G_T0} to the emergent PC case.

Indeed, the factor $|J_{sd}|^2/(|J_{sd}|^2+J_{-}^2)$ appearing in Eq.~\ref{eq:G0_emPC_W0} reduces, as it must to $|V_{s}V_{d}|^2/(|V_{s}|^2+|V_{d}|^2)^2$ in the case of exact PC in the bare model, for which we have 
$|J_{sd}|^2=J_{ss}J_{dd}$ and $J_{ss}/J_{dd}=|V_{s}/V_{d}|^2$. This is equivalent 
to the well-known geometric factor $\Gamma_s \Gamma_d/(\Gamma_s + \Gamma_d)^2$ appearing in Eq.~\ref{eq:MW_G_T0} after cancelling factors of $\pi$ and $\rho_0$. 

%%%%%%%%%%%%%%%%%%%%
%%%%%%%%%%%%%%%%%%%%

%%%%%%%%%%%%%%%%
\begin{figure*}[t!]
\includegraphics[width=15cm]{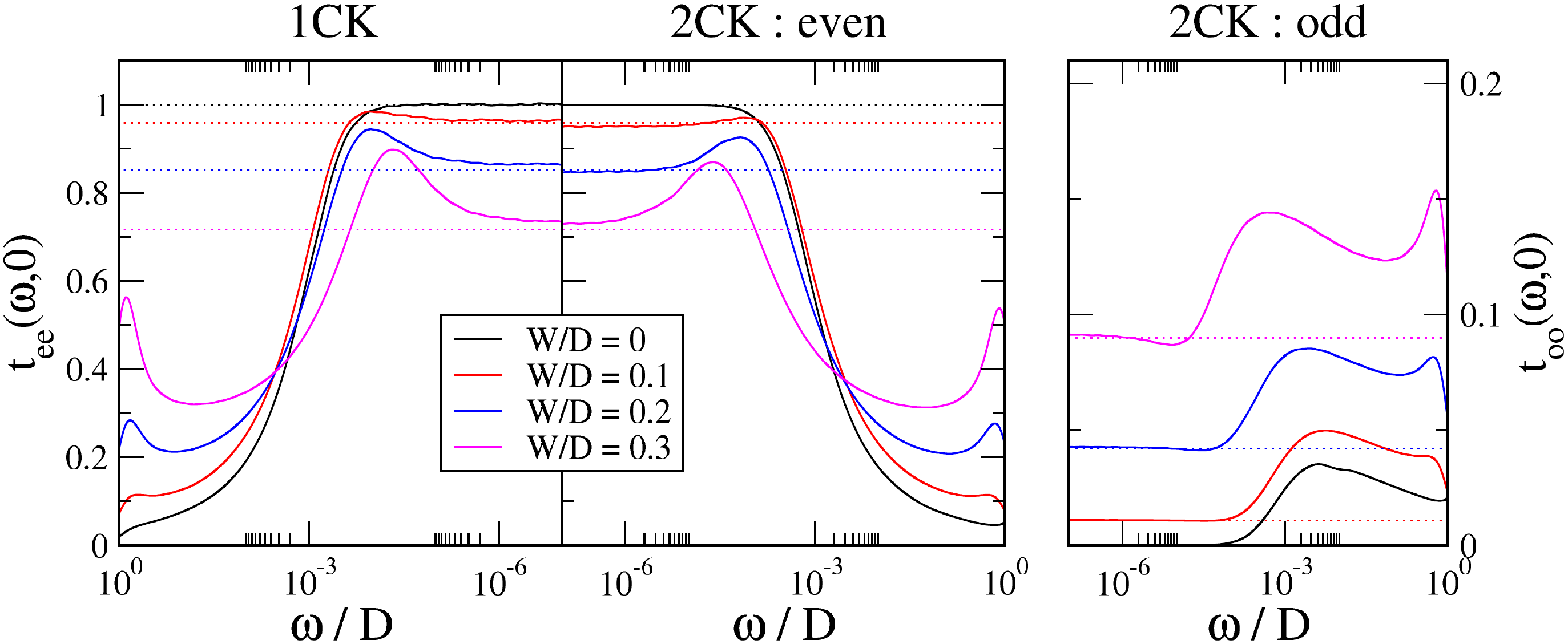}
  \caption{NRG results for the $T=0$ t-matrix spectrum of the effective 1CK model Eq.~\ref{eq:H1ck} (left panel), and 2CK model Eq.~\ref{eq:H2ck_eo_W} (even/odd channels in the centre/right panels). For the 2CK model, we take the $sd$-symmetric case $J_{-}=W_{-}=0$, with fixed $J/D=0.25$ and $J_{sd}/D=0.05$, varying $W=3W_{sd}$. For the 1CK model we use the same parameters as for the even channel of the corresponding 2CK model ($J_{ee}/D=0.3$ and $W_{ee}=\tfrac{4}{3}W$). Dotted lines in the left/centre panels are given by Eq.~\ref{eq:tm_ee_W_sd}, while Eq.~\ref{eq:tm_oo_W_sd} is used in the right panel.
  }
  \label{fig:tm_sym}
\end{figure*}
%%%%%%%%%%%%%%%%

\subsubsection{$sd$-symmetry and finite potential scattering}\label{sec:sd_symm}
The results of the previous section hold when $W_{\alpha\beta}=0$. Although an important simple limit, the resulting behavior is not generic since even at particle-hole symmetry ($W_{ss}=W_{dd}=0$), the through-nanostructure cotunneling term $W_{sd}$ is typically finite. In this section we consider finite potential scattering $W_{\alpha\beta}\ne 0$ in the $sd$-symmetric limit -- meaning $J_{ss}=J_{dd}\equiv J$ and $W_{ss}=W_{dd}\equiv W$ (or equivalently $J_{-}=W_{-}=0$). In this case, the full model has an overall mirror symmetry with respect to exchanging $s$ and $d$ leads 

At high temperatures $T \gg T_{\rm K}$, renormalization due to the Kondo effect is weak, and the contribution to conductance is dominated by finite source-drain cotunneling, $W_{sd}$. Conductance in this limit is therefore given approximately by Eq.~\ref{eq:cond_qpc}, obtained for $J_{\alpha\beta}=0$. However, richer and more complex behavior results at lower temperatures, where there is a subtle interplay between $\boldsymbol{W}$ and $\boldsymbol{J}$. We therefore focus on the low-$T$ physics in the following.

First, we note that in the $sd$-symmetric limit, the even/odd channels are simply the symmetric/antisymmetric combinations of source and drain leads,
\begin{eqnarray}\label{eq:eo}
c_{e\sigma} = \tfrac{1}{\sqrt{2}} [c_{s\sigma} + c_{d\sigma} ] ~~~;~~~ c_{o\sigma} =\tfrac{1}{\sqrt{2}} [c_{s\sigma} - c_{d\sigma} ] 
\end{eqnarray}
which yields,
\begin{subequations}
\begin{align}
J_{ee}&=J+J_{sd} &;\qquad  J_{oo}&=J-J_{sd} \\
W_{ee}&=W+W_{sd} &; ~~~~~ W_{oo}&=W-W_{sd} 
\end{align}
\end{subequations}
and $J_{eo}=J_{oe}=W_{eo}=W_{oe}=0$. Note that with $sd$ symmetry, the lead transformation diagonalizing the exchange matrix $\boldsymbol{J}$ also diagonalizes the potential scattering matrix $\boldsymbol{W}$. We write the effective model,
\begin{equation}\label{eq:H2ck_eo_W}
    H_{\rm eff} = \tilde{H}_{\rm leads}  + J_{ee} \hat{\boldsymbol{S}} \cdot \hat{\boldsymbol{s}}_{ee} + J_{oo} \hat{\boldsymbol{S}} \cdot \hat{\boldsymbol{s}}_{oo} \;,
\end{equation}
where we have incorporated the local potentials $W_{ee}$ and $W_{oo}$ into the definition of the lead Hamiltonian, $\tilde{H}_{\rm leads} = \sum_{\alpha=e,o} [H_{\rm leads}^{\alpha} + W_{\alpha\alpha}^{\phantom{\dagger}} \sum_{\sigma} c_{\alpha\sigma}^{\dagger} c_{\alpha\sigma}^{\phantom{\dagger}}]$. 

Analysis of Eq.~\ref{eq:H2ck_eo_W} proceeds similarly to that of Eq.~\ref{eq:H2ck_eo}, except now we account for the potential scattering by rediagonalizing the lead Hamiltonian $\tilde{H}_{\rm leads}$. This can be done implicitly using Green's function techniques, which allow us to relate the local lead Green's functions including the potential to those of the bare leads. We find  $[\tilde{\mathcal{G}}^0_{\alpha\alpha,\sigma}(\omega)]^{-1}=[\mathcal{G}^0_{\alpha\alpha,\sigma}(\omega)]^{-1} - W_{\alpha\alpha}$ for $\alpha=e,o$. This implies that the renormalized Fermi level density of states is $\tilde{\rho}_{\alpha}=-\tfrac{1}{\pi}{\rm Im} \tilde{\mathcal{G}}^0_{\alpha\alpha,\sigma}(0) = \rho_0/(1+\tilde{W}_{\alpha\alpha}^2)$, where we again use the notation $\tilde{W}_{\alpha\beta}=\pi \rho_0 W_{\alpha\beta}$.

In fact, the quantities controlling the Kondo physics and RG flow in such problems are the dimensionless parameters $j_{\alpha}=\tilde{\rho}_{\alpha}J_{\alpha\alpha}$, rather than the bare exchange couplings $J_{\alpha\alpha}$. For $j_e>j_o$, the even lead flows to strong coupling, while the odd lead decouples; the opposite applies for $j_e<j_o$. Indeed, even with $J_{ee}>J_{oo}$ and $W_{ee}>W_{oo}$, one may be able to realize $j_e=j_o$, as required for 2CK criticality. It may therefore be possible to tune across the 2CK quantum phase transition in a given system by tuning gate voltages.

We now explore the effect of the local potentials $W_{ee}$ and $W_{oo}$ on the dynamics and conductance. For $W_{\alpha\beta}=0$ considered in the previous section, we argued that the emergent decoupling of the odd lead at low energies and temperatures $\omega,T\ll T^{\rm FL}$ produced an effective PC condition, with $t_{ee,\sigma}(0,0)$ and hence $G_C(0)$ then being controlled by an effective \emph{single} channel Kondo (1CK) model, Eq.~\ref{eq:H1ck}. With finite $W_{ee}$ and $W_{oo}$, the lead with the smaller $j_{\alpha}$ still decouples asymptotically. However, $t_{ee,\sigma}(0,0)<1$ and $t_{oo,\sigma}(0,0)>0$ due to the potential scattering.

This physical picture is confirmed in Fig.~\ref{fig:tm_sym} where we plot the $T=0$ spectra of the even and odd t-matrix for the 2CK model Eq.~\ref{eq:H2ck_eo_W} in the centre and right panels, for fixed exchange couplings $J_{ee}$ and $J_{oo}$, but with different potential scattering strengths $W=3W_{sd}$. Below an emergent scale $T^{\rm FL} \sim T_{\rm K}^e$ (here $\sim 10^{-3}D$), the t-matrix of the even channel exhibits a Kondo resonance, while the incipient RG flow of the odd channel is arrested. 

Since the odd channel is decoupled at $\omega=T=0$, the t-matrix $t_{oo,\sigma}(0,0)$ is determined completely by $W_{oo}$. From the t-matrix equation, Eq.~\ref{eq:tmatrix}, together with the bare and renormalized Green's functions 
$\mathcal{G}^0_{oo,\sigma}(0)=-i\pi\rho_0$ and $\tilde{\mathcal{G}}^0_{oo,\sigma}(0,0)=-i\pi\tilde{\rho}_o$, we immediately obtain,
\begin{equation}\label{eq:tm_oo_W_sd}
    t_{oo,\sigma}(0,0)=\frac{\tilde{W}_{oo}^2}{1+\tilde{W}_{oo}^2} \;.
\end{equation}
This result is equivalent to a phase shift $\delta_o= \tan^{-1}[\tilde{W}_{oo}]$ in the odd channel, since $t_{oo,\sigma}(0,0)=\sin^2(\delta_o)$ \cite{langreth1966friedel}. The full odd-channel t-matrix is $\pi\rho_0 T_{oo,\sigma}(0,0)=(\tilde{W}^{-1}_{oo}-i)\times t_{oo,\sigma}(0,0)$.

For the systems considered in Fig.~\ref{fig:tm_sym}, we plot the corresponding values of $t_{oo,\sigma}(0,0)$ obtained from Eq.~\ref{eq:tm_oo_W_sd} in the right panel as the dotted horizontal lines. The full odd-channel t-matrix obtained by NRG is seen to saturate to these values for $|\omega|\ll T^{\rm FL}$, as anticipated.

More interestingly, comparison between the left and centre panels of Fig.~\ref{fig:tm_sym} shows that the low-energy dynamics of the \emph{even} channel can be accurately understood in terms of an effective single-channel Kondo model with the same $J_{ee}$ and $W_{ee}$. This is again a consequence of the emergent decoupling of the odd channel. Indeed, provided $\delta\equiv J_{ee}-J_{oo} \gg T_{\rm K}^e$ (as is fairly standard), the correct Kondo scales are also well reproduced.

For the bare Andersonian nanostructure model, calculating the low-energy behaviour of the scattering t-matrix is deeply nontrivial. Even for the single-impurity Anderson model, $t_{ee,\sigma}(0,0)=\sin^2(\delta_e)$ requires knowledge of the even-channel phase shift $\delta_e$. In this case, the Friedel sum rule dictates that $\delta_e=\tfrac{\pi}{2} n_{\rm imp}$ in terms of the excess charge due to the impurity, $n_{\rm imp}$.\cite{logan2014common} But the latter is renormalized by interactions and must itself be calculated using many-body techniques away from particle-hole symmetry. On the other hand, the above physical arguments and NRG results in Fig.~\ref{fig:tm_sym} establish that $t_{ee,\sigma}(0,0)$ can be obtained from an effective 1CK model, Eq.~\ref{eq:H1ck}. Treating the exchange coupling $J_{ee}$ as a perturbation to the even-channel charge, we may then write $n_{\rm imp} \simeq \tfrac{2}{\pi}\tan^{-1}[\tilde{W}_{ee}]+1$ (it was shown in Ref.~\onlinecite{cragg1979universality} that the phase shifts due to $J_{ee}$ and $\tilde{W}_{ee}$ are additive when the Hamiltonian is transformed in the potential scattering eigenbasis), where the first term is the excess charge induced in a free bath due to a boundary potential $W_{ee}$, and we add $1$ for the singly-occupied impurity local moment in the Kondo model. This approximation gives $\delta_e \simeq \tan^{-1}[\tilde{W}_{ee}]+\tfrac{\pi}{2}$ and hence from the Friedel sum rule,
\begin{equation}\label{eq:tm_ee_W_sd}
    t_{ee,\sigma}(0,0)\simeq \frac{1}{1+\tilde{W}_{ee}^2} \;.
\end{equation}
Using the relation between the phase shift and the argument of the t-matrix \cite{hewson1997kondo} $\delta_e = \arg[T_{ee,\sigma}(0,0)]$ the full even-channel t-matrix can be determined to be $\pi\rho_0 T_{ee,\sigma}(0,0)=-(\tilde{W}_{ee}+i)\times t_{ee,\sigma}(0,0)$.

%%%%%%%%%%%%%%%%
\begin{figure}[t!]
\includegraphics[width=8.5cm]{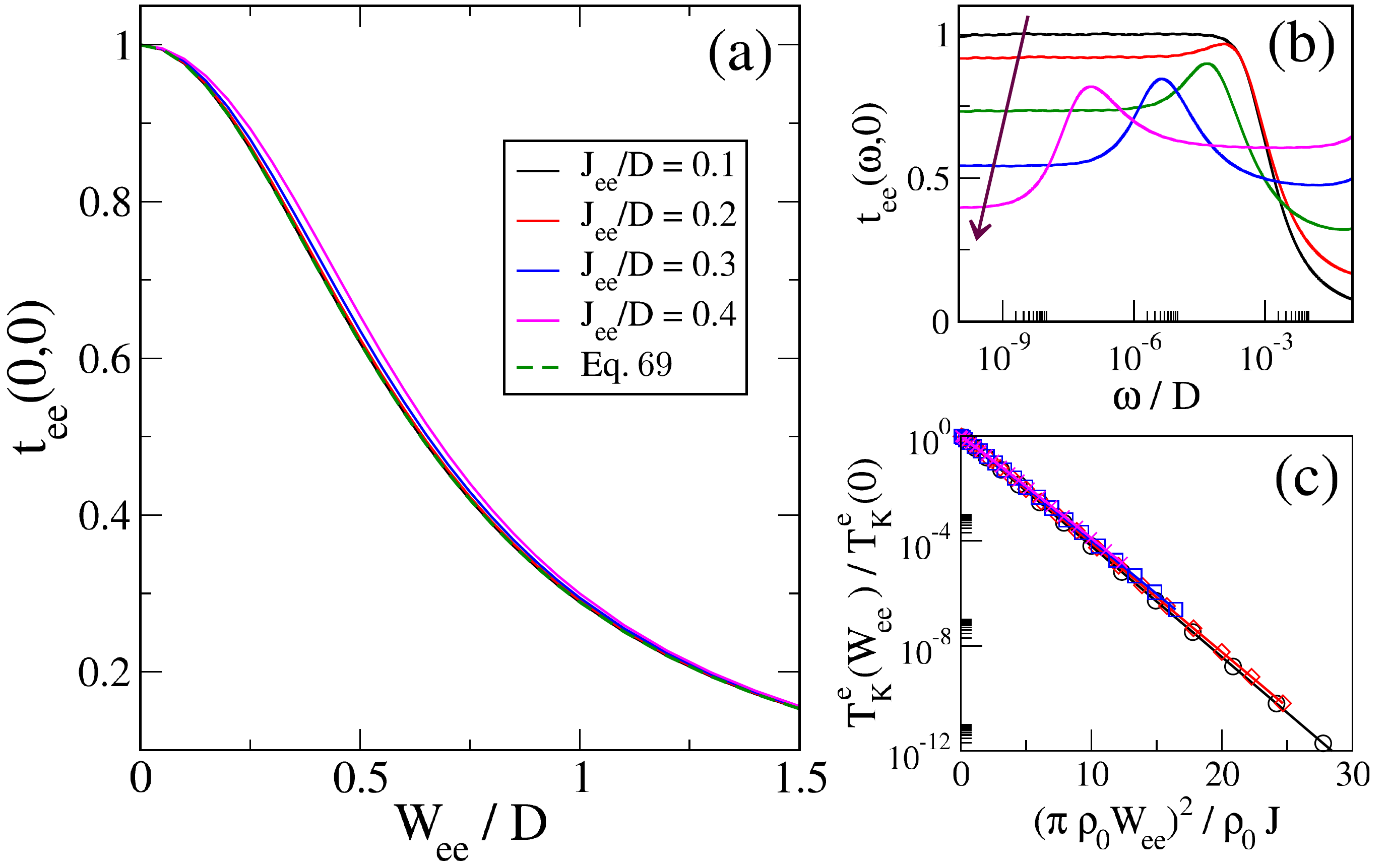}
  \caption{NRG results for the effective 1CK model, Eq.~\ref{eq:H1ck}. (a) $T=0$ spectrum of the t-matrix at the Fermi energy $t_{ee}(0,0)$, as a function of the local potential $W_{ee}$, for different $J_{ee}$. Dashed line is Eq.~\ref{eq:tm_ee_W_sd}, showing only small deviations even at rather large $J_{ee}$. (b) The full energy-dependence of the t-matrix $t_{ee}(\omega,0)$ is shown for $J_{ee}/D=0.3$ with $W_{ee}/D=0, 0.2, 0.4, 0.6, 0.8$ increasing in the direction of the arrow. (c) Dependence of the Kondo temperature $T_{\rm K}^e$ on $W_{ee}$ for the same $J_{ee}$ used in (a), confirming Eq.~\ref{eq:tk_1ck}.
  }
  \label{fig:tm_1ck}
\end{figure}
%%%%%%%%%%%%%%%%

These predictions are quantitatively substantiated in Fig.~\ref{fig:tm_1ck}, which shows NRG results for a pure 1CK model with exchange coupling $J_{ee}$ and potential scattering $W_{ee}$. The t-matrix $t_{ee,\sigma}$ at $\omega=T=0$ is plotted in panel (a) as a function of $W_{ee}$ for different $J_{ee}$. Even for rather large $J_{ee}/D=0.4$, the exact results agree accurately with Eq.~\ref{eq:tm_ee_W_sd} (dashed line) over the entire range of $W_{ee}$. 

The full energy dependence of the t-matrix is shown in panel (b) for fixed $J_{ee}$ with increasing $W_{ee}$. The asymptotic $\omega=0$ value reported in (a) can be extracted for $|\omega|\ll T^e_{\rm K}$, with the Kondo scale $T^e_{\rm K}$ decreasing as the potential scattering strength $W_{ee}$ increases. Since $T^e_{\rm K} \sim D \exp[-1/\tilde{\rho}_e J_{ee}]$ and the Fermi level density of states incorporating $W_{ee}$ is $\tilde{\rho}_e=\rho_0/(1+\tilde{W}_{ee}^2)$ as before, for the 1CK model we may write,
\begin{equation}\label{eq:tk_1ck}
T^e_{\rm K}(W_{ee}) \simeq T^e_{\rm K}(0)\times \exp[-\tilde{W}^2_{ee}/\rho_0 J_{ee}] \;.
\end{equation}
This is confirmed in Fig.~\ref{fig:tm_1ck}(c) for the same systems as in panel (a).

Eq.~\ref{eq:tm_ee_W_sd} can therefore be used to find the low-energy behaviour of the even-channel t-matrix in the generalized 2CK model,\cite{Kondo1964} while Eq.~\ref{eq:tk_1ck} provides an estimate of the regime of applicability of this result. Remarkably, the low-energy scattering behaviour can therefore be extracted purely from a knowledge of the effective model parameters $J_{ee}$ and $W_{ee}$, and full solution of the generalized 2CK model is not required. 

%%%%%%%%%%%%%%%%
\begin{figure}[t!]
\includegraphics[width=8.5cm]{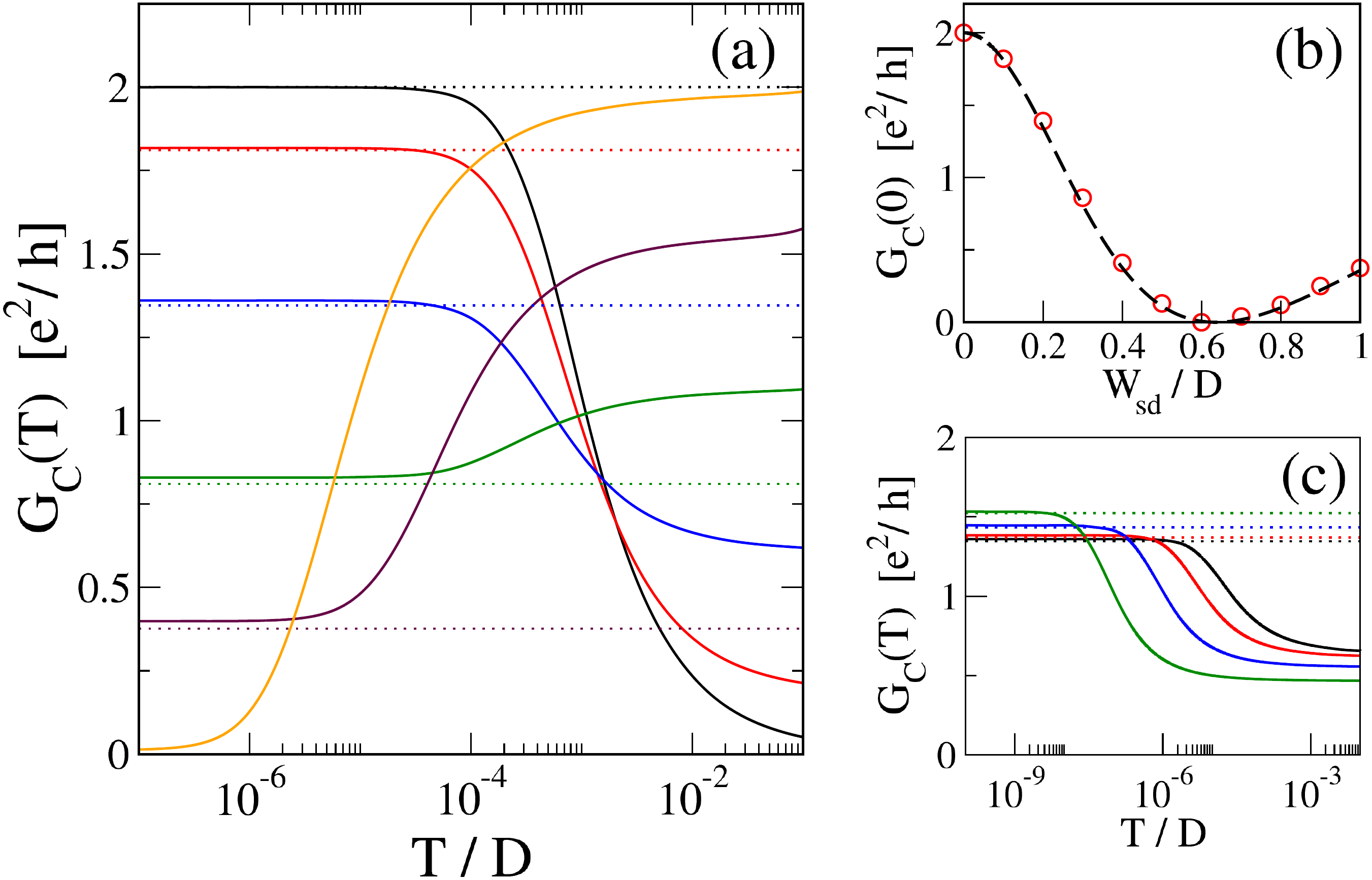}
  \caption{Conductance of the generalized 2CK model Eq.~\ref{eq:H2ck} at $sd$-symmetry, obtained by NRG using the improved Kubo method. (a) $G_C(T)$ vs $T$ for fixed $J/D=0.25$, $J_{sd}/D=0.05$ and $W=0$ with different $W_{sd}/D=0, 0.1, 0.2, 0.3, 0.4, 0.6$ for the solid black, red, blue, green, violet and orange lines. Dotted lines are the low-$T$ predictions of Eq.~\ref{eq:cond_2ck_sym}. (b) $G_C(0)$ vs $W_{sd}$, for systems with the same parameters as in (a), showing the conductance node near $\tilde{W}_{sd}=1$, compared with Eq.~\ref{eq:cond_2ck_sym} (dashed line). (c) $G_C(T)$ vs $T$ for fixed $J/D=0.15$, $J_{sd}/D=0.05$, $W/D=0.2$, varying $W_{sd}/D=0, 0.1, 0.2, 0.3$ for the black, red, blue and green solid lines (again compared with Eq.~\ref{eq:cond_2ck_sym}, dotted lines). 
  }
  \label{fig:cond_2ck_sym}
\end{figure}
%%%%%%%%%%%%%%%%

Finally, we use this information on the dynamics to extract the generic low-temperature conductance properties. 
We use Eq.~\ref{eq:oguri_tm} in the even/odd basis of Eq.~\ref{eq:eo} to write $G_C(0)=(2e^2/h)\times |\pi\rho_0 T_{ee,\sigma}(0,0)-\pi\rho_0 T_{oo,\sigma}(0,0)|^2$. Using the form of the even/odd t-matrices derived in this section, we obtain,
\begin{equation}\label{eq:cond_2ck_sym}
G_{C}(0)=\left ( \frac{2e^2}{h}\right ) \frac{(1+\tilde{W}_{ee}\tilde{W}_{oo})^2}{(1+\tilde{W}_{ee}^2)(1+\tilde{W}_{oo}^2)}\;.
\end{equation}
This is the main result of this section: it allows the Kondo-renormalized low-$T$ conductance to be calculated purely from a knowledge of the effective model parameters. Note that the result holds approximately for all $T \ll T^{\rm FL}$. 

In the case of $sd$-symmetry as considered here, only $W_{ee}=W+W_{sd}$ and $W_{oo}=W-W_{sd}$ are required. Furthermore,  Eq.~\ref{eq:cond_2ck_sym} implies the existence of an exact quantum-interference effect conductance node, 
\begin{equation}\label{eq:QInode}
G_C(0)=0\qquad\text{when}\qquad W_{sd}^2=W^2+(1/\pi\rho_o)^2 
\end{equation}
This condition can in principle be satisfied on tuning gate voltages in multi-orbital nanostructures. Although conductance nodes due to single-particle quantum interference in effective non-interacting systems are well-known, their many-body counterparts are comparatively poorly understood -- even though strong electron interactions are ubiquitous in e.g.~semiconductor quantum dot devices and single molecule transistors. We emphasize that in such systems, electron interactions are crucial to understand the low-temperature conductance properties. Indeed, the $T=0$ conductance node, Eq.~\ref{eq:QInode}, is driven by the Kondo effect (a finite conductance is expected at higher temperatures, $T\gg T_{\rm K}$). This ``Kondo Blockade'' has been proposed as a mechanism for efficient quantum interference effect transistors.\cite{mitchell2017kondo} The above analysis puts such phenomena in the context of a more general framework. 

The quantitative accuracy of Eq.~\ref{eq:cond_2ck_sym} is established by NRG results in Fig.~\ref{fig:cond_2ck_sym}. 
The dc conductance $G_C(T)$ of the full 2CK model,  Eq.~\ref{eq:H2ck}, under $sd$ symmetry is obtained within NRG using the improved Kubo approach, Eqs.~\ref{eq:kubo} and \ref{eq:newK}. 
In Fig.~\ref{fig:cond_2ck_sym}(a) we plot $G_C$ as a function of $T$ for fixed $J$, $J_{sd}$ and $W=0$, with different $W_{sd}$. Although the high-$T$ conductance is rather accurately reproduced by \ref{eq:cond_qpc}, Kondo correlations strongly affect the behaviour at low temperatures $T\ll T_{\rm K}$, with $T_{\rm K}$ the emergent Kondo temperature. The horizontal dotted lines are the result of Eq.~\ref{eq:cond_2ck_sym}, and are seen to agree very well with the low-temperature conductance from NRG. In particular, the Kondo resonance at $W_{sd}=0$ (black line) is inverted to a Kondo blockade for $W_{sd} = 0.6D$ (orange line). $G_C(0)$ is plotted for the same parameters in Fig.~\ref{fig:cond_2ck_sym}(b) as a function of $W_{sd}$, with NRG results (points) compared with Eq.~\ref{eq:cond_2ck_sym} (dashed line), showing the conductance node predicted by Eq.~\ref{eq:QInode}. Fig.~\ref{fig:cond_2ck_sym}(c) confirms the more general form of Eq.~\ref{eq:cond_2ck_sym} for finite $W$ and $W_{sd}$.

%%%%%%%%%%%%%%%%%%%%
%%%%%%%%%%%%%%%%%%%%

\subsubsection{finite potential scattering and broken $sd$-symmetry}
Finally, we consider the most general case, in which both $\boldsymbol{J}$ and $\boldsymbol{W}$ are finite, and $sd$-symmetry is broken ($J_{-}\ne 0$ and $W_{-}\ne 0$). 

First, we note that for arbitrary $J_{\pm}$, $J_{sd}$ and $W_{\pm}$, $W_{sd}$ the lead basis rotation $\boldsymbol{U}$ that diagonalizes the exchange matrix $\boldsymbol{J}$ will not in general diagonalize the potential scattering matrix $\boldsymbol{W}$ (that is, although $\boldsymbol{U}^{\dagger} \boldsymbol{J}\boldsymbol{U}$ is constructed such that $J_{eo}=J_{oe}^*=0$, $\boldsymbol{U}^{\dagger} \boldsymbol{W}\boldsymbol{U}$ still yields finite $W_{eo}=W^*_{oe}\ne 0$). This means that the even/odd channels defined in this way are not distinct; and in particular the odd channel does not completely decouple at low energies, as required for the previous analysis.

The first step is therefore to define a new basis, in terms of which we can identify \emph{distinct} conduction electron channels, labelled $a$ and $b$. We define $\boldsymbol{\mathcal{W}}=\boldsymbol{P}^{\dagger} \boldsymbol{W}\boldsymbol{P}$ to be diagonal, such that,
\begin{subequations}\label{eq:WeeWoo}
\begin{align}
\mathcal{W}_{aa} &=\tfrac{1}{2} (W_{+}+\sqrt{W_{-}^2+4|W_{sd}|^2}) \label{eq:Wee}\\
\mathcal{W}_{bb} &=\tfrac{1}{2}(W_{+}-\sqrt{W_{-}^2+4|W_{sd}|^2}) \label{eq:Woo}
\end{align}
\end{subequations}
and where $\mathcal{W}_{ab}=\mathcal{W}^*_{ba}=0$ by construction. Eq.~\ref{eq:WeeWoo} is as such the potential scattering analog of Eq.~\ref{eq:JeeJoo} for the exchange couplings. Incorporating these local potentials into the definition of the free lead Hamiltonian in the even/odd basis $\mathcal{H}_{\rm leads}$, we have renormalized Fermi level densities of states $\tilde{\rho}_{\alpha}=\rho_0/[1+(\pi\rho_0\mathcal{W}_{\alpha\alpha})^2]$. The same transformation $\boldsymbol{P}$ is applied to the exchange matrix $\boldsymbol{J}$, noting that $\boldsymbol{\mathcal{J}}=\boldsymbol{P}^{\dagger} \boldsymbol{J}\boldsymbol{P}$ is not in general diagonal.

We may now identify the relevant dimensionless quantities to analyze the RG flow as $j_{\alpha\beta}=\sqrt{\tilde{\rho}_{\alpha}\tilde{\rho}_{\beta}} \mathcal{J}_{\alpha\beta}$. The $2\times 2$ matrix of these couplings is denoted $\boldsymbol{j}$. We perform a final rotation of the lead basis to diagonalize this matrix, $\boldsymbol{j}'=\boldsymbol{Q}^{\dagger} \boldsymbol{j}\boldsymbol{Q}$, to yield the desired even/odd exchange couplings $j'_{ee}$ and $j'_{oo}$. It is in \emph{this} basis that we can say that
the more strongly coupled even channel flows to strong coupling and the more weakly coupled odd channel decouples,\cite{note_eo} analogous to the analysis of the previous section. Potential scattering in this basis is given by $\boldsymbol{W}'=\boldsymbol{Q}^{\dagger} \boldsymbol{\mathcal{W}}\boldsymbol{Q}$, and we define $\tilde{W}'_{\alpha\beta}=\pi\rho_0 W'_{\alpha\beta}$.

The emergent decoupling of the odd channel in this basis at low energies and temperatures $\omega, T \ll T_{\rm K}$ yields the non-trivial result $t'_{eo}(0,0)=t'_{oe}(0,0)=0$. This is a consequence of Kondo renormalization, and scattering between even and odd channels is finite at higher energies. In the fully asymmetric case, there is no lead rotation that yields $t_{eo}(\omega,T)=0$ at all energies and temperatures. The above prescription, although somewhat complicated, is needed in order to identify the specific lead combination where inter-channel scattering vanishes at low energies. We have confirmed this result explicitly with NRG. 

In this basis, the analog of Eqs.~\ref{eq:tm_oo_W_sd} and \ref{eq:tm_ee_W_sd} for the t-matrix spectra are $t'_{oo,\sigma}(0,0)=(\tilde{W}'_{oo})^2/(1+(\tilde{W}'_{oo})^2)$ and $t'_{ee,\sigma}(0,0)=1/(1+(\tilde{W}'_{ee})^2)$, with the corresponding complex t-matrices given by $\pi\rho_0 T'_{oo,\sigma}(0,0)=[(\tilde{W}')^{-1}_{oo}-i]\times t'_{oo,\sigma}(0,0)$ and $\pi\rho_0 T'_{ee,\sigma}(0,0)=-[\tilde{W}'_{ee}+i]\times t'_{ee,\sigma}(0,0)$. The low-$T$ conductance then follows from Eq.~\ref{eq:oguri_tm},
\begin{equation}\label{eq:cond_gen}
G_C(0) = \left (\frac{2e^2}{h}\right)  | \Omega_e^{\phantom{'}}  T'_{ee,\sigma}(0,0) + \Omega_o^{\phantom{'}}  T'_{oo,\sigma}(0,0)|^2 \;,
\end{equation}
where the t-matrix $T_{ds,\sigma}$ in the physical basis of $d$ and $s$ leads is brought first into the basis of $a$ and $b$ channels using the transformation $\boldsymbol{P}$, and then into the even/odd basis via $\boldsymbol{Q}$. This yields the coefficients,
\begin{equation}
    \Omega_{\gamma} = 2\pi\rho_0\sum_{\alpha,\beta \in a,b} P_{d\alpha}Q_{\alpha \gamma} P^*_{s\beta}Q^*_{\beta \gamma} \;.
\end{equation}
%\begin{subequations}
%\begin{align}
%x&=(P_{da} Q_{ae} + P_{db} Q_{be}) (P_{sa} Q_{ae} + P_{sb} Q_{be}) \\
%y&=(P_{da} Q_{ao} + P_{db} Q_{bo}) (P_{sa} Q_{ao} + P_{sb} Q_{bo})
%\end{align}
%\end{subequations}
The low-temperature conductance obtained from NRG is found to agree well with these predictions. 

We emphasize that the benefit of the above framework is that the low-temperature conductance can be accurately estimated, purely from a knowledge of the effective 2CK model parameters $\boldsymbol{J}$ and $\boldsymbol{W}$. The 2CK model itself does not need to be solved. Furthermore, since any microscopic multi-orbital system hosting a net spin-$\tfrac{1}{2}$ coupled to source and drain leads can in principle be mapped to such a model, the problem of calculating low-$T$ transport becomes one of determining effective model parameters. These can be estimated perturbatively using Eq.~\ref{eq:BWPT}, or non-perturbatively using more sophisticated techniques, such as model machine learning.\cite{rigo2020machine}

%%%%%%%%%%%%%%%%%%%%%%%%%

\subsection{$S=1$}\label{sec:S1}
When the nanostructure hosts an even number of electrons, a high-spin state with $S=1$ may result. This is especially prevalent in single molecule junctions. The effective model in this case is again Eq.~\ref{eq:H2ck}, but now with $\hat{\boldsymbol{S}}$ a spin-1 operator ($\hat{\boldsymbol{s}}_{\alpha\beta}$ are still spin-$\tfrac{1}{2}$ operators for the leads).

In the case with exact PC on the level of the bare model, an effective single-channel spin-1 Kondo model results (Eq.~\ref{eq:H1ck} with $\hat{\boldsymbol{S}}$ a spin-1 operator). The low-energy physics is that of the underscreened Kondo effect \cite{nozieres1980kondo} -- a singular Fermi liquid with a residual free spin-$\tfrac{1}{2}$ local moment surviving down to $T=0$. Conductance is again Kondo-enhanced in this case, but with logarithmic corrections to the low-temperature approach to the fixed point conductance.\cite{mehta2005regular} This situation can arise in parallel double quantum dots; although the exact PC condition on the hybridization matrix is challenging to realize in practice.

The more generic scenario is when the effective nanostructure spin-1 is delocalized and the PC condition is not satisfied. In this case, both $J_{ee}$ and $J_{oo}$ remain finite in the even/odd lead basis. The nanostructure spin-1 is exactly screened on the lowest temperature scales\cite{Coleman_conductanceS1_PRL2005} (assuming both exchange couplings are antiferromagnetic). However, since $J_{ee}>J_{oo}$ we have $T_{\rm K}^e > T_{\rm K}^o$ and a `two-stage' Kondo effect results: the odd lead participates in screening of the nanostructure spin-1, and remains coupled down to $T=0$, in contrast to the spin-$\tfrac{1}{2}$ case discussed above. The nanostructure spin-1 is underscreened by the even lead to an effective spin-$\tfrac{1}{2}$ on the scale of $T_{\rm K}^e$, and then fully screened down to  $S=0$ by the odd lead on the scale of $T_{\rm K}^o$.  Correspondingly, the conductance is Kondo enhanced for $T_{\rm K}^o < T < T_{\rm K}^e$ but suppressed on the lowest temperature scales $T\ll T_{\rm K}^o$. These scenarios have been explored experimentally in e.g.~Refs.~\onlinecite{paaske2006non,roch2009observation,parks2010mechanical,kurzmann2021kondo}.

%###########################
%###########################

\section{Emergent Proportionate Coupling: Charge Degeneracy}\label{sec:emPC_MV}
In this section we consider the generic electrical conductance properties of systems near a Coulomb peak, corresponding to the charge degeneracy point between $N$ and $N+1$ electrons on the nanostructure. Typically, one can tune to the Coulomb peaks in experiment by applying suitable plunger gate voltages. We expect enhanced cotunneling conductance at such points due to strong nanostructure charge fluctuations (mixed-valence).\cite{DeFranceschi_ElectronCotunneling2001Exp} By contrast with the Coulomb blockade situation of the previous section,\cite{NazarovCoTun1990} here electrons can tunnel from source to drain leads without leaving the nanostructure ground state manifold; although electron interactions still typically renormalize the conductance.

Our aim in this section is to understand the low-temperature conductance properties of arbitrary systems in this regime, in terms of simple effective models. By studying the generic behaviour of such models, we again ultimately relate the conductance directly to the effective model parameters, avoiding the need to solve the effective models each time for every new system considered.

Effective models near charge-degeneracy points can be obtained perturbatively to \emph{first} order in the hybridization $H_{\rm hyb}$, which connects $N$ and $N+1$ electron states of the isolated nanostructure. We define $\hat{1}_{\rm gs} = \sum_{j=1}^{m_N} |N;j\rangle\langle N;j| + \sum_{k=1}^{m_{N+1}} |N+1;k\rangle\langle N+1;k|$ as a projector onto the manifold of \emph{retained} states in the $N$ and $N+1$ electron sectors of $H_{\rm nano}$. We only keep the $m_N$ lowest energy states of the $N$ electron sector and the $m_{N+1}$ lowest states of the $N+1$ electron sector. Here $H_{\rm nano} |n;l\rangle = E^n_l |n;l\rangle$, where the index $l=1,2,3,...$ labels eigenstates of the isolated $n$-electron nanostructure in order of increasing energy $E_l^n$. The truncated set of states included in $\hat{1}_{\rm gs}$ will be sufficient to describe the transport at low temperatures $T\ll \Delta E_{\rm min}$, where $\Delta E_{\rm min}$ is the energy of the lowest lying state not included in $\hat{1}_{\rm gs}$. The typically large charging energy in small nanostructures means that removing an electron from an $N$ electron ground state, or adding an electron to an $N+1$ electron ground state, yields a highly excited state that can be safely neglected. However, one should also check that the gap to excitations within the $N$ and $N+1$ electron sectors is large enough to justify the specific truncation scheme used. Therefore we take $\Delta E_{\rm min} = \min [E_1^{N-1}, E_1^{N+2}, E_{m_N+1}^N, E_{m_{N+1}+1}^{N+1} ]$. In practice, the $T\to 0$ transport is typically controlled by the lowest spin multiplet in $N$ and $N+1$ electron sectors, and we now focus on these situations. The effective model is then defined as,
\begin{equation}\label{eq:Heff_mv}
    H_{\rm eff} = H_{\rm leads} + \sum_{n=N}^{N+1} \sum_{l=1}^{m_n} E_l^n |n;l\rangle\langle n;l|+ \hat{1}_{\rm gs} H_{\rm hyb} \hat{1}_{\rm gs}
\end{equation}
We consider specific prominent cases below, and give example applications for the TQD in Sec.~\ref{sec:tqd}.

%%%%%%%%%%%%%%

\subsection{Spinless case}\label{sec:spinless}
When the nanostructure and leads are effectively spinless (which may arise physically in the presence of a strong magnetic field, for example) Eq.~\ref{eq:Heff_mv} takes a simple form. In particular, the standard non-degenerate case where a single state can be retained in each of the $N$ and $N+1$ electron sectors yields an effective resonant level model,\cite{hewson1997kondo}
\begin{equation}\label{eq:Heff_mv_spinless}
    H_{\rm eff} = H_{\rm leads} + \epsilon f^{\dagger}f + \sum_{\alpha=s,d} \left ( t_{\alpha}^{\phantom{\dagger}} f^{\dagger}c^{\phantom{\dagger}}_{\alpha} + t_{\alpha}^* c_{\alpha}^{\dagger}f^{\phantom{\dagger}} \right )\;,
\end{equation}
where $f^{\dagger}=|N+1;1\rangle\langle N;1|$ and $f=(f^{\dagger})^{\dagger}$ are canonical fermionic operators with the proper anticommutation relations. Here the effective model parameters are 
$\epsilon=E_1^{N+1}-E_1^{N}$ and $t_{\alpha}=V_{\alpha}\langle N+1;1 | \bar{d}_{\alpha}^{\dagger} | N;1\rangle$, which can be obtained straightforwardly from the diagonal representation of the isolated $H_{\rm nano}$. 

Importantly, note that the effective model Eq.~\ref{eq:Heff_mv_spinless} is in PC, even though the bare model is in general not. This emergent PC property greatly facilitates quantum transport calculations at low energy and temperature scales where the effective model provides a valid description. Furthermore, the effective model is non-interacting, despite the strong electron correlations typically present on the physical nanostructure. This means we can write down the effective $f$-level Green's function exactly,
\begin{equation}\label{eq:MV_G_spinless}
    G_{\textit{ff}}(\omega)=\frac{1}{\omega^+ - \epsilon - \sum_{\alpha}|t_{\alpha}|^2\mathcal{G}^0_{\alpha\alpha}(\omega) }\;.
\end{equation}
Electrical and heat conductances then follow immediately from the Landauer formula, Eq.~\ref{eq:land}, with the transmission function being the analog of Eq.~\ref{eq:TlandauerPC},
\begin{equation}\label{eq:MV_T_spinless}
    \mathcal{T}_{\rm L}(\omega) = 4 \tilde{\Gamma}_s\tilde{\Gamma}_d |G_{\textit{ff}}(\omega)|^2 \simeq \frac{4 \tilde{\Gamma}_s\tilde{\Gamma}_d }{(\omega-\epsilon)^2 + (\tilde{\Gamma}_s+\tilde{\Gamma}_d)^2} \;,
\end{equation}
where $\tilde{\Gamma}_{\alpha}=\pi\rho_0 |t_{\alpha}|^2$ and we have used the wide flat band approximation $|t_{\alpha}|^2\mathcal{G}^0_{\alpha\alpha}(\omega)\simeq -i\tilde{\Gamma}_{\alpha}$. Eq.~\ref{eq:MV_T_spinless} is as such a Lorentzian of width $(\tilde{\Gamma}_s+\tilde{\Gamma}_d)$ and centred on $\omega=\epsilon$. The effective parameters $\epsilon$, $t_s$ and $t_d$ are therefore sufficient to calculate the transport properties for $T\ll \Delta E_{\rm min}$. In particular, note that the maximum low-$T$ conductance is obtained at the charge degeneracy point (where $\epsilon=0$) and when $t_s=t_d$ (which does not necessarily imply the symmetry $V_s=V_d$ in the bare model). 

%%%%%%%%%%%%%%

\subsection{Singlet-doublet transition}\label{sec:MVsd}
A more generic and interesting situation arises in spinful systems where states of $H_{\rm nano}$ form a spin multiplet structure. Here we focus on the most common case where the ground state of the nanostructure with an even number of electrons $N$ is a non-degenerate spin-singlet, while with (odd) $N+1$ electrons we have a spin-doublet. Both singlet and doublet must be taken into account in the effective model near the charge degeneracy point between $N$ and $N+1$ electrons.

The three accessible states in the low-energy nanostructure manifold can be represented by a single spinful fermionic degree of freedom with operators $f_{\sigma}^{\dagger}=|N+1;\sigma;1\rangle\langle N;0;1|$ and $f_{\sigma}^{\phantom{\dagger}}=(f_{\sigma}^{\dagger})^{\dagger}$, together with a constraint to exclude spurious double occupancy of this effective site. Here we have additionally labelled states of $H_{\rm nano}$ by their total $S^z$ quantum number $|n;S^z;l\rangle$, where $\sigma=\uparrow,\downarrow$ denotes $S^z=+\tfrac{1}{2},-\tfrac{1}{2}$, and $l$ now labels the spin multiplet. The corresponding energies are $E^{n}_{S^z;l}$. Since we consider only the $l=1$ ground state multiplets, we suppress the $l$ labels below. The effective model Eq.~\ref{eq:Heff_mv} then reads,
\begin{equation}\label{eq:Heff_mv_s-d}
\begin{split}
    H_{\rm eff} = H_{\rm leads} + \epsilon \hat{n}_f  &+ B_{\rm eff} \hat{S}^z_f + U_c \hat{n}_{f\uparrow}\hat{n}_{f\downarrow}\\
    &+ \sum_{\substack{\alpha=s,d\\ \sigma=\uparrow,\downarrow}} \left ( t_{\alpha\sigma}^{\phantom{\dagger}} f^{\dagger}_{\sigma}c^{\phantom{\dagger}}_{\alpha\sigma} + \rm{H.c.} \right )\;,
\end{split}
\end{equation}
where $\hat{n}_{f\sigma}= f_{\sigma}^{\dagger}f_{\sigma}^{\phantom{\dagger}}$, $\hat{n}_f=\hat{n}_{f\uparrow}+\hat{n}_{f\downarrow}$ and $\hat{S}^z_f=\tfrac{1}{2}(\hat{n}_{f\uparrow}-\hat{n}_{f\downarrow})$. Here $U_c\to \infty$ implements the hard-core constraint, while the effective parameters are given by $\epsilon=\tfrac{1}{2}[E^{N+1}_{\uparrow}+E^{N+1}_{\downarrow}]-E^N_{0}$, $B_{\rm eff}=[E^{N+1}_{\uparrow}-E^{N+1}_{\downarrow}]$ and $t_{\alpha\sigma}=V_{\alpha}\langle N+1;\sigma | \bar{d}_{\alpha\sigma}^{\dagger} | N;0\rangle$. Eq.~\ref{eq:Heff_mv_s-d} is as such an infinite-$U$ single impurity Anderson model in the mixed-valent regime,\cite{hewson1997kondo} which possesses an emergent PC (independent of whether or not the bare model is in PC). This implies that a single effective conduction electron channel couples to the impurity $f$ level, $c_{e\sigma}=\tfrac{1}{t_{\sigma}^2}[t_{s\sigma}c_{s\sigma}+t_{d\sigma}c_{d\sigma}] $, where $t_{\sigma}^2=|t_{s\sigma}|^2+|t_{d\sigma}|^2$, and with effective hybridization $\tilde{\Gamma}_{\sigma}=\pi\rho_0 t_{\sigma}^2$.

Another consequence of the emergent PC in the effective model is that quantum transport at low temperatures $T\ll \Delta E_{\rm min}$ may be obtained using the MW formula Eq.~\ref{eq:MW_cond} with transmission function,
\begin{eqnarray}\label{eq:T_MV_sd}
\mathcal{T}_{\rm MW}(\omega,T) = \frac{4t_s^2 t_d^2}{(t_s^2+t_d^2)^2} \sum_{\sigma} t_{\sigma}(\omega,T) \;,\;
\end{eqnarray}
where $t_{\sigma}(\omega,T)={\rm Im}[-\tilde{\Gamma}_{\sigma} G_{\textit{ff};\sigma}(\omega,T)]$ is the t-matrix spectrum and $G_{\textit{ff};\sigma}(\omega,T)\equiv \langle\langle f_{\sigma}^{\phantom{\dagger}} ; f_{\sigma}^{\dagger}\rangle\rangle_{\omega,T}$ is the Green's function of the effective $f$ level. 
Unlike Eq.~\ref{eq:MV_G_spinless} for the spinless case, here the dynamics are nontrivial due to the constraint $U_c\to \infty$. However, $G_{\textit{ff};\sigma}(\omega,T)$ can be calculated numerically using e.g.~NRG, and hence the quantum transport properties can be obtained. This requires only the solution of the effective model, rather than the full bare model. Reference results for the effective model can therefore be calculated and reused for any microscopic system.

In the absence of a magnetic field in the bare model, we have $E^{N+1}_{\uparrow}=E^{N+1}_{\downarrow}$ such that the effective field vanishes, $B_{\rm eff}=0$, and the effective tunneling amplitudes $t_{\alpha\sigma}\equiv t_{\alpha}$ become independent of the spin label $\sigma$ (hence $t_{\sigma}\equiv t$ and $\tilde{\Gamma}_{\sigma}\equiv \tilde{\Gamma}$). We focus on this case below.

\begin{figure}[t!]
\includegraphics[width=8.7cm]{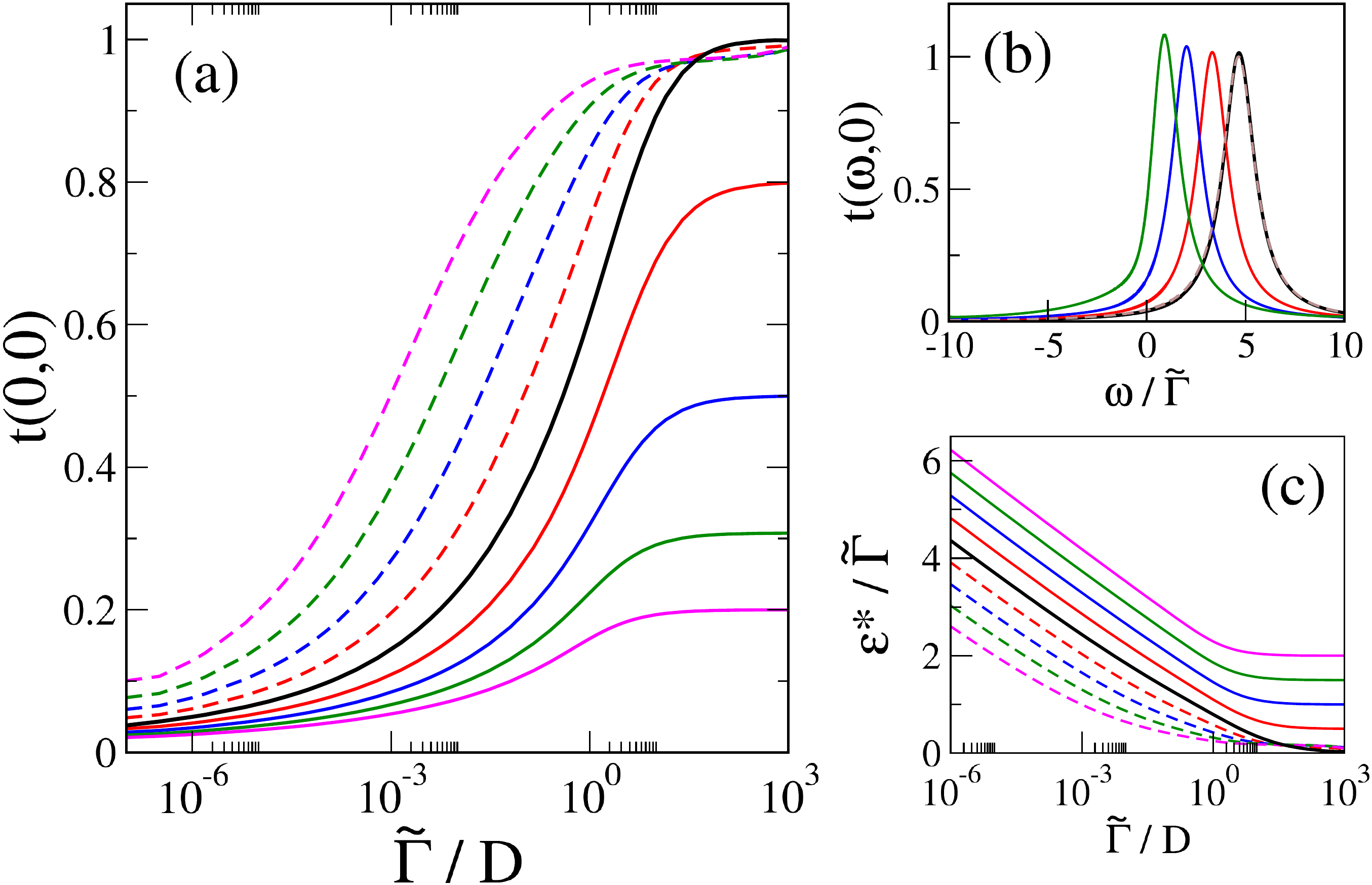}
  \caption{t-matrix for the effective model Eq.~\ref{eq:Heff_mv_s-d} at $T=0$, describing the dynamics in the vicinity of a singlet-doublet charge-degeneracy point, calculated from NRG. (a) $t(0,0)$ as a function of $\tilde{\Gamma}$ for $\epsilon=0$ (black line) and $|\epsilon|/\tilde{\Gamma}=0.5, 1, 1.5, 2$ (red, blue, green and magenta lines) with $\epsilon>0$ shown as solid lines and $\epsilon<0$ as dashed lines. (b) $t(\omega,0)$ vs $\omega/\tilde{\Gamma}$ at $\epsilon=0$ for $\tilde{\Gamma}/D=10^{-1},10^{-3},10^{-5},10^{-7}$ for the green, blue, red and black lines. Eq.~\ref{eq:MV_sd_lor} is shown as the dashed line for comparison to black line. (c) Behaviour of $\epsilon^*/\tilde{\Gamma}$ as a function on $\tilde{\Gamma}$ for the same parameters as in panel (a), showing small $\tilde{\Gamma}/D\ll 1$ asymptotes following Eq.~\ref{eq:MV_sd_eps}.
  }
  \label{fig:MV_sd}
\end{figure}

In Fig.~\ref{fig:MV_sd} we consider the spectrum of the t-matrix $t_{\sigma}(\omega,T)\equiv t(\omega,T)$ for the effective model Eq.~\ref{eq:Heff_mv_s-d} with $B_{\rm eff}=0$ and $U_c\to \infty$, obtained by full NRG calculations. For given effective parameters $\epsilon$ and $\tilde{\Gamma}$, the low-$T$ transport can then be obtained via Eqs.~\ref{eq:MW_cond}, \ref{eq:T_MV_sd}. Panel (b) shows the $T=0$ spectra at the charge degeneracy point $\epsilon=0$ for different $\tilde{\Gamma}$. At small $\tilde{\Gamma}/D\ll 1$ the spectrum is accurately approximated by that of an equivalent non-interacting resonant level model\cite{hewson1997kondo} ($U_c=0$) with the same $\tilde{\Gamma}$, but with an interaction-renormalized level energy $\epsilon^*=\epsilon+{\rm Re}\Sigma_{\textit{ff}}(\omega=0)$, where $\Sigma_{\textit{ff}}(\omega)$ is the $f$-level interaction self-energy,
\begin{equation}\label{eq:MV_sd_lor}
    t(\omega,0) \simeq \frac{1}{1+(\omega-\epsilon^*)^2/\tilde{\Gamma}^2} \;.
\end{equation}
This is shown for comparison as the dashed line in Fig.~\ref{fig:MV_sd}(b).

%%%%%%%%%%%%%%%%
%%%%%%%%%%%%%%%%

At larger $\tilde{\Gamma}$, discrepancies from Eq.~\ref{eq:MV_sd_lor} at finite frequencies become more pronounced, due mainly to the finite ${\rm Im}\Sigma_{\textit{ff}}(\omega)$, leading to an enhanced peak height and asymmetric shape. However, since ${\rm Im}\Sigma_{\textit{ff}}(0)=0$ by Fermi liquid theory, $t(0,0)$ is always given exactly by Eq.~\ref{eq:MV_sd_lor}, once $\epsilon^*$ is known. 

Only $t(0,0)$ is needed for the $T\to 0$ conductance, and reference NRG results for it as a function of $\tilde{\Gamma}$ are presented in Fig.~\ref{fig:MV_sd}(a) for different $\epsilon/\tilde{\Gamma}$. The behaviour of $\epsilon^*$, which yields $t(0,0)$ exactly from Eq.~\ref{eq:MV_sd_lor}, is given for the same parameters in Fig.~\ref{fig:MV_sd}(c). In the physically-relevant regime $\tilde{\Gamma}/D\ll 1$, the behaviour is found to be simply,
\begin{equation}\label{eq:MV_sd_eps}
    \epsilon^*/\tilde{\Gamma} \simeq a\epsilon/\tilde{\Gamma} - b\log(\tilde{\Gamma}/D) +c \;,
\end{equation}
where numerically we find $a\simeq 0.93$, $b\simeq 0.29$ and $c\simeq 0.35$.
 By contrast, at large $\tilde{\Gamma}/D\gg 1$, the self-energy correction can be neglected for $\epsilon>0$, such that $\epsilon^* \to \epsilon$; while for $\epsilon<0$ the Kondo effect gives $\epsilon^*\to 0$. This is confirmed directly by the saturation values in Fig.~\ref{fig:MV_sd}(c).

Equipped with these results, the low-$T$ conductance of a system in the vicinity of a singlet-doublet charge-degeneracy point can be straightforwardly predicted, once the effective model parameters $\epsilon$ and $t_{s,d}$ have been determined. 

%%%%%%%%%%%%%%

\subsection{Doublet-triplet transition}\label{sec:MVdt}
The charge degeneracy point between spin-doublet and spin-triplet states also yields nontrivial strongly-correlated electron behaviour -- yet it can again be captured in a simple effective model in PC. The doublet-triplet transition scenario is more common in the context of single molecule junctions, where Hund's Rule or other complex multi-orbital interactions can yield a high spin $S=1$ ground state in the even electron sector of the isolated nanostructure (by contrast to an even-electron $S=0$ singlet, as considered in the previous section). 

Here we shall take $N$ to be odd, with the $N$-electron nanostructure hosting a net $S=\tfrac{1}{2}$ doublet state; while in the (even) $N+1$ electron sector, the nanostructure ground state is an $S=1$ triplet. Focusing again on the low-$T$ transport, we construct an effective model in which only the lowest energy spin multiplet of each sector is retained. In this case the effective impurity Hamiltonian comprises five states: two for the $N$ electron doublet, and three for the $N+1$ electron triplet. For simplicity we consider below the SU(2) spin-symmetric case arising when no external magnetic field acts, such that states $|n;S^z;l\rangle$ of a given spin multiplet $l$ are degenerate ($E_{S^z;l}^n \equiv E_l^n$ is independent of $S^z$). We also set $l=1$ (ground state) and suppress $l$ labels in the following.

We construct an effective model using Eq.~\ref{eq:Heff_mv}, casting the result in terms of one effective fermionic site $f$, and one effective spin-$\tfrac{1}{2}$ degree of freedom $\hat{\boldsymbol{S}}_g$, together with carefully chosen constraints (which are implemented through additional hard-core terms in the Hamiltonian). This allows us to capture the various spin-multiplet transitions from doublet to triplet by adding a single electron. Our effective model reads,
\begin{equation}\label{eq:Heff_mv_d-t}
\begin{split}
H_{\rm eff} = H_{\rm leads} &+ \delta_c \hat{n}_f - J_c \hat{\boldsymbol{S}}_f\cdot \hat{\boldsymbol{S}}_g + \epsilon (\hat{n}_f-1)^2 \\
    &+ \sum_{\substack{\alpha=s,d\\ \sigma=\uparrow,\downarrow}} \left ( t_{\alpha}^{\phantom{\dagger}} f^{\dagger}_{\sigma}c^{\phantom{\dagger}}_{\alpha\sigma} + \rm{H.c.} \right )\;,
\end{split}
\end{equation}
where $\hat{\boldsymbol{S}}_g$ is a spin-$\tfrac{1}{2}$ operator for the effective local moment, $\hat{\boldsymbol{S}}_f=\tfrac{1}{2}\sum_{\nu,\nu'} f_{\nu}^{\dagger} \boldsymbol{\sigma}_{\nu\nu'}f_{\nu'}^{\phantom{\dagger}}$ is the spin density and  $\hat{n}_f=\sum_{\sigma} f_{\sigma}^{\dagger}f_{\sigma}^{\phantom{\dagger}}$ is the number operator for the effective $f$ level, and
\begin{align*}
    f^{\dagger}_{\uparrow} &= |N\text{+1;}S^z\text{=1}\rangle\langle N\text{;}S^z\text{=}\tfrac{1}{2} | + \tfrac{1}{\sqrt{2}} |N\text{+1;}S^z\text{=0}\rangle\langle N\text{;}S^z\text{=-}\tfrac{1}{2} |\\
    f^{\dagger}_{\downarrow} & = |N\text{+1;}S^z\text{=-1}\rangle\langle N\text{;}S^z\text{=-}\tfrac{1}{2} | + \tfrac{1}{\sqrt{2}} |N\text{+1;}S^z\text{=0}\rangle\langle N\text{;}S^z\text{=}\tfrac{1}{2} |
\end{align*}
with $f_{\sigma}=(f_{\sigma}^{\dagger})^{\dagger}$ as usual. Here, the $\tfrac{1}{\sqrt{2}}$ factors are Clebsch-Gordon coefficients implied by the Wigner-Eckart theorem. 

Of the 8 effective states of the $fg$ system in Eq.~\ref{eq:Heff_mv_d-t}, we wish to retain only the 5 that represent the $N$ electron doublet and the $N+1$ electron triplet of the bare model: 
\begin{align*}
    |N;S^z=\tfrac{1}{2}\rangle &\equiv |0\rangle_f\otimes|\uparrow\rangle_g \;,\\ |N;S^z=-\tfrac{1}{2}\rangle &\equiv |0\rangle_f\otimes|\downarrow\rangle_g \;,\\ 
 |N+1;S^z=1\rangle &\equiv |\uparrow\rangle_f\otimes|\uparrow\rangle_g \;,\\ 
|N+1;S^z=0\rangle &\equiv \tfrac{1}{\sqrt{2}}( |\uparrow\rangle_f\otimes|\downarrow\rangle_g + |\downarrow\rangle_f\otimes|\uparrow\rangle_g) \;,\\ 
|N+1;S^z=-1\rangle &\equiv |\downarrow\rangle_f\otimes|\downarrow\rangle_g \;.
\end{align*}
The spurious states $|\uparrow\downarrow\rangle_f\otimes|\uparrow\rangle_g$, $|\uparrow\downarrow\rangle_f\otimes|\downarrow\rangle_g$ and $\tfrac{1}{\sqrt{2}}( |\uparrow\rangle_f\otimes|\downarrow\rangle_g - |\downarrow\rangle_f\otimes|\uparrow\rangle_g)$ are eliminated by the constraints in Eq.~\ref{eq:Heff_mv_d-t} by setting $\delta_c=\tfrac{1}{4}J_c$ and sending $J_c\to \infty$.

With the above definitions, the effective tunneling amplitudes in Eq.~\ref{eq:Heff_mv_d-t} can be obtained from the extremal weight matrix elements of the isolated nanostructure,
\begin{equation*}
    t_{\alpha}=V_{\alpha}\langle N+1;S^z=1 | \bar{d}_{\alpha\uparrow}^{\dagger} | N;S^z=\tfrac{1}{2}\rangle \;,
\end{equation*} 
although by SU(2) spin symmetry we may also write 
$t_{\alpha}=\sqrt{2} V_{\alpha}\langle N\text{+1;}S^z\text{=0}| \bar{d}_{\alpha\uparrow}^{\dagger} | N\text{;}S^z\text{=-}\tfrac{1}{2}\rangle$. Deviations from the precise charge-degeneracy point are captured by the parameter $\epsilon=E^{N+1}-E^{N}$. The effective model is therefore a mixed-valent single impurity Anderson model, side-coupled to an additional spin-$\tfrac{1}{2}$ local moment by a ferromagnetic interaction.

%%%%%%%%%%%%%%%%
\begin{figure}[t!]
\includegraphics[width=7cm]{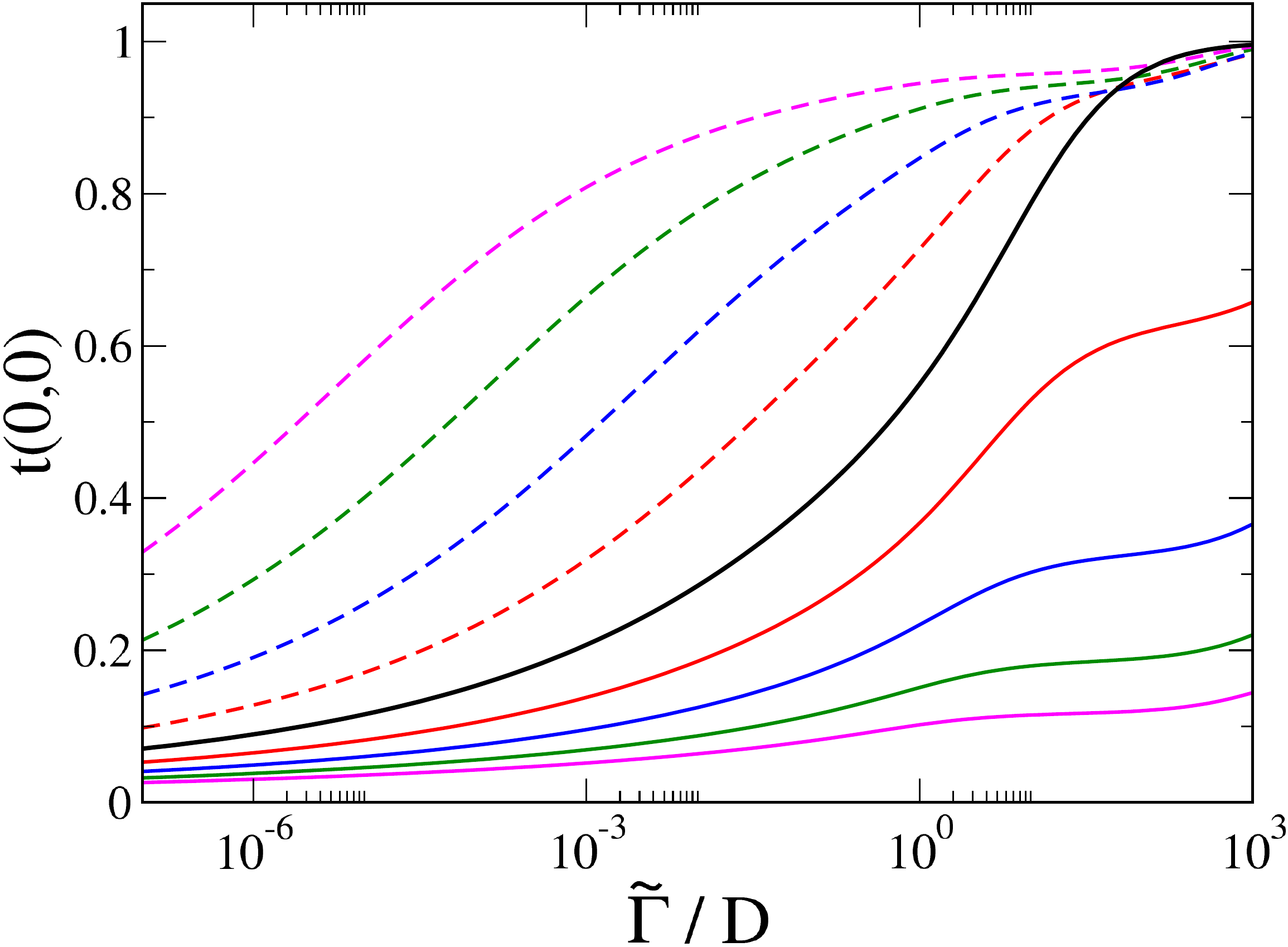}
  \caption{t-matrix $t(0,0)$ at $T=0$ as a function of the hybridization $\tilde{\Gamma}$ near a doublet-triplet charge-degeneracy point, described by the effective model Eq.~\ref{eq:Heff_mv_d-t}, obtained by NRG. Plotted for $\epsilon=0$ (black line) and $|\epsilon|/\tilde{\Gamma}=0.5, 1, 1.5, 2$ (red, blue, green and magenta lines) with $\epsilon>0$ shown as solid lines and $\epsilon<0$ as dashed lines. 
  }
  \label{fig:MV_dt}
\end{figure}
%%%%%%%%%%%%%%%%

Turning now to the low-temperature quantum transport, the PC form of the effective model implies that we may use the MW formula Eq.~\ref{eq:MW_cond}, with transmission coefficient Eq.~\ref{eq:T_MV_sd} given in terms of the t-matrix  $t_{\sigma}(\omega,T)\equiv t(0,0)={\rm Im}[-\tilde{\Gamma} G_{\textit{ff};\sigma}(\omega,T)]$, where $G_{\textit{ff};\sigma}(\omega,T)\equiv \langle\langle f_{\sigma}^{\phantom{\dagger}} ; f_{\sigma}^{\dagger}\rangle\rangle_{\omega,T}$ is the (spin-independent) Green's function of the effective $f$ level and $\tilde{\Gamma}=\pi\rho_0(|t_s|^2+|t_d|^2)$ is the effective hybridization as before.

The $T\to 0$ conductances of the physical system therefore depend on the dynamics of the effective model Eq.~\ref{eq:Heff_mv_d-t} as $T\to 0$ and $\omega\to 0$, as encoded in $t(0,0)$. In Fig.~\ref{fig:MV_dt} we plot $t(0,0)$ as a function of $\tilde{\Gamma}$ for different $\epsilon/\tilde{\Gamma}$, obtained from NRG. This is the analogous plot for the doublet-triplet transition as Fig.~\ref{fig:MV_sd}(a) for the singlet-doublet transition. Although we find qualitatively similar behaviour to that of the singlet-doublet case, the details are different. Indeed, here the full spectrum $t(\omega,0)$ shows much more pronounced asymmetries than the simple Lorentzian Eq.~\ref{eq:MV_sd_lor}, even at very small $\tilde{\Gamma}$.  However, in the physically-relevant regime $\tilde{\Gamma}\ll 1$ we find $t(0,0)$ for the effective model is still accurately approximated by Eqs.~\ref{eq:MV_sd_lor} and \ref{eq:MV_sd_eps} but with modified $a\simeq 1.15$, $b\simeq 0.18$ and $c\simeq 0.7$. 

With these NRG results the low-$T$ conductances for real systems near a doublet-triplet charge degeneracy point can be obtained simply from the effective model parameters $\epsilon$ and $t_{s,d}$. Comparison of Fig.~\ref{fig:MV_sd}(a) and Fig.~\ref{fig:MV_dt} shows that for a given $\tilde{\Gamma}$ and $\epsilon$, the conductance is larger in the vicinity of the doublet-triplet charge-degeneracy point that than near the singlet-doublet charge-degeneracy point, which we attribute to enhanced spin fluctuations in the former.

Finally, we remark that the above analysis of the effective model Eq.~\ref{eq:Heff_mv_d-t}, obtained perturbatively to leading order in the nanostructure-lead hybridization, accounts for the dominant contributions to conductance. However, higher-order corrections will in general result in an effective coupling to the neglected odd conduction electron channel, taking the effective model out of PC. Although irrelevant for the singlet-doublet case considered in the previous section (whose many-body ground state is always a non-degenerate singlet due to the Kondo effect), in the doublet-triplet case the involvement of the odd channel is expected to induce a second-stage Kondo screening, quenching the $\ln(2)$ residual entropy found for  Eq.~\ref{eq:Heff_mv_d-t}.  Since such terms arise only at higher order in perturbation theory, the associated Kondo scale for the second-stage screening is expected to be very small, and hence may be neglectable at experimentally-relevant temperatures. We therefore argue that the effective model Eq.~\ref{eq:Heff_mv_d-t}, and results of this section, remain valid close to the charge degeneracy point where there is a clear separation of scales.

%%%%%%%%%%%%%%

\subsection{General spinful case}\label{sec:MVgen}
We briefly touch upon the general spinful situation near a charge degeneracy point. Since $N$ and $N+1$ electron sectors of the nanostructure are connected to first order in $H_{\rm hyb}$ by tunneling of a single electron, nontrivial spin-charge entanglement arises near the charge-degeneracy point between spin-$S$ and $S\pm \tfrac{1}{2}$ ground states. The two most common such scenarios (doublet to singlet and doublet to triplet) were considered explicitly above.

Generalizing now to the charge-degeneracy point between an $N$ electron spin-$S$ multiplet ground state, and an $N+1$ electron spin-$S+\tfrac{1}{2}$ multiplet ground state, it is easily seen that an effective model with the same form as Eq.~\ref{eq:Heff_mv_d-t} must arise -- but with $\boldsymbol{\hat{S}}_g$ now a spin-$S$ operator, and with 
$ t_{\alpha}=V_{\alpha}\langle N+1;S^z=S+\tfrac{1}{2} | \bar{d}_{\alpha\uparrow}^{\dagger} | N;S^z=S\rangle $. However $\epsilon=E^{N+1}-E^N$ as before.

$4S+3$ physical states of the spin $S$ and $S+\tfrac{1}{2}$ multiplets are retained in the ground state manifold of the effective model, meaning that $4S+1$ spurious states must be eliminated in Eq.~\ref{eq:Heff_mv_d-t} using the constraints $\delta_c$ and $J_c$. This implies setting $\delta_c=\tfrac{1}{2}S J_c$ and sending $J_c\to \infty$.

Low-$T$ quantum transport can be obtained in the same way as above from a knowledge of $t(0,0)$, which can be calculated for the effective model using e.g.~NRG.

%###########################
%###########################

\section{Auxiliary field formulation}\label{sec:aux}
Recently in Ref.~\onlinecite{sen2020mott},  topological properties of the Mott metal-insulator transition in the Hubbard model were uncovered by mapping the interaction self-energy of the effective impurity problem within dynamical mean field theory, to auxiliary non-interacting degrees of freedom. 

Here we explore the consequences of the same mapping for quantum transport. We start with an application to the paradigmatic single-impurity Anderson model, Eq.~\ref{eq:aim}.

The auxiliary field mapping is an exact representation of the single-particle dynamics for an interacting system in terms of a completely non-interacting one.\cite{sen2020mott} The Dyson equation, Eq.~\ref{eq:dyson}, for the Anderson impurity reads,
\begin{eqnarray}\label{eq:Gaim}
G_{\sigma}(\omega)\equiv \langle\langle d_{\sigma}^{\phantom{\dagger}} ; d_{\sigma}^{\dagger}\rangle\rangle = \frac{1}{\omega^+ -\epsilon_d -\sum_{\alpha}\Delta_{\alpha}(\omega)- \Sigma_{\sigma}(\omega)} \;,\qquad
\end{eqnarray}
where $\Delta_{\alpha}(\omega)=V_{\alpha}^2\mathcal{G}^0_{\alpha\alpha}(\omega)$ is the hybridization between the impurity and the physical lead $\alpha=s,d$, and $\Sigma_{\sigma}(\omega)$ is the interaction self-energy. In the absence of a magnetic field, we have SU(2) spin symmetry, and so we write $\Sigma_{\sigma}(\omega)\equiv \Sigma(\omega)$. For convenience we absorb the static contribution to the self-energy into the definition of the renormalized level $\epsilon_d^*=\epsilon_d+{\rm Re}~\Sigma(0)$, and work with the dynamical part of the self-energy $\tilde{\Sigma}(\omega)=\Sigma(\omega)-{\rm Re}~\Sigma(0)$.

%%%%%%%%%%%%%%%%
\begin{figure}[t!]
\includegraphics[width=9cm]{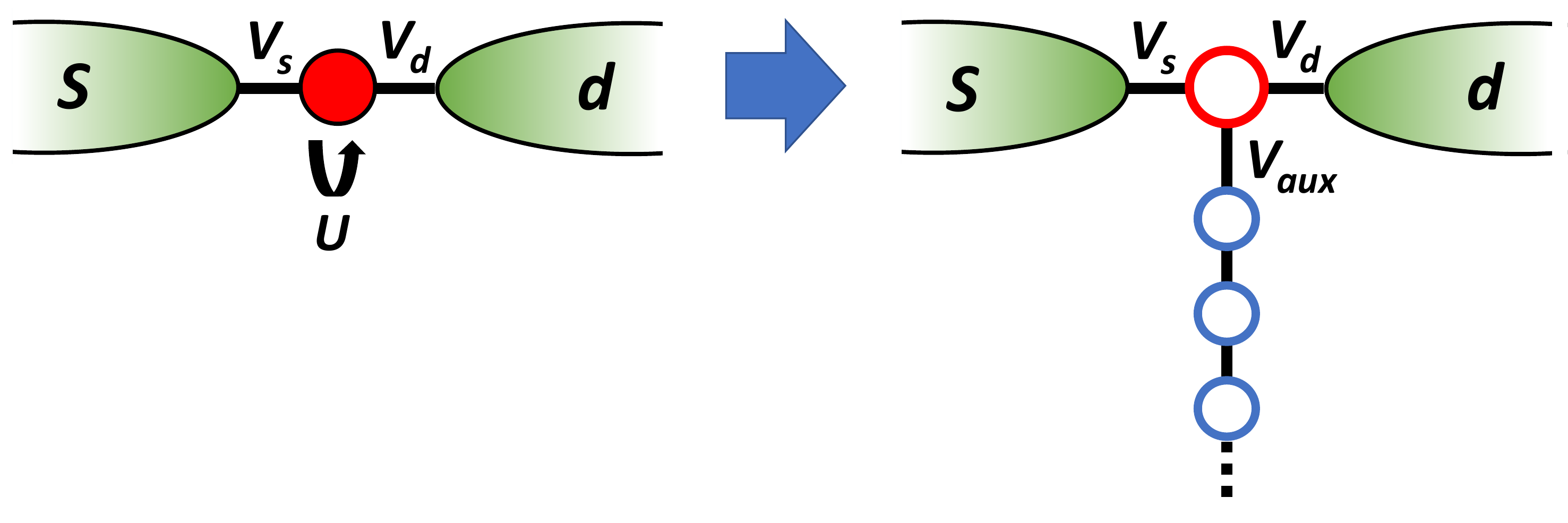}
  \caption{Schematic of the auxiliary field mapping. The interaction self-energy $\tilde{\Sigma}(\omega)$ of the single-impurity Anderson model (left) is mapped to an auxiliary non-interacting tight-binding chain (right). The current between physical source and drain leads due to a bias voltage in the interacting system is reproduced in the mapped non-interacting system through a zero-current constraint for the auxiliary `lead' in the 3-terminal Landauer-B\"uttiker formula.
  }
  \label{fig:aux}
\end{figure}
%%%%%%%%%%%%%%%%

Following Ref.~\onlinecite{sen2020mott} we now interpret $\tilde{\Sigma}(\omega)\to \Delta^{\rm aux}(\omega)$ as a hybridization to a bath of auxiliary non-interacting degrees of freedom. 
The effect of the electronic scattering due to the Coulomb interaction on the impurity is reproduced exactly by proper choice of the auxiliary bath. Here we describe the auxiliary system as a semi-infinite linear chain,
\begin{eqnarray}\label{eq:Haux}
H_{\rm aux} = \sum_{n=0}^{\infty}\sum_{\sigma} \left ( t_n^{\phantom{\dagger}} f_{n\sigma}^{\dagger} f_{(n+1)\sigma}+{\rm H.c.} \right ) \;,
\end{eqnarray}
where for simplicity we have now assumed particle-hole symmetry, $\epsilon_d=-U_d/2$ (such that $\epsilon_d^*=0$ and Eq.~\ref{eq:Haux} does not require inclusion of onsite potentials). The auxiliary chain is coupled at one end to the impurity,
\begin{eqnarray}\label{eq:Hhyb_aux}
H_{\rm imp-aux}= \sum_{\sigma}\left ( V^{\phantom{\dagger}}_{\rm aux} d_{\sigma}^{\dagger}f_{0\sigma}^{\phantom{\dagger}} + {\rm H.c.} \right ) \;.
\end{eqnarray}
The mapping is illustrated schematically in Fig.~\ref{fig:aux}.

The impurity-auxiliary chain hybridization function is thus given by $\Delta^{\rm aux}(\omega)=V_{\rm aux}^2 \mathcal{G}^0_{\rm aux}(\omega)$, where 
$\mathcal{G}^0_{\rm aux}(\omega) =\langle\langle f_{0\sigma}^{\phantom{\dagger}} ; f_{0\sigma}^{\dagger}\rangle\rangle^0$ 
is the boundary Green's function of the isolated auxiliary system. The latter can be expressed simply as a continued fraction\cite{ContinuedFraction_Mori1965} in terms of the tight-binding parameters $\{t_n\}$ in Eq.~\ref{eq:Haux} as $\mathcal{G}^0_{\rm aux}(\omega)=1/[\omega^+ -\Delta^{\rm aux}_0(\omega)]$ with $\Delta^{\rm aux}_n(\omega)=t_n^2/[\omega^+-\Delta^{\rm aux}_{n+1}(\omega)]$. 

The auxiliary parameters $V_{\rm aux}$ and $\{t_n\}$ are uniquely determined\cite{sen2020mott} by setting
$\Sigma(\omega)\to \Delta^{\rm aux}(\omega) =V_{\rm aux}^2 \mathcal{G}^0_{\rm aux}(\omega)$.
Specifically, $V_{\rm aux}^2=-\tfrac{1}{\pi}{\rm Im}\int d\omega \Sigma(\omega)$ initializes an iterative scheme in which successive $t_n$ are obtained by $t_{n}^2=-\tfrac{1}{\pi}{\rm Im}\int d\omega \Delta_n^{\rm aux}(\omega)$. At half-filling, $V_{\rm aux}=U_d/2$. The other auxiliary parameters are regular and well-behaved, with the recursion being efficient and numerically stable. Since the original self-energy $\Sigma(\omega)$ is a continuous  function, the recursion does not terminate (the auxiliary chain is semi-infinite); however the $t_n$ settle down to a regular pattern after a finite number of steps.

The key point for the present discussion is not the specific form of these parameters for a given model, but rather the fact that this mapping exists and is unique.

For the mapped non-interacting system (Fig.~\ref{fig:aux}, right) we have \emph{three} effective leads, with a resonant level impurity Green's function,
\begin{eqnarray}\label{eq:auxG}
G_{\sigma}(\omega)= \frac{1}{\omega^+ -\epsilon_d^* -\sum_{\gamma}\Delta^{\gamma}(\omega)} \;,\qquad
\end{eqnarray}
where $\gamma=s,d$ or ${\rm aux}$.

We now wish to calculate the current $I^d$ flowing from source lead to drain lead due to a source-drain bias voltage $\Delta V_b$. In the physical interacting system (Fig.~\ref{fig:aux}, left), the linear response conductance follows from the MW formula, Eq.~\ref{eq:MW_G}, with the interacting impurity Green's function, Eq.~\ref{eq:Gaim}. For the mapped non-interacting system (Fig.~\ref{fig:aux}, right), we may use the 3-terminal linearlized Landauer-B\"uttiker (LB) formula for the current into lead $\gamma$,
\begin{equation}\label{eq:LB}
I^{\gamma}(T) = \frac{e}{h} \sum_{\beta\ne\gamma}\int d\omega ~[\partial_{\omega} f_{\rm eq}(\omega)](\mu_{\gamma}-\mu_{\beta})\mathcal{T}_{\text{L}\:\gamma\beta}(\omega) \;,
\end{equation}
where $\mu_{\gamma}$ is the chemical potential of lead $\gamma$ and $\mathcal{T}_{\text{L} \; \gamma\beta}(\omega)=4\Gamma_{\gamma}(\omega)\Gamma_{\beta}(\omega)\sum_{\sigma}|G_{\sigma}(\omega)|^2$ is the analogue of Eq.~\ref{eq:TlandauerPC}, 
with $G_{\sigma}(\omega)$ the effective non-interacting Green's function given in Eq.~\ref{eq:auxG}, and $\Gamma_{\gamma}(\omega)=-{\rm Im}~\Delta^{\gamma}(\omega)$. Eq.~\ref{eq:LB} is a generalization of the usual LB formula to the case with inequivalent leads with arbitrary density of states (Appendix~\ref{app:inequiv}). This is important because the auxiliary `lead' has a specific form that must be accounted for. We will however assume for simplicity that the source and drain leads are equivalent: $\Gamma_s(\omega)=\pi V_s^2\rho(\omega)$ and $\Gamma_d(\omega)=\pi V_d^2\rho(\omega)$ with the same (but otherwise arbitrary) density of states $\rho(\omega)$.

Of course, the auxiliary lead is not a physical lead and so we do not apply a voltage to it ($\mu_{\rm aux}=0$), and no current flows into or out of it ($I^{\rm aux}=0$). The latter property is also required by current conservation in the physical system, $I^s=-I^d$. From Eq.~\ref{eq:LB}, these constraints imply that $V^2_s\mu_s+V^2_d\mu_d=0$. The voltage bias $e\Delta V_b \equiv \mu_s-\mu_d$ must be split across source and drain leads in a specific way to satisfy this constraint, with $\mu_s=e\Delta V_b(1+V^2_s/V^2_d)^{-1}$ and $\mu_d=-e\Delta V_b(1+V^2_d/V^2_s)^{-1}$. We note however that the $\Delta V_b \to 0$ linear response conductance does not depend on the details of this splitting. 

Substituting in Eq.\ref{eq:LB}, we obtain,
\begin{equation}\label{eq:LBaux}
\begin{split}
I^{d}(T) = \Delta V_b \frac{e^2 }{h} ~\frac{4\pi V^2_dV^2_s}{V^2_d+V^2_s} \int d\omega& ~[-\partial_{\omega} f_{\rm eq}(\omega)]\rho(\omega) \\
&\times\sum_{\gamma,\sigma}\Gamma_{\gamma}(\omega)|G_{\sigma}(\omega)|^2 \;.
\end{split}
\end{equation}
This reduces correctly to Eq.~\ref{eq:Kubo_PC_rho_dc} since $\sum_{\gamma}\Gamma_{\gamma}(\omega)|G_{\sigma}(\omega)|^2=-{\rm Im}~G_{\sigma}(\omega)$ from Eq.~\ref{eq:auxG}. Furthermore, in the wide flat band limit of the leads (when $\rho(\omega)\to \rho_0$ and $\Gamma_{\alpha}(\omega)\to \Gamma_{\alpha} = \pi \rho_0 V_{\alpha}^2$ for $\alpha=s,d$), Eq.~\ref{eq:LBaux} recovers the standard PC form of the MW formula, Eq.~\ref{eq:MW_G}, as expected.

The above arguments generalize trivially to any multi-orbital two-lead system in PC. The Green's function for the PC frontier orbital $\bar{\bar{d}}_{\sigma}$ coupling to the leads can always be expressed as,
\begin{eqnarray}\label{eq:G_PC_aux}
\overline{\overline{G}}_{\sigma}(\omega)\equiv \langle\langle \bar{\bar{d}}_{\sigma}^{\phantom{\dagger}} ; \bar{\bar{d}}_{\sigma}^{\dagger}\rangle\rangle = \frac{1}{\omega^+ -\epsilon_d -\sum_{\alpha}\Delta_{\alpha}(\omega)- \Sigma'_{\sigma}(\omega)} \;,\qquad
\end{eqnarray}
where $\Sigma'_{\sigma}(\omega)$ includes the effect of scattering from coupling of $\bar{\bar{d}}_{\sigma}$ to the other nanostructure degrees of freedom, as well as accounting for electronic interactions. Following the same steps as before, $\Sigma'_{\sigma}(\omega)$ can be mapped to a single non-interacting auxiliary chain (Eq.~\ref{eq:Haux}) coupled at one end to a single resonant level (Eq.~\ref{eq:Hhyb_aux}), which is also coupled to the physical source and drain leads. Schematically, the mapped system is identical to that depicted in Fig.~\ref{fig:aux}(b).

The non-PC case is more subtle, since the equivalent non-interacting form of $\overline{G}_{sd,\sigma}(\omega)$ for use in Eq.~\ref{eq:land_I} must be determined. In general this requires mapping the effective self-energies to \emph{two} auxiliary chains, attached to both frontier orbitals $\bar{d}_{s\sigma}$ and $\bar{d}_{d\sigma}$. We leave this for future investigation. 

In summary, quantum transport for interacting systems can be understood in terms of the non-interacting Landauer-B\"uttiker formula, in which the self-energy plays the role of an additional fictitious lead, subject to a zero-current constraint. 
This formulation provides a simple way of viewing the correction to quantum transport due to interactions. 

Furthermore, the auxiliary chain representation may provide a route to simple approximations, given its convenient structure and well-defined asymptotic form. For example, at $T=0$ in the metallic Kondo screened case,\cite{sen2020mott}
\begin{equation}\label{eq:auxtn}
t_n \sim  \frac{D}{2}\sqrt{1-\frac{2(-1)^n}{n+d}} \;,
\end{equation}
for large $n$, where $D$ is the effective bandwidth and $d\sim 1/Z$ is related to the quasiparticle weight $Z$. The conductance formulae can be expressed in terms of the auxiliary chain parameters.

An application of the auxiliary field method is given for the TQD system in the next section.

%###########################
%###########################

\section{Applications and Comparison of techniques}\label{sec:apps}
The goal of this section is to demonstrate the results of the previous sections by application to specific
 multi-orbital strongly-correlated systems. 
 Our chosen test systems are the two-lead triangular triple quantum dot (TQD) and the serial multilevel double quantum dot (MLDQD).
 %!!!!!!!!!!!!!!!!
 
 \subsection{Triple quantum dot}\label{sec:tqd}
  
 %%%%%%%%%%%%%%%%
\begin{figure}[t!]
\includegraphics[width=8.7cm]{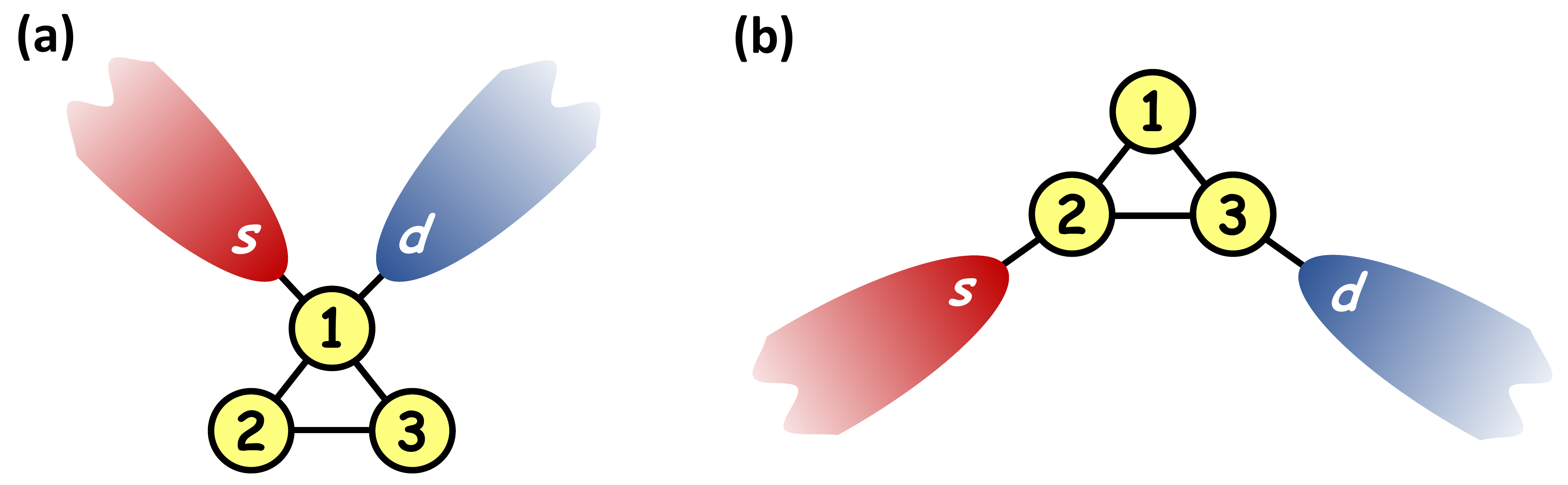}
  \caption{Illustration of the triple quantum dot devices studied. (a) PC coupling geometry; (b) non-PC geometry.  }
  \label{fig:tqd}
\end{figure}
%%%%%%%%%%%%%%%%

 Here we consider the TQD model $H=H_{\rm leads}+H_{\rm TQD}+H_{\rm hyb}$, with $H_{\rm leads}$ given in Eq.~\ref{eq:Hleads} and,
\begin{equation}\label{eq:tqd}
\begin{split}
    &H_{\rm TQD} = \\ &\sum_{j=1,2,3} \left ( \epsilon_j \hat{n}_j + U_j \hat{n}_{j\uparrow}\hat{n}_{j\downarrow} \right )  +\sum_{i\ne j} \left (U_{ij}'\hat{n}_i\hat{n}_j   + \sum_{\sigma} t_{ij}^{\phantom{\dagger}} d^{\dagger}_{i\sigma}d^{\phantom{\dagger}}_{j\sigma} \right) \;,
\end{split}
\end{equation}
where $\hat{n}_j=\sum_{\sigma}\hat{n}_{j\sigma}$, and the inter-dot tunneling matrix elements satisfy $t_{ij}=t_{ji}^*$. We take for simplicity equivalent dots with $\epsilon_j\equiv \epsilon$, $U_j\equiv U$, $U'_{ij}\equiv U'$, and consider the mirror symmetric case  $t_{12}=t_{13}\equiv t$ but $t_{23}\equiv t'$. We consider two geometries for the TQD-lead hybridization,
\begin{subequations}
\begin{align}
    H_{\rm hyb}^{I}  &= \sum_{\alpha=s,d} \sum_{\sigma} \left ( V_{\alpha}^{\phantom{\dagger}}d^{\dagger}_{1\sigma}c_{\alpha\sigma}^{\phantom{\dagger}} +{\rm H.c.} \right ) \;,\label{eq:tqd_hyb_1}\\
    H_{\rm hyb}^{II} &= \sum_{\sigma} \left (V_s^{\phantom{\dagger}} d^{\dagger}_{2\sigma}c_{s\sigma}^{\phantom{\dagger}}+V_d^{\phantom{\dagger}} d^{\dagger}_{3\sigma}c_{d\sigma}^{\phantom{\dagger}} +{\rm H.c.} \right ) \;. \label{eq:tqd_hyb_2}
\end{align}
\end{subequations}
For simplicity we also take $V_s=V_d$.
Importantly, note that $H_{\rm hyb}^{I}$ in Eq.~\ref{eq:tqd_hyb_1} (illustrated in Fig.~\ref{fig:tqd}a) satisfies the PC geometry condition while $H_{\rm hyb}^{II}$ in Eq.~\ref{eq:tqd_hyb_2} (Fig.~\ref{fig:tqd}b) is non-PC and therefore irreducibly 2-channel.

The behavior of TQD systems is notoriously rich, with aspects of their complex physics having been uncovered in both experiments (see e.g.~Refs.~\onlinecite{vidan2004triple,schroer2007electrostatically,rogge2008two,granger2010three,gaudreau2012coherent,seo2013charge}) and theory (see e.g.~Refs.~\onlinecite{mitchell2009quantum,mitchell2010two,mitchell2013local,bonvca2008numerical,vernek2009kondo,numata2009kondo,*oguri2011kondo,saraga2003spin,wrzesniewski2018dark,koga2013field,wang2013enhancement,ramos2017spin}). Our purpose here is to examine the quantum transport properties in a systematic and consistent fashion, applying and comparing the techniques discussed above. We solve the underlying quantum impurity problems using NRG. 

%%%%%%%%%%%%%%%%
\begin{figure}[t!]
\includegraphics[width=8.7cm]{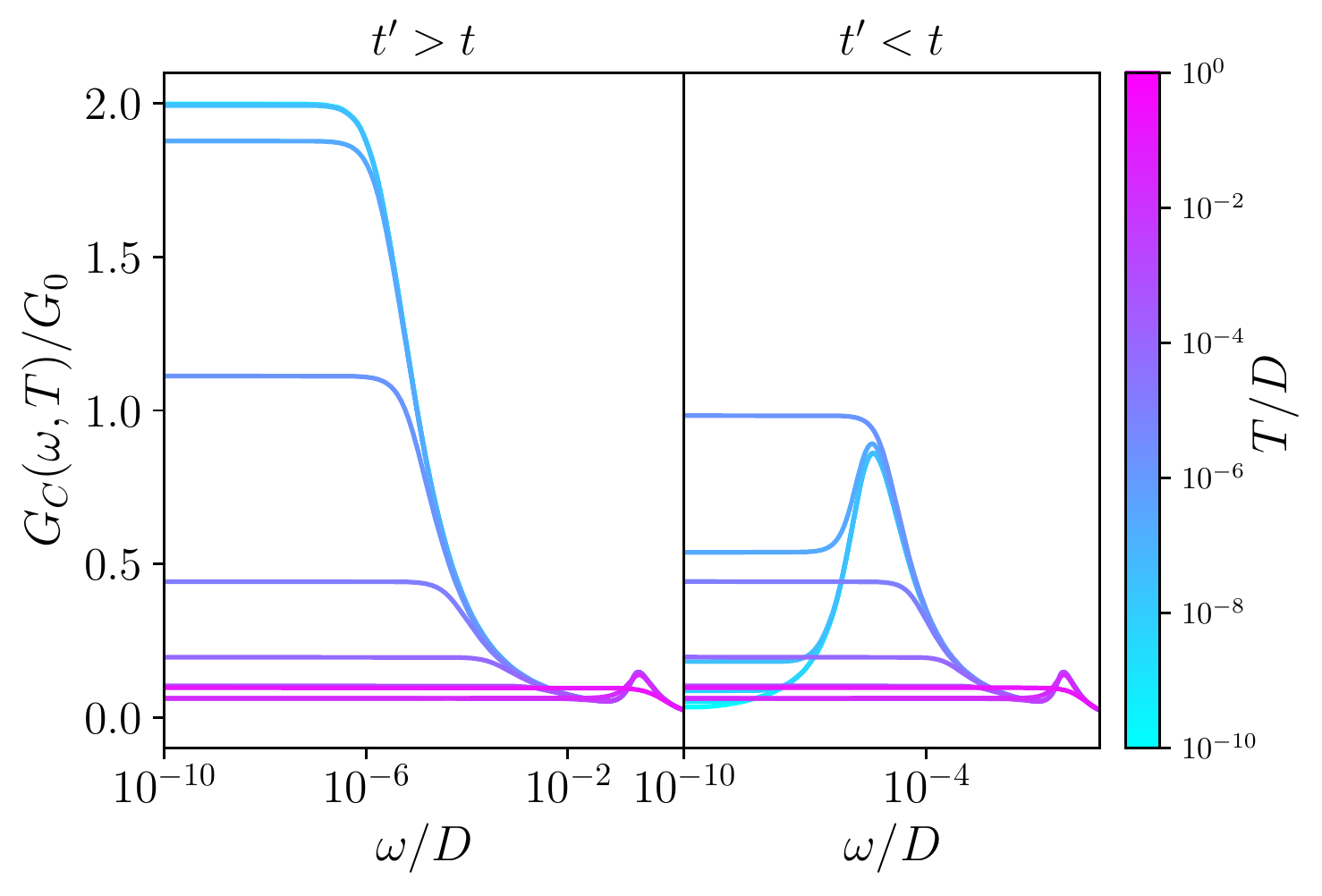}
  \caption{AC conductance $G_C(\omega,T)$ as a function of driving frequency $\omega$ at different temperatures $T$ as indicated in the color scale, for the TQD model in the PC geometry (see Fig.~\ref{fig:tqd}(a) and Eq.~\ref{eq:tqd_hyb_1}). Results shown for the Kondo screened phase $t'>t$ (left) and local moment phase $t'<t$ (right). NRG results are obtained using the ac-generalized MW formula, Eq.~\ref{eq:Kubo_PC}, within the effective single-channel model. Model parameters: $U=0.4D$, $\epsilon=-U/2$, $U'=0$, $V_s=V_d=0.071D$, $t=5\times10^{-4}D$ and $t'=10^{-2}D$ (left) or $t'=0$ (right). NRG parameters: $\Lambda=2.5$, $M_K=3000$.
  }
  \label{fig:Gac}
\end{figure}
%%%%%%%%%%%%%%%%

\subsubsection{Proportionate Coupling}
First we consider the PC case, Eq.~\ref{eq:tqd_hyb_1} (Fig.~\ref{fig:tqd}a) with $U'=0$. The behavior of the linear dc electrical conductance $G_C(T)$ for this system was reported in Ref.~\onlinecite{mitchell2009quantum}. In Fig.~\ref{fig:Gac} we consider instead the \textit{dynamical} conductance $G_C(\omega,T)$, plotted vs ac driving frequency $\omega$ at different temperatures $T$.
The TQD in this geometry supports a quantum phase transition driven by geometric frustration in the triangular arrangement of dots.\cite{mitchell2009quantum} The transition, tuned by the ratio $t/t'$,  embodies the competition between an antiferromagnetic-Kondo Fermi liquid with a screened spin-singlet ground state, and a ferromagnetic-Kondo singular Fermi liquid with a degenerate local moment ground state. \cite{VarmaSingularNFL2002} We show how this plays out in the ac conductance in Fig.~\ref{fig:Gac}, which we calculate via Eq.~\ref{eq:Kubo_PC}. As such, the conductance is controlled by the equilibrium retarded Green's function for dot 1 (interacting, lead-coupled), since in this geometry $\overline{\overline{G}}_{\sigma}(\omega,T)=G_{11,\sigma}(\omega,T)\equiv  \langle\langle d_{1\sigma}^{\phantom{\dagger}} ; d_{1\sigma}^{\dagger}\rangle\rangle$. This is in turn obtained from the Dyson equation Eq.~\ref{eq:dyson}, with the TQD self-energy matrix calculated in NRG via Eq.~\ref{eq:UFG}.

For $t'>t$ (left panel) we see a Kondo resonance, with strongly enhanced ac conductance for $|\omega|, T \ll T_{\rm K}$. As the temperature is increased, the resonance is lost. To understand this dynamical transport behavior in the TQD, the screening mechanism\cite{mitchell2009quantum} must be understood. At temperatures $T\lesssim U$, the dots become essentially singly-occupied (assuming $t, t'\ll U$ as here). On the scale $J\sim t^2/U\ll U$, effective antiferromagnetic exchange interactions then bind the three spins of the TQD into a combined doublet ground state,
\begin{eqnarray}\label{eq:plusstate}
|+;S_{\rm TQD}^z=\tfrac{1}{2}\rangle  = \tfrac{1}{\sqrt{2}}[|\uparrow\uparrow\downarrow\rangle - |\uparrow\downarrow\uparrow\rangle ] \;,
\end{eqnarray}
where the basis states $|\sigma_1,\sigma_2,\sigma_3\rangle$ are labelled by the spins on the three dots, and the other component of the doublet with $S_{\rm TQD}^z=-\tfrac{1}{2}$ is obtained by replacing $\uparrow \leftrightarrow \downarrow$. 
This doublet then couples to the leads with an effective exchange interaction $J_{\rm eff}^+\sim +V^2/U$, which generates spin-flip scattering of conduction electrons, and results in the Kondo effect below the Kondo temperature\cite{hewson1997kondo} $T^+_{\rm K} \sim D \exp(-1/\rho_0 J^+_{\rm eff})$. For the parameters used, we have $T^+_{\rm K}\approx 10^{-5}D$, with $D$ the conduction electron bandwidth.

For $T\ll T^+_{\rm K}$ the conduction electron scattering rate is on the order of $T^+_{\rm K}$. Therefore, when a small dc bias voltage is applied, the enhanced electronic scattering boosts the net current flowing from source to drain.\cite{pustilnik2004kondo,pustilnik2004quantum}
The conductance in this case can reach its maximum value, $G_C=2e^2/h$. In the ac case, conductance is still strongly enhanced, provided that the electronic scattering time is much shorter than the period of the voltage bias oscillations, or $|\omega| \ll T^+_{\rm K}$. As the frequency increases, the current does not have time to build up fully before the bias voltage changes sign. At large frequencies, the TQD sees an `averaged' bias and the conductance is low. The resonance condition is therefore $|\omega|\sim T^+_{\rm K}$ when $T\ll T^+_{\rm K}$. This is seen in the low-$T$ limit (turquoise line) in the left panel of Fig.~\ref{fig:Gac}, which is found to have asymptotic behavior in the Kondo-screened phase following,
\begin{subequations}\label{eq:Gtpgtrt}
\begin{align}
G_C(\omega,0)/G_0 &= 2 - \bar{a}(\omega/T^+_{\rm K})^2 ~~~~ &:~|\omega|\ll T^+_{\rm K}~\\
&=\bar{b}/\ln^2(|\omega|/T^+_{\rm K}) ~~~~&:~|\omega|\gg T^+_{\rm K}~
\end{align}
\end{subequations}
with $\bar{a},\bar{b}$ constants of order 1. 

As the temperature is increased, the Kondo effect is destroyed, the resonance condition is lost, and the low-frequency ac conductance decreases. For $T\gg T^+_{\rm K}$, the conductance is low, being the result of incoherent sequential tunneling only (it is not boosted by spin-flip scattering from the Kondo effect). This is seen in the left panel of Fig.~\ref{fig:Gac} by the sequence of lines from turquoise to purple on increasing temperature. 

Upon tuning $t'$ from $>t$ to $<t$, the isolated TQD ground state changes, and this results in a quantum phase transition in the lead-coupled system.\cite{mitchell2009quantum}   For $t'<t$ (right panel, Fig.~\ref{fig:Gac}), the collective TQD doublet ground state (in the singly-occupied limit) is,
\begin{eqnarray}
|-;S_{\rm TQD}^z=\tfrac{1}{2}\rangle  = \tfrac{1}{\sqrt{6}}[2|\downarrow\uparrow\uparrow\rangle - |\uparrow\uparrow\downarrow\rangle-|\uparrow\downarrow\uparrow\rangle ] \;.\label{eq:minusstate}
\end{eqnarray}
This doublet state again forms on the scale of $J\sim t^2/U$, but its effective coupling to the leads is now  \emph{ferromagnetic}, $J^-_{\rm eff} <0$. This results in the \textit{suppression} of spin-flip scattering on reducing the temperature\cite{hewson1997kondo} below $T \sim J$, and hence a suppression of the conductance. In the dc limit, we have  $G_C(0,0)=0$, embodying the emergent decoupling of the TQD from the leads. However, singular Fermi liquid corrections to the local moment fixed point at finite energy \cite{koller2005singular} lead to logarithmic corrections to the low-frequency ac conductance \cite{mehta2005regular},
\begin{eqnarray}
G_C(\omega,0)/G_0 =\bar{\alpha}_1/\ln^2(\bar{\alpha}_2|\omega|/J) \qquad &:~|\omega|\ll J~~~
\end{eqnarray}
%Since the underscreened TQD doublet has a finite energy $J$, it has a finite lifetime due to the energy-time uncertainty principle. Even at $T=0$, scattering from this state can therefore be probed at finite frequencies. 

At energies $\gg J$, the collective TQD doublet $|-;S_{\rm TQD}^z\rangle$ has not developed, and lead conduction electron scattering is dominated by the direct coupling between the leads and dot 1. In this regime the physics is controlled by an effective single-impurity Kondo temperature $T^1_{\rm K} \simeq T_{\rm K}^+$. If $J\gg T_{\rm K}^1$, then scattering is incoherent and weak at energies $\gg J$. Correspondingly, the ac conductance for $|\omega| \gg J$ is small. However, if $J\ll T_{\rm K}^1$ then Kondo-enhanced scattering with a rate $\sim T_{\rm K}^1$ controls conductance at frequencies $|\omega|\gg J$, following Eq.~\ref{eq:Gtpgtrt}. In particular, we expect in this regime Kondo resonant ac conductance for $|\omega| \sim T^1_{\rm K} \gg J$, but suppressed conductance for $|\omega|\ll J$. 

This is precisely what is observed in the right panel of Fig.~\ref{fig:Gac} at low-$T$ (turquoise line), where we have chosen $J\sim T_{\rm K}^1$. This gives rise to a non-monotonic behavior as the temperature is increased, with the Kondo effect suppressed at $T\ll J$ by ferromagnetic correlations, activated around $T\sim T_{\rm K}^1$ due to incipient screening of dot 1, and then again suppressed thermally for $T\gg T_{\rm K}^1$.

The dynamical ac conductance $G_C(\omega,T)$ therefore contains much richer information on the scattering and screening processes than the steady-state dc conductance $G_C(T)$.

%%%%%%%%%%%%%%%%
\begin{figure}[t!]
\includegraphics[width=8.7cm]{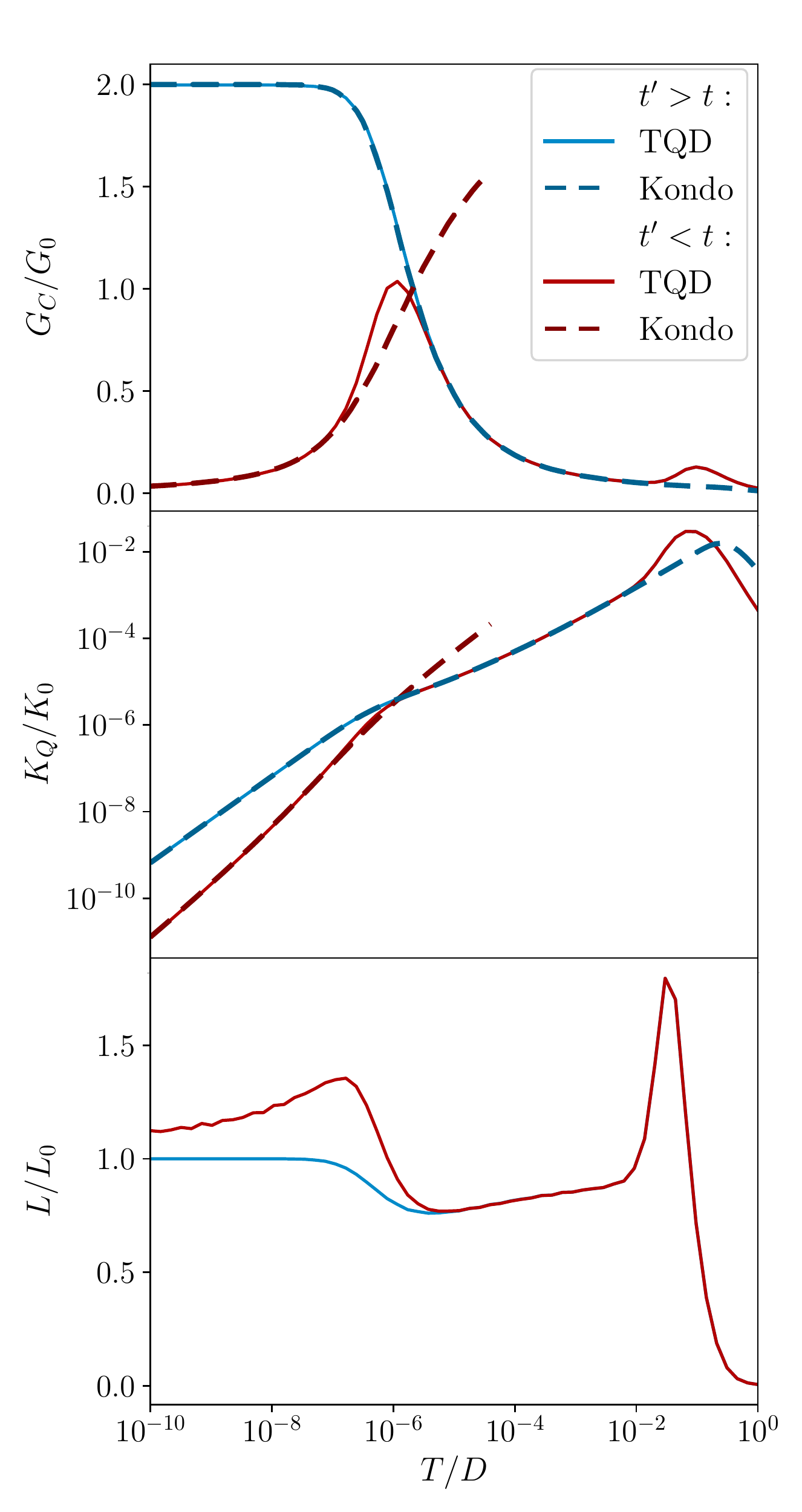}
  \caption{DC electrical conductance $G_C(T)$ (top), heat conductance $K_Q(T)$ (middle), and Lorenz ratio $L(T)$ (bottom) as a function of temperature, for the same TQD system as in Fig.~\ref{fig:Gac}. NRG results obtained using the MW formulae Eq.~\ref{eq:MW_cond} (solid lines) for both the Kondo screened phase ($t'>t$, blue) and the unscreened local moment phase ($t'<t$, red). Dashed lines show the result of the effective single-impurity, single-channel Kondo model, with effective exchange coupling $J_{\rm eff}$ obtained from model machine learning methods, Eq.~\ref{eq:ML}.  }
  \label{fig:MWthermoelectric}
\end{figure}
%%%%%%%%%%%%%%%%

In Fig.~\ref{fig:MWthermoelectric} we investigate the temperature-dependence of dc thermoelectric properties in the same system. The solid lines in the top panel for the dc electrical conductance $G_C(T)$ are obtained by taking the  $\omega\to 0$ limit of the $G_C(\omega,T)$ shown in Fig.~\ref{fig:Gac}. In practice, the dc limit is realized for $|\omega| \ll \max(T_{\rm K},T)$ and the NRG data are well-converged in this regime. This is formally and numerically equivalent to using the MW formula Eq.~\ref{eq:MW_G}. We again show the representative behavior in each phase (blue line for $t'>t$, red line for $t'<t$). The middle panel shows the corresponding curves for the heat conductance $K_Q(T)$, obtained from the MW formula Eq.~\ref{eq:MW_K}, with transmission function Eq.~\ref{eq:MW_PC_T}, and t-matrix $t_{ee,\sigma}(\omega,T)=-\Gamma~{\rm Im}G_{11,\sigma}(\omega,T)$. The lower panel shows the Lorenz ratio $L=\kappa/T G_C$ in units of the limiting WF value,\cite{franz1853ueber} $L_0=\pi^2 k_{\rm B}^2/3e^2$.

We consider first the behavior in the Kondo-screened phase for $t'>t$ (blue lines). At low temperatures $T\ll T_{\rm K}^+$, the physics of the TQD is typical of a Fermi liquid,\cite{costi2010thermoelectric} with quadratic corrections to the strong coupling fixed point. Here we find,
\begin{subequations}\label{eq:tpgtrtdcFL}
\begin{align}
G_C(T)/G_0 &= 2 - a(T/T^+_{\rm K})^2 \\
K_Q(T)/T K_0 &= \frac{2\pi^2}{3}-b(T/T^+_{\rm K})^2 \\
L(T)/L_0 &=1-c(T/T^+_{\rm K})^2
\end{align}
\end{subequations}
where $a,b,c$ are constants of order 1 depending on the specific definition of $T_{\rm K}^+$ used. Fitting to the NRG data yields $b/a \approx 9$, while $c=3b/2\pi^2-a/2$. At the lowest temperatures, the TQD behaves like a renormalized non-interacting system with $G_c \sim e^2/h$ and $K_Q\sim k_{\rm B}T/h$ while $L \to L_0$. Note that the geometric factor $4\Gamma_s\Gamma_d/(\Gamma_s+\Gamma_d)^2$ in Eq.~\ref{eq:MW_PC_TM} describing the relative coupling strength of the TQD to source and drain leads drops out here for $V_s=V_d$. In general for $V_s\ne V_d$, both $G_C$ and $K_Q$ are multiplied by this overall constant (independent of temperature), such that the Lorenz ratio $L$ is always unaffected by the coupling geometry in PC. This is not necessarily the case for non-PC systems.

At higher temperatures $T\gg T_{\rm K}^+$ in the Kondo phase, spin-flip scattering of conduction electrons from the TQD show up in thermoelectric properties as,
\begin{subequations}\label{eq:tpgtrtdcLM}
\begin{align}
G_C(T)/G_0 &= \frac{\alpha_1}{\ln^2(\alpha_2 T/T_{\rm K}^+)} \\
K_Q(T)/T K_0 &= \frac{\gamma_1}{\ln^2(\gamma_2 T/T_{\rm K}^+)} \\
L(T)/L_0 &=1-\frac{\xi}{\ln(\gamma_2 T/T_{\rm K}^+)}\label{eq:L1}
\end{align}
\end{subequations}
where we find $\gamma_1/\alpha_1=\pi^2/3$ and $\gamma_2/\alpha_2\approx 1.7$. It then follows that $\xi=\ln(\gamma_2/\alpha_2)>0$.

The numerical results in Fig.~\ref{fig:MWthermoelectric} show the full crossover behavior. Note also that although $G_C(\omega,0)$ and $G_C(0,T)\equiv G_C(T)$ appear similar, the details of their functional forms are different. The Lorenz ratio $L$ is found to approach the WF value $L_0$ from below for both $T\ll T_{\rm K}^+$ and $T\gg T_{\rm K}^+$ in the Kondo phase, although in the latter case the limit is saturated logarithmically slowly. At intermediate temperatures, especially $T\sim T_{\rm K}^+$, we see strong violations of the WF law, driven by electronic scattering from interactions.\cite{wang2013enhancement} This phenomenology is also known experimentally in single-electron transistors.\cite{kubala2008violation} Such violations are in general expected for interacting nanostructures. Note however that at intermediate temperatures, $L\ne L_0$ does not imply non-Fermi liquid physics: the WF law strictly applies\cite{franz1853ueber} only in the limit $T \to 0$, where most conventional systems exhibit Fermi liquid behavior and satisfy the WF law.

Turning now to the ferromagnetic local moment phase realized in the TQD for $t'<t$ (red lines in Fig.~\ref{fig:MWthermoelectric}), singular Fermi liquid corrections appear in the thermoelectric properties at low temperatures $T\ll J$,
\begin{subequations}\label{eq:tplesst}
\begin{align}
G_C(T)/G_0 &= \frac{\alpha'_1}{\ln^2(\alpha'_2 T/J)} \\
K_Q(T)/T K_0 &= \frac{\gamma'_1}{\ln^2(\gamma'_2 T/J)} \\
L(T)/L_0 &=1-\frac{\xi}{\ln(\gamma'_2 T/J)}\label{eq:L2}
\end{align}
\end{subequations}
where the ratios $\gamma'_1/\alpha'_1=\pi^2/3$ and $\gamma'_2/\alpha'_2\approx 1.7$ as well as $\xi=\ln(\gamma'_2/\alpha'_2)$ are the same as for Eq.~\ref{eq:tpgtrtdcLM}. In this local moment phase, the electrical and heat conductances are suppressed relative to the Kondo phase; in particular for $T\to 0$ we have $G_C\to 0$, and $K_Q/T \to 0$ corresponding to a transmission node. However, their ratio remains finite and $L \to L_0$ as $T\to 0$. The fact that the effective ferromagnetic Kondo effect in the TQD asymptotically satisfies the WF law is a signature of its low-energy singular Fermi liquid correlations. By contrast, conventional transmission nodes typically yield strongly enhanced Lorenz ratios.\cite{bergfield2009thermoelectric}

For $T\gg J$ the incipient Kondo screening of dot 1 results in behavior described by Eq.~\ref{eq:tpgtrtdcLM}. The crossover between Eqs.~\ref{eq:L1} and \ref{eq:L2} upon lowering the  temperature from $T\gg J \sim T_{\rm K}^+ $ to $T\ll J$ is therefore predicted to give a sign-change in $L(T)-L_0$, and this fingerprint of the ferromagnetic Kondo effect is observed in the lower panel of Fig.~\ref{fig:MWthermoelectric}.

The above results support the perturbative prediction\cite{mitchell2009quantum} that the physics of the half-filled TQD in the PC coupling geometry is always described by an effective single-channel, single spin-$\tfrac{1}{2}$ Kondo model (Eq.~\ref{eq:H1ck}) at the lowest temperatures.
This is to be expected from RG theory, since the Kondo model is the \textit{minimal} effective model, comprising only the most RG-relevant terms.\cite{hewson1997kondo} The resulting universality implies that the low-energy behavior of dynamic observables in the TQD is the same as that of the Kondo model when rescaled. 

Since the observables of interest here are the thermoelectric transport coefficients, we expect that $G_C(T)$ and $K_Q(T)$ calculated via Eqs.~\ref{eq:MW_PC_T}, \ref{eq:MW_PC_TM} for the effective Kondo model will precisely match those of the full TQD at low temperatures. This is because the electrical and heat conductances are controlled by the spectrum of the scattering t-matrix $t_{ee,\sigma}(\omega,T)$, which should agree in bare and effective models at low temperatures and energies -- even though the single-channel Kondo model itself has no such capacity for quantum transport measurements. This illustrates a broader point that low-energy effective models with greatly reduced complexity can still be used to obtain the correct transport properties at low temperatures -- provided the effective model parameters have been accurately determined.

 However, the perturbative prediction of the coupling constant $J_{\rm eff}^{\pm}$ via Eq.~\ref{eq:BWPT} is naturally approximate, and is in fact rather inaccurate for $U < D$ (this is seen already in the Schrieffer-Wolff mapping of the AIM to the Kondo model\cite{krishna1980renormalization}). 

A better estimation of $J_{\rm eff}^{\pm}$ can be obtained by techniques borrowed from machine learning,\cite{rigo2020machine} where the `distance' between the bare TQD model and the low-energy effective Kondo model is encoded in a loss function, which is then minimized by gradient descent. Since our target observables are determined by the  t-matrix $t_{ee,\sigma}(\omega,T)$, we define a loss function based on the symmetrized Kullback-Leibler divergence,\cite{kullback1951information,KLDnote}
\begin{subequations}\label{eq:ML}
\begin{align}
    L(J_{\rm eff}) &=\tilde{D}_{\rm KL}(t^{\rm TQD}_{ee,\sigma} \Vert t^{\rm K}_{ee,\sigma}) +\tilde{D}_{\rm KL}(t^{\rm K}_{ee,\sigma} \Vert t^{\rm TQD}_{ee,\sigma}) \;, \label{eq:loss}\\
    \tilde{D}_{\rm KL}(P \Vert Q) &= \int^{\Lambda_U}_{\Lambda_L} d\omega ~\frac{P(\omega)}{\omega} \ln\left(\frac{P(\omega)}{Q(\omega)}\right) \;,\label{eq:KLD}
\end{align}
\end{subequations}
where the roles of un-normalized distributions $P(\omega)$ and $Q(\omega)$ in Eq.~\ref{eq:KLD} are here played by the zero-temperature energy-resolved electronic scattering amplitudes in the bare TQD model and the effective Kondo model, given by $t_{ee,\sigma}(\omega,0)$. The Kullback-Leibler divergence therefore compares the t-matrices of the TQD and Kondo models over the desired energy range $\Lambda_L \le \omega \le \Lambda_U$. The loss function $L(J_{\rm eff})$ is minimized by tuning the coupling in the Kondo model. The best-fit effective model is obtained for $\partial_{J_{\rm eff}}L(J_{\rm eff})=0$ by gradient descent. We implemented a simple numerical finite-difference approximation to the gradient of $L(J_{\rm eff})$ for the optimization procedure, since finding the derivative of dynamical correlation functions is in general a hard problem; but we note that the exact gradient can in principle be calculated using the recently-developed `differentiable NRG' technique.\cite{rigo2021automatic}

Using the above systematic approach, we extracted the numerically-exact effective Kondo couplings $J_{\rm eff}$ for the TQD systems considered in Fig.~\ref{fig:MWthermoelectric}. For the Kondo screened case $t'>t$, the effective model pertains for all $T\ll U$, and so we chose an upper cutoff for the fitting $\Lambda_U=0.1\times U$. To capture the Kondo resonance we choose a low-energy cutoff $\Lambda_L=10^{-10}D \ll T_{\rm K}^+$. The model machine learning technique yielded $J^+_{\rm eff}=0.1624D$. This should be compared with the perturbative result\cite{mitchell2009quantum} for these bare parameters, $J^+_{\rm eff}=0.2D$ (but recall the exponential sensitivity of the Kondo temperature on $J_{\rm eff}$). The t-matrix of the resulting single-channel Kondo model is then used to compute the full temperature-dependence of the conductances $G_C(T)$ and $K_Q(T)$. These are plotted as the blue dashed lines in Fig.~\ref{fig:MWthermoelectric}, and are seen to agree perfectly over the full range of applicability for the effective model, $T\ll U$.

More interesting is the behavior in the local moment phase obtained in the TQD for $t'<t$. The perturbative prediction\cite{mitchell2009quantum} $J^-_{\rm eff}\simeq -0.07D$ fails to capture the first-stage renormalization from the incipient screening of dot 1 at temperatures $T\gg J$, and does not yield even qualitatively correct results for the low-$T$ physics. Using the model machine learning approach however, we find a surprisingly strong ferromagnetic coupling $J^-_{\rm eff}=-19.826D$. Since the effective Kondo model is here valid only for $T\ll J$ we set the upper cutoff to $\Lambda_U=10^{-6}D$ (while $\Lambda_L=10^{-10}D$ again). The t-matrix for the optimized effective model was used to calculate $G_C(T)$ and $K_Q(T)$ -- see red dashed lines in Fig.~\ref{fig:MWthermoelectric}. As expected, it agrees quantitatively with the exact results for the TQD in the low-temperature regime $T\ll J$.

Ultimately, the quantum phase transition in the TQD on tuning $t'/t$ can be traced to the quantum phase transition in the effective Kondo model on changing the sign of the coupling $J_{\rm eff}$. Specifically, we have antiferromagnetic coupling $J_{\rm eff}^+>0$ for $t'>t$ and ferromagnetic coupling $J_{\rm eff}^-<0$ for $t'<t$. However,  to obtain quantitatively accurate results from the effective model requires a more sophisticated non-perturbative estimation of the effective parameters, such as the method described above.\\

%%%%%%%%%%%%%%%%
\begin{figure}[t!]
\includegraphics[width=8.7cm]{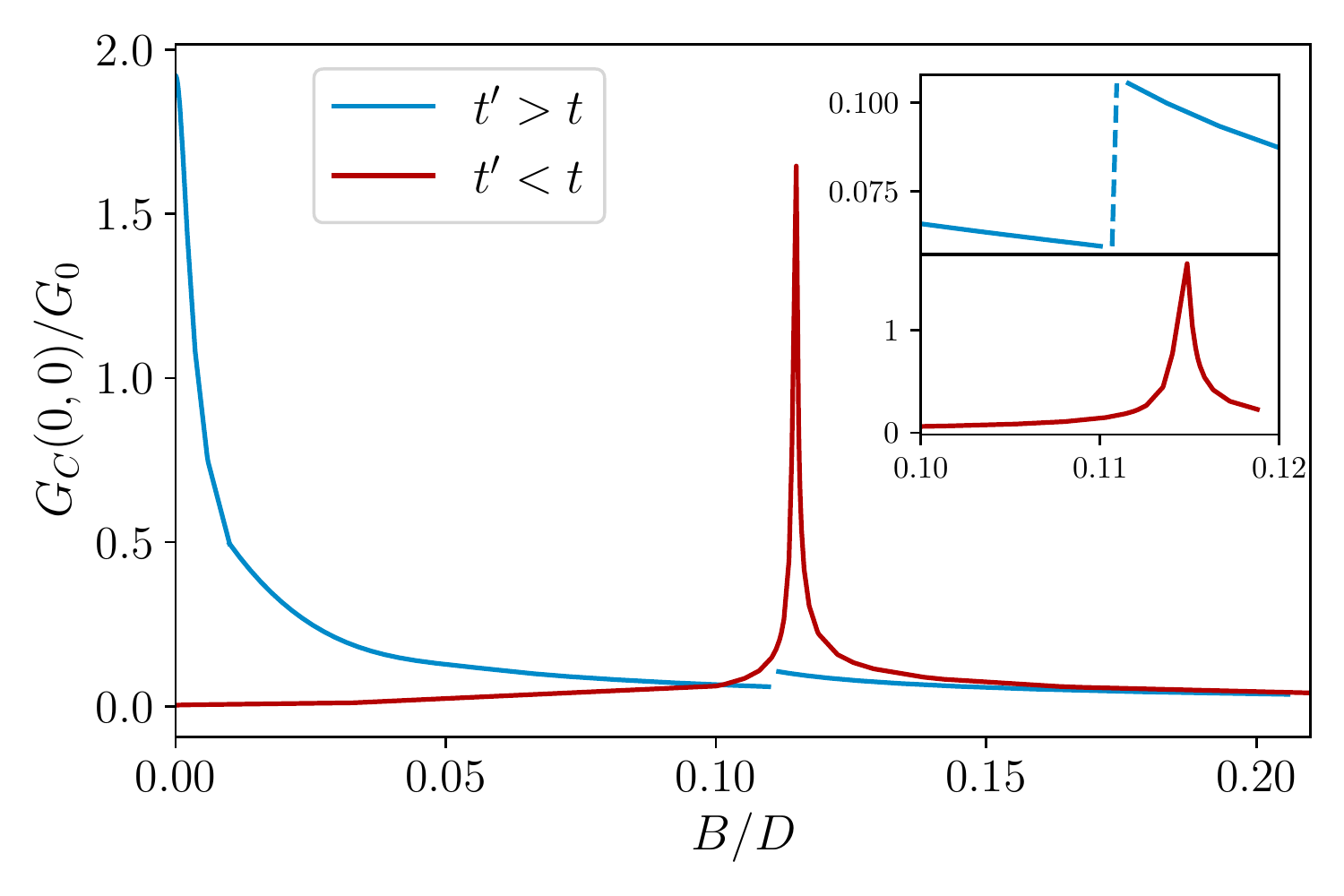}
  \caption{DC electrical conductance $G_C(0,0)$ at $T=0$ for the TQD model in PC geometry, as a function of applied magnetic field $B$. Blue line for Kondo screened phase $t'>t$, red line for local moment phase $t'<t$. Inset shows magnified view around the field-induced ground-state transition in the isolated TQD, which results in a level-crossing QPT in the lead-coupled system for $t'>t$ but re-entrant Kondo for $t'<t$. Model parameters: $U=0.4D$, $U'=0$, $\epsilon=-U/2$, $V_s=V_d=0.1D$ and $(t,t')=(0.094,0.04)$ and $(0.08,0.1)$ for red and blue lines, respectively. NRG parameters: $\Lambda=2.5$, $M_K=2000$. 
  }
  \label{fig:bfield}
\end{figure}
%%%%%%%%%%%%%%%%

We consider now transport through the TQD in the presence of a finite local magnetic field $B$. To do this we add a term to the TQD Hamiltonian $H_{\rm field}=-B\sum_j \hat{S}^z_j$. The t-matrix controlling quantum transport in this coupling geometry becomes spin-dependent $t_{ee,\uparrow}(\omega,T)\ne t_{ee,\downarrow}(\omega,T)$, and therefore the current for up and down spin electrons is unequal. Within the MW formalism, the total (spin-summed) linear conductances follows from Eqs.~\ref{eq:MW_PC}, \ref{eq:MW_PC_TM}. 

Focusing on the $T=0$ dc linear electrical conductance $G_C(0)$ as a function of magnetic field strength $B$ in Fig.~\ref{fig:bfield}, we again compare the behavior in the Kondo phase ($t'>t$, blue line) and the local moment phase ($t'<t$, red line). For $B=0$, we see from Fig.~\ref{fig:MWthermoelectric} that $G_C(0)=0$ for $t'<t$ but $G_C(0)\simeq 2e^2/h$ for $t'>t$ (deviations from the ideal unitarity limit for a single-electron transistor arise if particle-hole symmetry is broken). For $t'>t$, the degenerate TQD doublet ground state $|+;S^z_{\rm TQD}=\pm\tfrac{1}{2}\rangle$ at $B=0$ gets split at finite $B$. The Kondo effect is still operative, providing the Zeeman splitting $\sim |B|$ is much less than the Kondo binding energy $T_{\rm K}^+$. However, the Kondo effect is completely destroyed\cite{hewson1997kondo} and the low-$T$ conductance collapses\cite{pustilnik2004kondo} for $B \gg T_{\rm K}^+$, since then the unique TQD ground state $|+;+\tfrac{1}{2}\rangle$ is singled out and spin-flip excitations involving $|+;-\tfrac{1}{2}\rangle$ become suppressed. The blue line in Fig.~\ref{fig:bfield} shows the expected\cite{koga2013field,hamad2015scaling} rapid decrease in the conductance as $B$ increases. 

As $B$ increases further to $B\sim J$, the ground state of the isolated TQD with $t'>t$ changes from $|+;+\tfrac{1}{2}\rangle$ to the extremal weight state of the spin-quartet $|\uparrow\uparrow\uparrow\rangle$. Exactly at the transition $B=B^*$, the $|+;+\tfrac{1}{2}\rangle$ and $|\uparrow\uparrow\uparrow\rangle$ TQD states are precisely degenerate. However, they cannot be interconverted by tunneling to the leads because they have different \emph{parity} under the mirror symmetry operation\cite{mitchell2009quantum} that swaps the labels on dots 2 and 3 ($|+;S^z_{\rm TQD}\rangle$ is symmetric while $|\uparrow\uparrow\uparrow\rangle$ is antisymmetric). This means that $\langle \uparrow\uparrow\uparrow | \hat{S}^+_1 | +;+\tfrac{1}{2}\rangle=0$, as can be seen directly from the form of Eq.~\ref{eq:plusstate} (although the symmetry argument holds to all orders and does not require the half-filled limit approximation). As a consequence, there can be no field-induced Kondo effect\cite{sasaki2000kondo,pustilnik2001kondo,PustilnikGlazman_realQD2001,costi2000KondoBfield,golovach2003kondo,Dias_2QD-SIAM_2008} in the TQD system for $t'>t$, and we see a first-order (level-crossing) quantum phase transition in the full lead-coupled system, with a change in many-body ground states at the critical field strength $B^*$. This shows up in the $T=0$ conductance as a small discontinuity -- see inset, Fig.~\ref{fig:bfield} (blue line). 

The $t'<t$ phase is more interesting because, although no Kondo effect occurs at $B=0$, we now have the possibility of a field-induced Kondo effect at $B=B^*$. This is because the ground state of the isolated TQD at finite $B$ crosses over from $| -;S^z_{\rm TQD}=+\tfrac{1}{2}\rangle$ to $|\uparrow\uparrow\uparrow\rangle$ as a function of $B$ for  $t'<t$, and both states have the same parity (they are both antisymmetric under $2\leftrightarrow 3$ exchange). At $B^*$, the two states are degenerate and can therefore be interconverted by coupling to the leads. The matrix element $\langle \uparrow\uparrow\uparrow | \hat{S}^+_1 | -;+\tfrac{1}{2}\rangle \ne 0$, as can be seen from Eq.~\ref{eq:minusstate}, and therefore an effective Kondo model of the form Eq.~\ref{eq:H1ck} arises at $B=B^*$, where the two components of the effective impurity spin-$\tfrac{1}{2}$ are played here by the degenerate TQD states $| -;+\tfrac{1}{2}\rangle$ and $|\uparrow\uparrow\uparrow\rangle$, similar to the finite-field singlet-triplet Kondo effect in quantum dots.\cite{sasaki2000kondo,pustilnik2001kondo} This novel prediction for a field-induced Kondo effect involving the crossing of spin doublet and spin quartet states in the TQD is borne out by the NRG results shown as the red lines in Fig.~\ref{fig:bfield}, where a dramatic conductance enhancement arises for $t'<t$ at $B\sim J$. There is no quantum phase transition as a function of $B$ for $t'<t$.

The TQD therefore hosts Kondo-boosted conductance at $B=0$ for $t'>t$, but at $B\sim J$ for $t'<t$. This provides sensitive control over the current through the TQD, with a high on-off ratio, by tuning magnetic fields. \\

%%%%%%%%%%%%%%%%
\begin{figure}[t!]
\includegraphics[width=9cm]{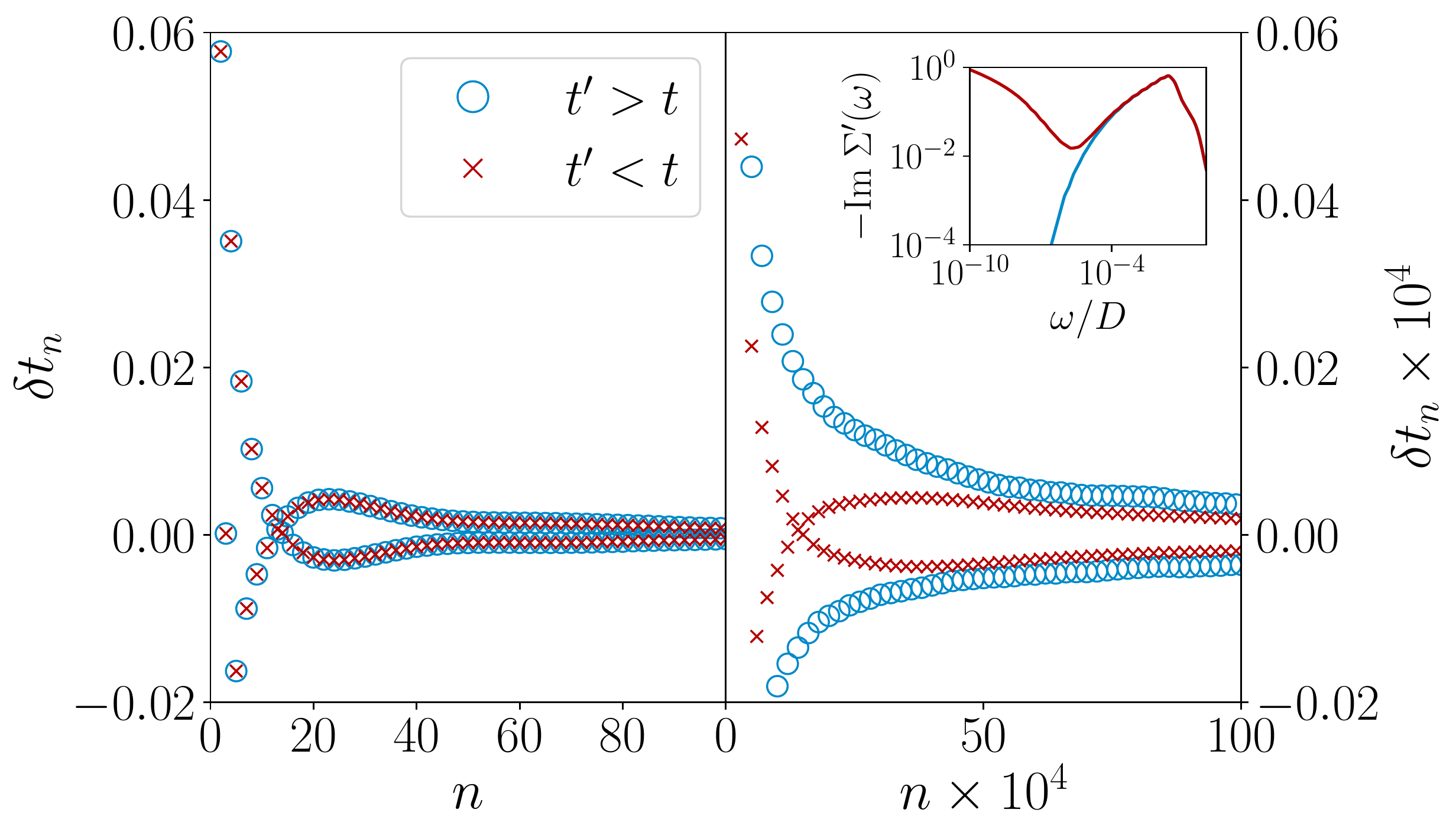}
  \caption{Mapping of the TQD self-energy in PC geometry to an auxiliary non-interacting chain Eq.~\ref{eq:Haux}, with effective tunneling matrix elements $\delta t_n=t_n-t_{\infty}$ plotted against chain site number $n$ close to the physical TQD (left) and far from it (right). Inset shows the input self-energy. Blue or red symbols or lines for Kondo screened or local moment phases, respectively. Parameters same as Fig.~\ref{fig:Gac}.
    }
  \label{fig:TQDaux}
\end{figure}
%%%%%%%%%%%%%%%%

Finally, we turn to the auxiliary field representation of the interacting TQD in Fig.~\ref{fig:TQDaux}. As discussed above, transport in the PC coupling geometry is controlled by the Green's function of dot 1, $\overline{\overline{G}}_{\sigma}(\omega) = G_{11,\sigma}(\omega)$. From Eq.~\ref{eq:G_PC_aux} we extract the effective single-impurity self-energy $\Sigma'_{\sigma}(\omega)$, which incorporates both the effect of dots 2 and 3 as well as the electronic interactions. Defining the dynamical part $\tilde{\Sigma}'_{\sigma}(\omega)=\Sigma'_{\sigma}(\omega) - {\rm Re} \Sigma'_{\sigma}(0)$, we map this to an effective non-interacting auxiliary chain Eq.~\ref{eq:Haux} by setting $\tilde{\Sigma}'_{\sigma}(\omega) \to \Delta_{\sigma}^{\rm aux}(\omega)$, as described in Sec.~\ref{sec:aux}. For simplicity we consider the case $B=0$ where $\tilde{\Sigma}'_{\sigma}(\omega)\equiv \tilde{\Sigma}'(\omega)$, and hence $\Delta_{\sigma}^{\rm aux}(\omega) \equiv \Delta^{\rm aux}(\omega)$, and half-filling. For the same TQD model parameters as in Fig.~\ref{fig:MWthermoelectric}, we extract the effective self-energy at $T=0$, perform the auxiliary chain mapping, and plot the chain hopping parameters $\delta t_n=t_n-t_{\infty}$ as a function of chain position $n$ in Fig.~\ref{fig:TQDaux}. As before, blue lines/points correspond to the Kondo phase $t'>t$ while red lines/points are for the local moment phase $t'<t$. Inset shows the behavior of the self-energy $\tilde{\Sigma}'(\omega)$ from which the $\delta t_n$ are extracted in each case. 

The overall behavior of the $\delta t_n$ is rather regular; the early part of the chain (small $n$) roughly corresponds to high-energy features in $\tilde{\Sigma}'(\omega)$, while low-energy information is encoded at large $n$. In particular, since the self-energy for both $t'>t$ and $t'<t$ is essentially equivalent down to $|\omega| \sim J$ (see inset), the resulting $\delta t_n$ at small $n$ are essentially identical for both phases -- see left panel of Fig.~\ref{fig:TQDaux}. On the other hand, the low-energy $|\omega|\to 0$ form of the self-energy is very different in the two phases, with $-{\rm Im}\tilde{\Sigma}'(\omega) \sim \omega^2$ characteristic of the Fermi liquid correlations in the $t'>t$ Kondo phase,\cite{hewson1997kondo} but with $-{\rm Im}\tilde{\Sigma}'(\omega) \sim \ln^2(|\omega|/J)$ in the $t'<t$ local moment phase, resulting from  singular Fermi liquid corrections.\cite{koller2005singular} The large $n$ behavior of $\delta t_n$ in the right panel of Fig.~\ref{fig:TQDaux} is therefore different in the two phases. In the Kondo phase we have asymptotic behavior $\delta t_n \sim (-1)^n/n$, satisfying Eq.~\ref{eq:auxtn} as expected, while in the local moment phase, we see a crossover in the envelope function of  $\delta t_n$ at $n \sim 10^5$ (red points), corresponding to the minimum in $-{\rm Im}\tilde{\Sigma}'(\omega)$ on the scale of $|\omega/D|\sim 10^{-5}$ (red line, inset). The asymptotic behavior is found to be $\delta t_n \sim (-1)^n/\ln^2(n)$.

In the auxiliary field representation of the TQD, the system is completely non-interacting, and transport properties can be determined using the LB formula, Eq.~\ref{eq:LBaux}. In general, one must of course determine the auxiliary chain parameters before applying the LB formula. We obtained them numerically exactly here from a knowledge of the self-energy to demonstrate the procedure; but the rather simple form of the coefficients $t_n$ might also justify the use of simple approximations to capture the essence of interaction effects on transport in a heuristic way.

%!!!!!!!!!!!!!!!!!!

%%%%%%%%%%%%%%%%
\begin{figure}[t!]
\includegraphics[width=8.7cm]{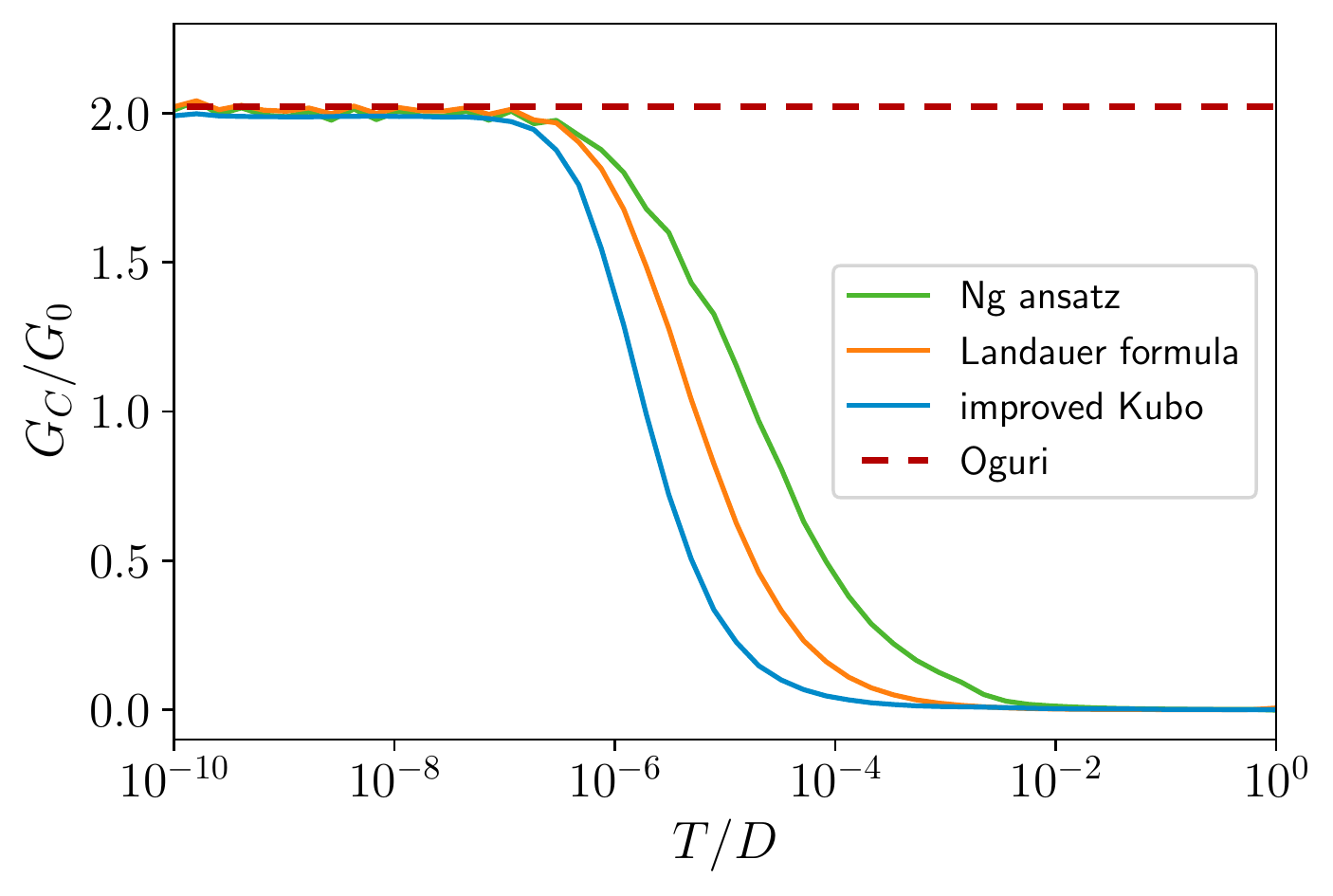}
  \caption{Serial dc electrical conductance $G_C(T)$ through the TQD in non-PC coupling geometry (see Fig.~\ref{fig:tqd}(b) and Eq.~\ref{eq:tqd_hyb_2}), comparing the numerically-exact improved Kubo formula (Eqs.~\ref{eq:kubo}, \ref{eq:kubo_dc}, \ref{eq:newK}, blue line) with common approximations based on Green's functions: Ng ansatz (Eqs.~\ref{eq:MW_G}, \ref{eq:ng_T}, green line), Landauer (Eqs.~\ref{eq:G_land}, \ref{eq:land_T} but with interacting Green's functions, orange line), and Oguri renormalized Landauer (Eq.~\ref{eq:G_oguri}, red dashed). Model parameters: $U = 0.4D$, $U'=0$,  $\epsilon=-U/2$, $V_s = V_d =0.12D$, $t=0.02D$, $t'=0.005D$. NRG parameters: $\Lambda=3$, $M_K = 10000$.
  }
  \label{fig:nonPC}
\end{figure}
%%%%%%%%%%%%%%%%

\subsubsection{Non-Proportionate Coupling}
We now turn to the more complex case in which leads are attached to dots 2 and 3 (see Fig.~\ref{fig:tqd}b and Eq.~\ref{eq:tqd_hyb_2}). Since the coupling geometry does not satisfy the PC condition, the MW formulation involving only retarded TQD Green's functions used in the previous section cannot be applied here. Instead, we obtain the linear electrical conductance by NRG from the Kubo formula Eq.~\ref{eq:kubo}, using the `improved' version of the current-current correlator, Eq.~\ref{eq:newK}.

In Fig.~\ref{fig:nonPC} we consider $G_C(T)$ for the TQD at half-filling with typical parameters, comparing the result of the improved Kubo formula (blue line, Eqs.~\ref{eq:kubo}, \ref{eq:kubo_dc}, \ref{eq:newK}) with approximations involving only retarded equilibrium single-particle Green's functions. The dc conductance shows a strong Kondo enhancement at low-temperatures $T\ll T_{\rm K}\sim 10^{-5}D$, signifying the involvement of collective TQD spin states in series transport.

The red dashed line is the Oguri result Eq.~\ref{eq:G_oguri}, for the asymptotic $T\to 0$ dc conductance $G_C(0)$. The agreement with the result from the Kubo formula for $T\ll T_{\rm K}$ is as expected for a system with a Fermi liquid ground state. The green line is obtained using the Ng Ansatz, in which the approximate transmission function Eq.~\ref{eq:ng_T} is employed in Eq.~\ref{eq:MW_G}. It reduces to the Oguri result, as expected, for $T\ll T_{\rm K}$, and captures the qualitative features of the true result, but significant deviations are observed for $T\sim T_{\rm K}$. Ironically, simply using the interacting retarded Green's functions in the non-interacting Landauer formula, Eq.~\ref{eq:G_land} (orange line) does a better job.

For linear response electrical transport in non-PC geometries, we conclude that the improved Kubo formula is the method-of-choice. Within NRG, the accuracy of the method was validated in Fig.~\ref{fig:KvsimpK}; it is in fact also simpler and less computationally expensive than the standard approximate alternatives.\\

%!!!

\begin{figure}[t!]
\includegraphics[width=8.7cm]{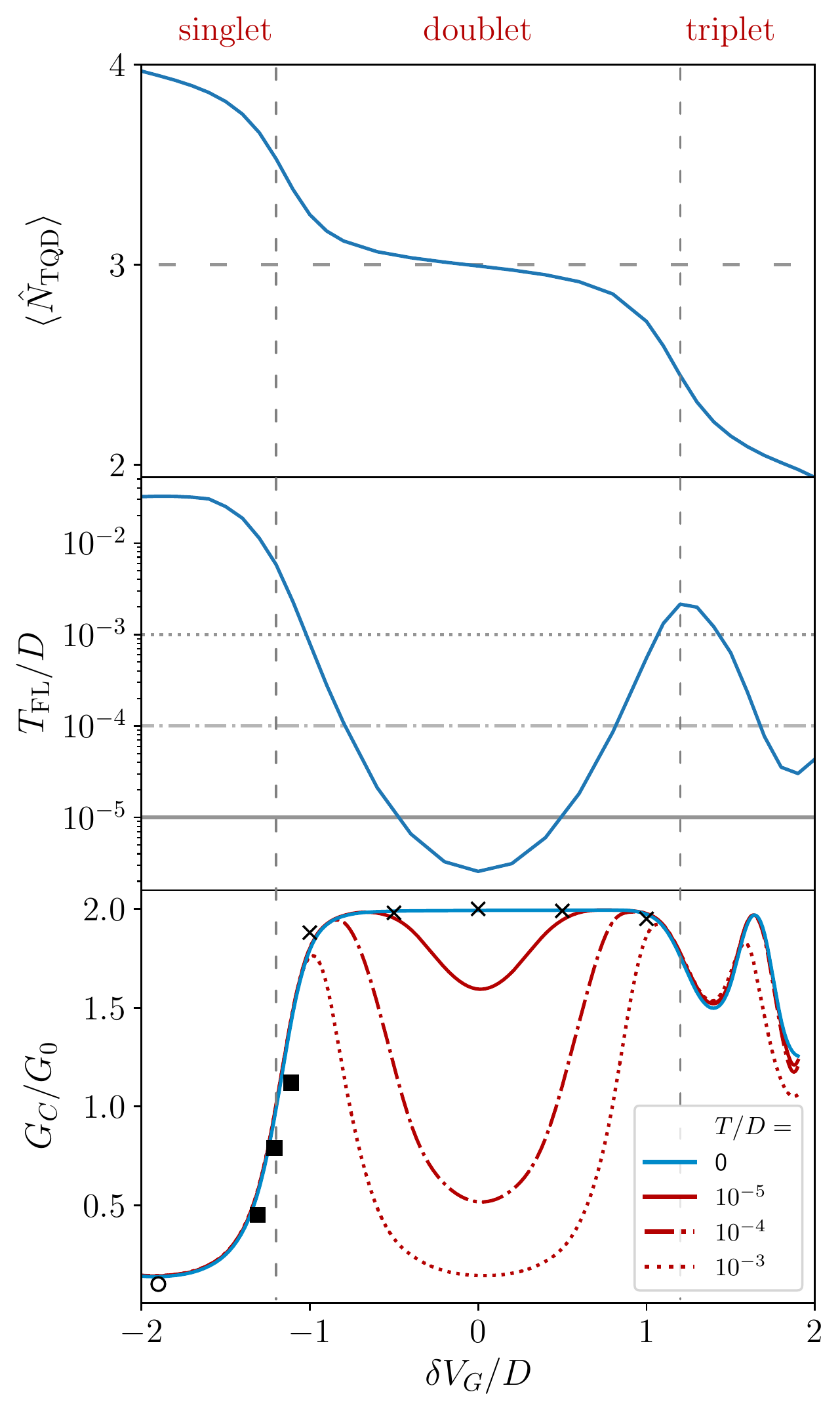}
  \caption{Gate-controlled TQD in the non-PC coupling geometry.\textit{ Top panel:} TQD occupancy $\langle \hat{N}_{\rm TQD}\rangle$ in the full lead-coupled, interacting system, showing Coulomb blockade staircase. \textit{Middle panel:} Low-energy Fermi liquid scale $T_{\rm FL}$, defined as the temperature below which TQD degrees of freedom are quenched (in practice via the TQD entropy $S_{\rm TQD}(T_{\rm FL})=0.1$). \textit{Bottom panel:} dc series electrical conductance, obtained from NRG using the improved Kubo formula, Eqs.~\ref{eq:kubo}, \ref{eq:kubo_dc}, \ref{eq:newK} (lines) at different temperatures given in the legend. Symbols show the $T=0$ effective model predictions: $S=1/2$ CB regime (cross points via Eq.~\ref{eq:cond_2ck_sym}), $S=0$ CB regime (circle point via Eq.~\ref{eq:Gqpc}), singlet-doublet MV crossover (square points via Eq.~\ref{eq:Heff_mv_s-d}). Vertical dashed lines show MV transitions in the isolated TQD for reference. 
 Model parameters: $U=3D$, $U'=D$, $t=0.3D$, $t'=0.1D$ and $V_s=V_d=0.3D$. Gate voltage defined such that $\delta V_G=0$ is at half-filling. NRG parameters: $\Lambda=3$, $M_K=6000$.
  }
  \label{fig:gateTQD}
\end{figure}

In the lower panel of Fig.~\ref{fig:gateTQD} we consider the series dc electrical conductance through the TQD as a function of gate voltage $\delta V_G=\epsilon-\epsilon_0$, defined such that $\delta V_G =0$ corresponds to half-filling. The conductance $G_C$ is calculated using NRG via the improved Kubo formula at different temperatures, corresponding to the different gate voltage trace lines. We use this to test the analytic predictions of Secs.~\ref{sec:emPC_CB} and \ref{sec:emPC_MV}, which are shown as the points in Fig.~\ref{fig:gateTQD}.

%%%%%%%%%%%%%%%%
%%%%%%%%%%%%%%%%

Before discussing transport, we comment on the underlying Kondo physics of this system.\cite{bonvca2008numerical,bonvca2007fermi} The number of electrons on the \textit{isolated} TQD is conserved and takes integer values. We define $\langle \hat{N}_{\rm TQD}\rangle$ in terms of the TQD number operator $\hat{N}_{\rm TQD}=\sum_{j,\sigma} d_{j\sigma}^{\dagger}d_{j\sigma}^{\phantom{\dagger}}$. With finite interdot repulsion $U'>0$, the half-filled $N_{\rm TQD}=3$ electron sector is well-separated in gate voltage from the $N_{\rm TQD}=2,4$ sectors. The transitions between these filling sectors are marked as the vertical grey dashed lines in Fig.~\ref{fig:gateTQD}, and correspond to the charge-degeneracy points in the isolated TQD (for the parameters used, they arise at $\delta V_G/D \approx \pm 1.2$). Analysis of the TQD shows that for $N_{\rm TQD}=3$ we have a spin-doublet ground state (effective impurity $S=\tfrac{1}{2}$), while for $N_{\rm TQD}=2$ we have an effective $S=1$ spin-triplet ground state, and for $N_{\rm TQD}=4$ the ground state is an effective $S=0$ spin-singlet. This results in a rich range of gate-tunable many-body physics on attaching the leads. In the top panel of Fig.~\ref{fig:gateTQD} we see how the sharp Coulomb blockade (CB) steps are now smoothed into a continuous evolution of $\langle \hat{N}_{\rm TQD}\rangle$. However, one can still identify clear CB regimes and mixed-valence (MV) crossovers. In particular, we have \textit{defined} $\langle \hat{N}_{\rm TQD}\rangle=3$ at $\delta V_G=0$, while, for the parameters used, the center of the triplet and singlet CB regimes with $\langle \hat{N}_{\rm TQD}\rangle=2,4$ occur at $\delta V_G/D \approx \pm 2$.

Deep in the spin-doublet CB regime, one can derive an effective $S=\tfrac{1}{2}$ two-channel Kondo model of the type given in Eq.~\ref{eq:H2ck}. Note that in standard situations, the cross terms $J_{sd}$ and $W_{sd}$ are finite; these are the processes in the effective model responsible for mediating a source-drain current. We consider explicitly the $sd$ symmetric case with $V_s=V_d$, and so $J_{ss}=J_{dd}\equiv J$ and $W_{ss}=W_{dd}\equiv W$. Deviations from $\langle \hat{N}_{\rm TQD}\rangle=3$ at finite $\delta V_G$ lead to finite potential scattering terms $W_{ss}$ and $W_{dd}$. 

From the results of Sec.~\ref{sec:sd_symm}, we expect an effective spin-$\tfrac{1}{2}$ single-channel Kondo effect on the lowest energy scales, forming with the even lead combination. The resulting Kondo temperature $T_{\rm K}$ in this regime is shown in the middle panel of Fig.~\ref{fig:gateTQD}, and exhibits the typical behavior of Eq.~\ref{eq:tk_1ck}. At $T=0$ we therefore expect Kondo-boosted conductance. This is shown in the lower panel of Fig.~\ref{fig:gateTQD} as the blue line, where the conductance is seen to reach up to $G_C(0) \simeq 2e^2/h$ at the particle-hole symmetric point $\delta V_G=0$. In fact, the conductance remains close to this maximum value throughout most of the doublet regime, despite significant deviations in $\langle \hat{N}_{\rm TQD}\rangle$ away from half-filling. The $T=0$ conductance only begins to attenuate when close to the mixed-valence points. 

This behavior can be understood from the effective low-energy model. 
With the chosen bare TQD parameters $U=3D$, $U'=D$ and $V_s=V_d=U/10$, the perturbative estimation of the effective model parameters via Eq.~\ref{eq:BWPT} is quite accurate. The low-$T$ conductance can then be estimated from Eq.~\ref{eq:cond_2ck_sym}. For strong interactions, we find that the effective potential scattering parameters $W_{ee}$ and $W_{oo}$ remain small until close to the MV points (where the applicability of Eq.~\ref{eq:H2ck} and hence Eq.~\ref{eq:cond_2ck_sym} breaks down anyway). The numerical values of $G_C(0)$ predicted in the doublet CB regime are shown in the lower panel of Fig.~\ref{fig:gateTQD} as the cross points, and agree rather well with the full NRG calculations. 

The red lines in Fig.~\ref{fig:gateTQD} show NRG results for the conductance at higher temperatures. Since the conductance remains Kondo-enhanced for $T\ll T_{\rm K}$ but is strongly reduced as the Kondo effect is destroyed for $T\gg T_{\rm K}$, we expect a non-trivial behaviour of $G_C(T)$ at a fixed $T$ on changing the gate voltage $\delta V_G$, because $T_{\rm K}$ itself depends on $\delta V_G$. We plot the conductance at $T/D=10^{-5}, 10^{-4}, 10^{-3}$ in the lower panel (these same temperatures are marked as the horizontal lines in the middle panel). As expected, we see reduced conductance as the ratio $T/T_{\rm K}$ increases.

Deep in the CB spin-singlet regime around $\delta V_G/D=-2$, the low-energy effective model is described by Eq.~\ref{eq:Heff_S0}, with conductance controlled simply by the effective cotunneling amplitude $W_{sd}$ via Eq.~\ref{eq:Gqpc}. There is no low-energy scale in this regime; the effective model is valid for all $T\ll U$ (we denote the scale where the spin singlet ground state forms as $T_{\rm K}$, but this should be understood as a Fermi liquid scale rather than a Kondo scale \textit{per se}). The perturbative estimation of $W_{sd}$ and hence $G_C(0)$ is shown as the circle point in the lower panel of Fig.~\ref{fig:gateTQD} and agrees well with NRG calculations in this regime. As anticipated, the conductance is small since electrons must tunnel through the interacting TQD unaided by the Kondo effect. 

In the CB spin-triplet regime, both even and odd conduction electron channels are involved in screening the collective TQD spin states (see Sec.~\ref{sec:S1}). The delicate interplay between the effective couplings $J_{ee}$ and $J_{oo}$ (and hence the Kondo scales $T_{\rm K}^e$ and $T_{\rm K}^o$) produces more complex behavior in the conductance. The Fermi liquid scale $T_{\rm FL}$  corresponds here to the temperature below which the TQD spin is exactly screened, $T_{\rm FL} \equiv \min(T_{\rm K}^e,T_{\rm K}^o) = T_{\rm K}^o$. From NRG we see a Kondo peak reaching up to the maximum $G_C=2e^2/h$ at around $\delta V_G \approx 1.7$ (that is, away from the center of the CB window at $\delta V_G \simeq 2$ where $\langle \hat{N}_{\rm TQD}\rangle=2$). This can be viewed as a many-body quantum interference effect resulting from different conductance pathways around the TQD ring.\cite{mitchell2017kondo,donarini2010interference}  The insensitivity of the conductance in this regime to the ratio $T/T_{\rm K}$ implies a two-stage screening with $T_{\rm K}^e\gg T_{\rm K}^o$. The first-stage partial screening arises at relatively high temperatures $T_{\rm K}^e >T$ and is chiefly responsible for the enhanced conductance.

We now consider the MV crossovers between the above regimes, where the TQD charge is strongly fluctuating and the CB approximation breaks down. For $\delta V_g/D \approx -1.2$ the isolated TQD has degenerate ground states with $N_{\rm TQD}=3$ and $4$ electrons, corresponding to a singlet-doublet MV transition. This is smoothed into a crossover on attaching the leads and the TQD occupancy varies smoothly as a function of gate; the many-body ground states in the Kondo-screened doublet CB regime and the CB singlet regime are continuously connected. In the vicinity of the singlet-doublet MV point we use the effective model Eq.~\ref{eq:Heff_mv_s-d}, calculating the effective parameters $\tilde{\Gamma}$ and $\epsilon$ from the isolated TQD eigenstates and energies. Since this effective model is in PC, the transmission function within the MW formalism then follows from t-matrix as plotted in Fig.~\ref{fig:MV_sd}. The low-$T$ conductance obtained using this approximation are shown as the solid square points in Fig.~\ref{fig:gateTQD}, and capture the behaviour of the exact $T=0$ result (blue line).

At $\delta V_g/D=+1.2$ the system is at the MV point between the $N_{\rm TQD}=3$ doublet and the $N_{\rm TQD}=2$ triplet. An effective model describing the coupling of these 5 TQD states to the leads to first order in the hybridization was derived in Eq.~\ref{eq:Heff_mv_d-t}. However, diagonalization of the isolated TQD shows multiple low-lying excited states in this regime. The analysis of Sec.~\ref{sec:MVdt} requires that the retained 5-fold-degenerate ground state TQD manifold be well-separated from excited states, and hence breaks down in this case. In addition, the coupling to the odd conduction channel is not captured by first-order perturbation theory, but 2-stage Kondo screening involving both even and odd channels is observed in NRG calculations for the full lead-coupled system. This becomes important when $T \lesssim T_{\rm K}^o$. Predictions of the  conductance using Eq.~\ref{eq:Heff_mv_d-t} in this regime (not shown) considerably underestimate the exact results. This demonstrates that care must be taken when using results of the effective models to check that the conditions used to derive them are satisfied for a given system. Of course, in such cases the effective models could be generalized to include more TQD states and/or go beyond first-order perturbation theory.

%###########################

%%%%%%%%%%%%%%%%
\begin{figure*}[t!]
\includegraphics[width=16cm]{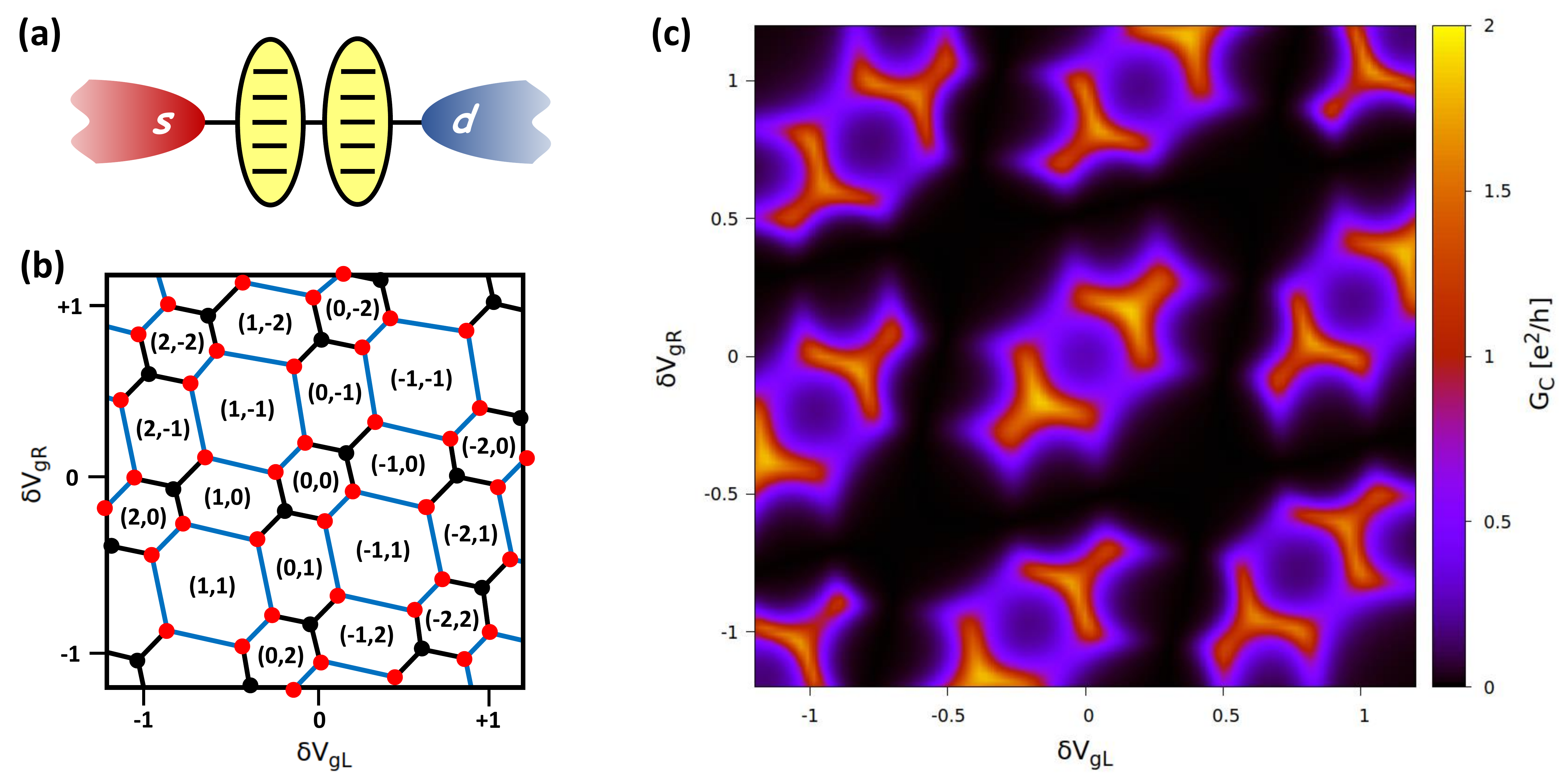}
  \caption{(a) Schematic illustration of the two-lead serial MLDQD model. (b) Charge stability diagram of the isolated DQD as a function of gate voltages $\delta V_{gL}$ and $\delta V_{gR}$. Charge configurations $(\delta N_L,\delta N_R)$ correspond to the left and right dot charges, relative to half-filling. 
 (c) NRG calculations of the series dc linear conductance through the MLDQD with $M=3$ levels per dot, as a function of gate voltages $\delta V_{gL}$ and $\delta V_{gR}$, obtained using the improved Kubo formula at finite temperature $T/D=10^{-3}$. Model parameters: $\delta\epsilon=0.2D$, $U=0.2D$, $V=0.2D$, $t=0.1D$. NRG parameters: $\Lambda=5$, $M_K=4096$. We exploit left/right symmetry and a particle-hole transformation to relate the quadrant $|\delta V_{gR}| \le \delta V_{gL}\ge 0$ to the other three quadrants. The periodic structure \emph{within} the calculated quadrant is a result of the multi-level nature of the dots (not enforced symmetrization). 
  }  \label{fig:dqd}
\end{figure*}
%%%%%%%%%%%%%%%%

\subsection{Serial multilevel double quantum dot}\label{sec:dqd}
In this section we consider the serial MLDQD model illustrated in Fig.~\ref{fig:dqd}(a), consisting of two equivalent multilevel dots, tunnel-coupled together and each coupled to its own lead. The Hamiltonian reads $H=H_{\rm leads}+H_{\rm DQD}+H_{\rm hyb}$, where $H_{\rm leads}$ is given in Eq.~\ref{eq:Hleads} as usual, and,
\begin{equation}\label{eq:dqd}
\begin{split}
    H_{\rm DQD} =& \sum_{\alpha=L,R}\left( \sum_{j=1}^M~\sum_{\sigma=\uparrow,\downarrow} \epsilon_{j}\hat{n}_{\alpha j \sigma}  +U \hat{N}_{\alpha}^2 \right ) + U' \hat{N}_L\hat{N}_R \\
    &+\sum_{j,k,\sigma} \frac{t}{M}\left ( d_{L j\sigma}^{\dagger} d_{R k\sigma}^{\phantom{\dagger}} + {\rm H.c.} \right ) + \sum_{\alpha}  V_{g\alpha} \hat{N}_{\alpha} \;, \\
    H_{\rm hyb} =& \sum_{\alpha,j,\sigma} \frac{V_j^{\alpha}}{\sqrt{M}}\left (  d_{\alpha j\sigma}^{\dagger}c_{\alpha \sigma}^{\phantom{\dagger}} + {\rm H.c.} \right ) \;,
\end{split}
\end{equation}
where $M$ is the number of levels per dot and $d_{\alpha j \sigma}^{\dagger}$ ($d_{\alpha j \sigma}$) is the creation (annihilation) operator for level $j=1,2,...,M$ of dot $\alpha=L,R$ with spin $\sigma=\uparrow,\downarrow$. We define number operators $\hat{n}_{\alpha j \sigma}= d_{\alpha j \sigma}^{\dagger}d_{\alpha j \sigma}^{\phantom{\dagger}}$ and $\hat{N}_{\alpha} =\sum_{j,\sigma} \hat{n}_{\alpha j \sigma}$. In this work we use $M=3$, and a constant level spacing $\delta \epsilon = \epsilon_{j+1}-\epsilon_j$ with $\epsilon_2=0$, set $U'=U/2$ and $V_j^{\alpha}\equiv V$, and define $\delta V_{g\alpha}=V_g^0+V_{g\alpha}$ such that $\delta V_{gL}=\delta V_{gR}=0$ is at half-filling. 

The charge stability diagram of the isolated DQD as a function of $\delta V_{gL}$ and $\delta V_{gR}$ for representative parameters is shown in Fig.~\ref{fig:dqd}(b). The charge configurations $(\delta N_L,\delta N_R)$ correspond to the left and right dot charges, relative to half-filling, which for $M=3$ means 3 electrons per dot. Lines denote degeneracies between two isolated DQD charge configurations; filled circles correspond to the triple points. The deformed hexagonal structure of the diagram is a ubiquitous feature of experimental double dot systems with both local interactions $U$ on each dot, and capacitive interactions $U'$ between dots.\cite{van2002electron,petta2004manipulation,keller2014emergent} Since typically $U'<U$, not all hexagons are equivalent, and pronounced even-odd (parity) effects are observed. For DQDs described by Eq.~\ref{eq:dqd} with many levels $M$, configurations $(n,m)$ are equivalent to $(n+2x,m+2y)$ with integer $x$ and $y$. For finite $M$ the periodic structure is truncated.

When the leads are attached, Kondo effects can arise at low temperatures -- provided either one or both of the dots host a net spin-$\tfrac{1}{2}$ local moment. The left dot may therefore participate in a Kondo effect when it is in the CB regime with an odd number of electrons (that is, deep inside a CB hexagon $(N_L,N_R)$ with odd $N_L$, independent of $N_R$). Similarly Kondo effects with the right dot require odd $N_R$. However, the inter-dot coupling competes with these individual dot-lead Kondo effects when both dots host local moments.\cite{jeong2001kondo} Therefore the Kondo effects arising in sectors $(N_L,N_R)$, with $N_L$ and $N_R$ both odd, are only partially developed (note that realistic Anderson-type models of double quantum dots do \textit{not} support a quantum phase transition between Kondo and RKKY phases,\cite{jayatilaka2011two} as would be predicted from the oversimplified two-impurity Kondo model\cite{jones1988low}). Strong DQD charge fluctuations arise in the vicinity of the lines separating two hexagons (MV lines). The lines connecting configurations $(N_L+1,N_R)$ and $(N_L,N_R+1)$ are particularly important:  such configurations are separated in gate-voltage space from other configurations by the finite inter-dot capacitance, and are interconverted by the inter-dot tunneling. Including dot-lead coupling, this can produce an orbital Kondo effect, which has an emergent SU(4) symmetry in the limit of strong capacitive interactions.\cite{galpin2005quantum,keller2014emergent,borda2003}

In terms of electronic transport through the device, we expect the highest series conductance at the triple points between three charge configurations [filled circles in Fig.~\ref{fig:dqd}(b)], since here an electron can pass between leads by traversing the DQD without leaving the ground state charge configurations. Conductance at the triple points can reach up to $G_C=2e^2/h$, even in the incoherent (sequential tunneling) regime. Enhanced conductance is also expected in CB regimes when both dots host a net spin-$\tfrac{1}{2}$ local moment, since enhanced dot-lead scattering processes due to the Kondo effect generated by finite $V_j^{\alpha}$ become correlated by the interdot coupling $t$. However, the finite interdot coupling necessary for a series current to flow favours interdot singlet formation and therefore weakens the dot-lead Kondo effects as discussed above (the phase shift of the leads must be less than $\pi/2$ for $|t|>0$). Thus the series conductance in this regime must be less than the maximum for a single electron transistor, $G_c < 2e^2/h$. By contrast, in the CB regimes inside hexagons where one or both dots host an even number of electrons and therefore have a net spin $S=0$, the Kondo effect is inoperative and conductance is strongly suppressed by the CB effect, $G_C \sim 0$. Finally, we distinguish between MV lines where spin or orbital Kondo effects can occur [black lines in Fig.~\ref{fig:dqd}(b)], and those where DQD charge fluctuations are not accompanied by Kondo [blue lines in Fig.~\ref{fig:dqd}(b)]. Conductance is expected to be enhanced when electronic scattering is Kondo-boosted. Conductance at triple points at the intersection of three such lines (black circles) will be high relative to triple points with only one such Kondo-boosted MV line (red circles).

In Fig.~\ref{fig:dqd}(c) we present full NRG results for the series dc linear conductance through the device at finite temperature, as a function of the dot gate voltages $\delta V_{gL}$ and $\delta V_{gR}$. Conductance maps of this type are commonly obtained from experimental transport measurements;\cite{bork2011tunable} our aim here is to simulate such results. NRG calculations for the irreducible two-channel MLDQD model utilize the improved Kubo formula (Eqs.~\ref{eq:kubo} and \ref{eq:newK}), which yields highly accurate transport properties at relatively low computational cost. This is important, because to generate a high-resolution conductance map as a function of gate voltages requires many NRG runs (in the case of Fig.~\ref{fig:dqd}(c), 2601 separate runs were performed). We verified that conductance results were fully converged at temperature $T=10^{-3}D$ with the NRG parameters used ($\Lambda=5$ and $M_K=4096$), which required about 30 minutes of computing time per NRG run on a standard desktop machine. To achieve similar quality results using other methods would be significantly more costly. Note that an equivalent single-channel formulation is not possible for such a model, and the standard PC form of the MW formula is inapplicable in non-PC coupling geometries. The Kubo formula is therefore required to make contact with experimental results, which are naturally at finite temperature.

The quantitative numerical results shown in Fig.~\ref{fig:dqd}(c) confirm the qualitative physical picture discussed above in relation to the charge stability diagram, Fig.~\ref{fig:dqd}(b). In particular, we see maximal conductance $G_C\sim 2e^2/h$ at triple points which lie at the intersection between three Kondo-boosted MV lines. Conductance is also relatively enhanced in CB hexagons where both dots host a net spin-$\tfrac{1}{2}$, although the competition between interdot singlet and Kondo singlet formation reduces the series conductance to $\sim 0.4 e^2/h$. CB hexagons where one or both dots are $S=0$ give extremely low conductance, $G_C \sim 0$. The periodic structure of the diagram owes to the multilevel nature of the dots; distortions to the pattern at high $|\delta V_{g\alpha}|$ are a physical feature of systems with a finite number of dot levels (in this case, $M=3$ per dot) since particle and hole excitations are asymmetrical.

%###########################
%###########################

\section{Conclusion}\label{sec:conc}
In this paper we reviewed and developed the theory of mesoscopic quantum transport through interacting nanostructures in the linear response regime. We presented new numerical techniques as well as analytical results which help overcome the limitations of conventional methodologies. This opens the door to a more systematic and rigorous theoretical treatment of quantum transport, bringing more complex systems within reach. In particular, our approach -- which treats orbital complexity, strong electron interactions, and nontrivial lead hybridization on an equal footing -- is essential for the realistic simulation of generic nanoelectronics devices and to make quantitative contact with experiments.

In linear response, the Kubo formula\cite{Kubo1956, Kubo1957} is arguably the most general and powerful formulation, but it can be challenging to extract accurate numerical results from it in practice, and it offers little analytical insight into underlying transport mechanisms. In Sec.~\ref{sec:newkubo} we derive an improved formulation of the Kubo formula for electrical conductance for use with NRG, demonstrating its practical advantages over the conventional implementation. 

The Meir-Wingreen approach\cite{meir1992landauer} is also rather general, but only takes a convenient form in terms of equilibrium correlation functions in the linear-response limit when the nanostructure-lead coupling geometry satisfies the `proportionate coupling' condition.
This turns out to be highly restrictive for realistic multi-orbital systems where source and drain leads are spatially separated (and the basic Landauer formula\cite{Landauer1957, *Landauer1970,*LandauerButtiker1985,*Buttiker1986} is inapplicable due to the presence of electron interactions).

Instead, in Secs.~\ref{sec:emPC_CB} and \ref{sec:emPC_MV} we introduce and exploit the concept of `emergent proportionate coupling', in which an effective combination of lead states decouples low temperatures, due to the renormalization from electron interactions. Despite the irreducible two-channel form of the bare model, under RG the system flows to an effective single-channel strong coupling description. This enables us to express low-temperature quantum transport coefficients directly in terms of effective model parameters. In addition to facilitating conductance calculations, the approach also provides physical insights into the underlying transport mechanisms at low temperatures.

We also derive connections between the various standard quantum transport formulations, in cases where equivalences can be made. In particular, we analyze the Kubo formula in the proportionate coupling limit, to obtain a generalized Landauer form valid in the ac frequency regime, and for structured (energy-dependent) leads in a multi-terminal setup (see Appendixes).

Finally, in Sec.~\ref{sec:apps}, we compare and contrast the various techniques discussed above, benchmarking against known results where possible. We do this for two nontrivial systems: the two-terminal double and triple quantum dot models. Our results and analysis indicate that numerical calculations for such systems are under control, and that  simple analytical predictions, although approximate, are nevertheless sufficiently accurate to capture the essential physics of interest in appropriate limits.

Finally, we emphasize that the results presented in this paper pertain to the linear-response regime. The full non-equilibrium problem at finite voltage bias, and/or with a finite temperature gradient between leads, is highly challenging. In many interacting, multi-orbital systems, non-linear transport calculations remain an open problem.

%###########################
%###########################

\begin{acknowledgments}
We are grateful to S.~Sen, P.~Wong, D.~Goldhaber-Gordon, W.~Pouse, for helpful discussions. We acknowledge funding from the Irish Research Council through the Laureate Award 2017/2018 grant IRCLA/2017/169 (AKM/JBR) and the Enterprise Partnership Scheme grant EPSPG/2017/343 (ELM). 
\end{acknowledgments}

%###########################
%###########################

\appendix
\section{Quantum transport with\\structured leads}\label{app:dos}
The formalism presented in Sec.~\ref{sec:LR} assumed that the leads were metallic, with a constant density of states $\rho_0$ inside a wide band of half-width $D$.\cite{sindel2005frequency} A more general formulation takes into account an arbitrary lead density of states $\rho(\omega)$. There are circumstances where this can become important; for example when $\rho(\omega)\sim |\omega|^r$ has a low-energy pseudogap \cite{gonzalez1998renormalization,fritz2006universal} as with graphene leads, whence $\rho_0\equiv \rho(\omega=0)$ is nominally zero.

The Kubo formulae Eqs.~\ref{eq:kubo} and \ref{eq:kubo_heat} remain unchanged, but the correlation functions $K(\omega,T)$ and $K'(\omega,T)$ must now be calculated in the presence of the structured leads.

For equivalent leads with proportionate coupling, the MW ac electrical conductance formula Eq.~\ref{eq:Kubo_PC} becomes,
\begin{equation}\label{eq:Kubo_PC_rho}
    \begin{split}
G_C(\omega,T) = \left(\frac{e^2}{h}\right) \frac{4\pi V_d^2 V_s^2}{V_d^2+V_s^2}\sum_{\sigma} \int_{-\infty}^{\infty}d\omega' ~{\rm Im}~ \overline{\overline{G}}_{\sigma}(\omega',T)&\\
\times \frac{1}{2\omega}\Big [ \rho(\omega'+\omega)\{f_{\rm eq}(\omega'+\omega)-f_{\rm eq}(\omega')&\} \\ 
- \rho(\omega'-\omega)\{f_{\rm eq}(\omega'-\omega)-f_{\rm eq}(\omega')&\}   \Big ]
    \end{split}
    \end{equation}
which in the dc limit $\omega\to 0$ reduces to,
\begin{equation}\label{eq:Kubo_PC_rho_dc}
    \begin{split}
G_C(T) = \left(\frac{e^2}{h}\right) \frac{4\pi V_d^2 V_s^2}{V_d^2+V_s^2}\sum_{\sigma} \int_{-\infty}^{\infty}d\omega' ~{\rm Im}~ \overline{\overline{G}}_{\sigma}(\omega',T)& \\
\times \rho(\omega')\partial_{\omega'}f_{\rm eq}(\omega')& \;.
    \end{split}
    \end{equation}
For the heat conductance, we similarly obtain,
\begin{equation}\label{eq:Kubo_PC_rho_dc_heat}
    \begin{split}
K_Q(T) = \left(\frac{1}{hT}\right) \frac{4\pi V_d^2 V_s^2}{V_d^2+V_s^2}\sum_{\sigma} \int_{-\infty}^{\infty}d\omega' ~{\rm Im}~ \overline{\overline{G}}_{\sigma}(\omega',T)& \\
\times\omega'^2\rho(\omega')\partial_{\omega'}f_{\rm eq}(\omega')& \;.
    \end{split}
    \end{equation}

Note in particular that Eqs.~\ref{eq:Kubo_PC_rho_dc} and \ref{eq:Kubo_PC_rho_dc_heat} are required in the pseudogap case, since the quantity $\rho(\omega){\rm Im}G_{\sigma}(\omega)$ appearing under the integral is subject to a generalized Friedel sum rule.\cite{logan2014common} In particular, $\rho(\omega){\rm Im}~G_{\sigma}(\omega)$ may remain finite as $\omega \to 0$ in Kondo phases, even though $\rho(\omega \to 0)\to 0$.

%###########################

\section{Multi-terminal transport}\label{app:LB}
We consider here briefly the situation for interacting nanostructures coupled to more than two leads. The system Hamiltonian is still given by Eqs.~\ref{eq:Hleads}--\ref{eq:Hhyb}, but with the lead index now $\alpha=1,2,3,...$ rather than just $s$ and $d$. 

Focusing on charge transport in linear response, we are interested in elements of the conductance tensor,
\begin{eqnarray}
G_{C}^{\alpha\beta} = \frac{dI_C^{\alpha}}{dV_b^{\beta}}\bigg |_{V_b^{\beta}\to 0} 
\end{eqnarray}
where, as before, $I^{\alpha}=-e\langle \dot{N}^{\alpha}\rangle$ is the electrical current into lead $\alpha$ and $V_b^{\beta}=\mu_{\beta}/e$ is the voltage bias applied to lead $\beta$. The corresponding dc Kubo formula reads,
\begin{equation}\label{eq:Kubo-MT}
    G_{C}^{\alpha\beta}(T) = \frac{e^2}{h} \lim_{\omega \to 0} \left [ \frac{-2\pi~{\rm Im}K_{\alpha\beta}(\omega,T)}{\omega} \right ] \;,
\end{equation}
where the current-current correlator $K_{\alpha\beta}(\omega)=\langle\langle \dot{N}^{\alpha} ; \dot{N}^{\beta} \rangle\rangle$ can be evaluated at any temperature $T$. This formulation is totally general and holds for any interacting or non-interacting system with any number of leads (which can also be structured and inequivalent).

For equivalent leads and a nanostructure in PC, further simplifications can be made. Eq.~\ref{eq:Kubo-MT} can then be cast in the Landauer-B\"{u}ttiker form. The first step to show this is a rotation of the lead basis, $\vec{f}_{\sigma}=\boldsymbol{U}\vec{c}_{\sigma}$, where $\vec{c}_{\sigma}^{~\dagger}=(c_{1\sigma}^{\dagger},c_{2\sigma}^{\dagger},c_{3\sigma}^{\dagger},...)$ and similarly for $\vec{f}_{\sigma}^{~\dagger}$, and $\boldsymbol{U}$ is unitary. We define $f_{1\sigma}=\sum_{\gamma}U_{1\gamma}c_{\gamma\sigma}$ specifically such that $U_{1\gamma}=V_{\gamma}/V$ where $V_{\gamma}$ are the nanostructure-lead couplings in Eq.~\ref{eq:Hhyb_PC} and $V^2=\sum_{\gamma}|V_{\gamma}|^2$. The nanostructure therefore couples only to the single transformed lead orbital $f_{1\sigma}$.

Note $\tilde{\mathcal{G}}_{\alpha\beta}^0(\omega)\equiv \langle\langle f_{\alpha\sigma}^{\phantom{\dagger}} ; f_{\beta\sigma}^{\dagger}\rangle\rangle^0=\sum_{\gamma}U_{\alpha\gamma}^{\phantom{*}}U_{\beta\gamma}^* \mathcal{G}_{\gamma\gamma}^0(\omega)$ is only diagonal if the original leads are equivalent ($-\tfrac{1}{\pi}{\rm Im}\mathcal{G}_{\gamma\gamma}^0(\omega) = \rho(\omega)$ independent of $\gamma$). In this case, $-\tfrac{1}{\pi}{\rm Im}\tilde{\mathcal{G}}_{\alpha\beta}^0(\omega) = \delta_{\alpha\beta}\rho(\omega)$, meaning that the subsystem containing the nanostructure and the lead $f_1$ is strictly decoupled from the other orthogonal lead combinations $f_2$, $f_3$, and so on. 

For concreteness, consider now $G_C^{12}$, for which we require $K_{12}(\omega)=\langle\langle \dot{N}^{1} ; \dot{N}^{2} \rangle\rangle$. Current conservation implies $\sum_{\gamma} \dot{N}^{\gamma} =0$. Furthermore, from Eq.~\ref{eq:Hhyb_PC}, we may write  $\dot{N}^{\gamma}\equiv i[\hat{H},\hat{N}^{\gamma}]=i\sum_{\sigma}(V_{\gamma}^{\phantom{\dagger}} \bar{\bar{d}}_{\sigma}^{\dagger}c_{\gamma\sigma}^{\phantom{\dagger}} -{\rm H.c.})$. This allows us to express $\dot{N}^1 \equiv \dot{N}^1+a\sum_{\gamma}\dot{N}^{\gamma}  =i\sum_{\sigma}(\tilde{V}_{12}^{\phantom{\dagger}} \bar{\bar{d}}_{\sigma}^{\dagger}f_{2\sigma}^{\phantom{\dagger}} -{\rm H.c.})$, with $f_{2\sigma}=\sum_{\gamma}U_{2\gamma} c_{\gamma\sigma}$ constructed to be orthogonal to $f_{1\sigma}$ via judicious choice of the constant $a$. With $f_{1\sigma}$ and $f_{2\sigma}$ so defined, $\dot{N}^2$ can now be expanded in the basis of the $f$ orbitals. We again exploit current conservation to eliminate the $f_{1\sigma}$ orbital, which yields  $\dot{N}^2 \equiv \dot{N}^2+b\sum_{\gamma}\dot{N}^{\gamma}  =i\sum_{\eta}'\sum_{\sigma}(\tilde{V}_{2\eta}^{\phantom{\dagger}} \bar{\bar{d}}_{\sigma}^{\dagger}f_{\eta\sigma}^{\phantom{\dagger}} -{\rm H.c.})$ where the primed sum excludes $\eta=1$.

Since $f_{2\sigma}$, $f_{3\sigma}$, ... are decoupled from each other as well as from $\bar{\bar{d}}_{\sigma}$, we may factorize the correlator $K_{12}(t)$ into two pieces corresponding to single-particle Green's functions for the nanostructure and the $f_2$ lead respectively. In the frequency domain this results in a convolution. After some manipulation, one finds,
\begin{equation}\label{eq:K12}
    \begin{split}
    {\rm Im}K_{12}(\omega)= \tilde{V}_{12}\tilde{V}_{22} \sum_{\sigma}\int_{-\infty}^{\infty} d\omega'  ~{\rm Im}~ \overline{\overline{G}}_{\sigma}(\omega',T)&\\
\times \Big [ \rho(\omega'+\omega)\{f_{\rm eq}(\omega'+\omega)-f_{\rm eq}(\omega')&\} \\ 
- \rho(\omega'-\omega)\{f_{\rm eq}(\omega'-\omega)-f_{\rm eq}(\omega')&\}   \Big ]
    \end{split}
    \end{equation}
where $\tilde{V}_{12}\tilde{V}_{22}=-V_1^2V_2^2/V^2$ follows directly from the above prescription together with the orthonormality condition $\boldsymbol{U}\boldsymbol{U}^{\dagger}=\boldsymbol{I}$.

Substituting Eq.~\ref{eq:K12} into Eq.~\ref{eq:Kubo-MT} and taking the $\omega \to 0$ limit yields the dc electrical conductance for the multi-terminal setup,
\begin{equation}
    \begin{split}
G_C^{\alpha\ne\beta}(T) = \left(\frac{e^2}{h}\right) \frac{4\pi V_{\alpha}^2 V_{\beta}^2}{V^2}\sum_{\sigma} \int_{-\infty}^{\infty}&d\omega' ~{\rm Im}~ \overline{\overline{G}}_{\sigma}(\omega',T) \\
&\times \rho(\omega')\partial_{\omega'}f_{\rm eq}(\omega') \;.
    \end{split}
    \end{equation}
which holds for interacting nanostructures coupled to an arbitrary number of equivalent structured leads. By current conservation, the diagonal components of the conductance tensor follow as $G_C^{\alpha\alpha}(T)=-\sum_{\beta\ne\alpha} G_C^{\alpha\beta}(T)$.

%###########################

\section{Inequivalent leads}\label{app:inequiv}
For systems in which the leads are inequivalent (meaning that the density of states $\rho_{\alpha}(\omega)$ is not the same for all leads $\alpha$), the PC condition cannot be satisfied. The strategy employed in the previous appendices to decouple orthogonal lead combinations cannot therefore be used in such cases. The Kubo formula Eq.~\ref{eq:kubo}, with the current-current correlator $K(\omega,T)$ given by Eq.~\ref{eq:K}, can still be used -- but for generic interacting nanostructures, it cannot be cast in the form of local retarded Green's functions for the nanostructure.

However, for non-interacting systems, an approximate  generalization of the Landauer formula can be used. We write  Eqs.~\ref{eq:land_I}, \ref{eq:land} but with a transmission function that now involves the full energy-dependent hybridizations, 
\begin{subequations}\label{eq:land_T_inequiv}
\begin{align}
\mathcal{T}_{\gamma\beta}(\omega)&=4\sum_{\sigma} {\rm Tr}[\boldsymbol{G}_{\sigma}^{(0)}(\omega)\boldsymbol{\Gamma}^{\gamma}(\omega)\boldsymbol{G}_{\sigma}^{(0)}(\omega)^* \boldsymbol{\Gamma}^{\beta}(\omega)] \\
&\equiv 4\Gamma_{\gamma}(\omega)\Gamma_{\beta}(\omega)\sum_{\sigma} \overline{G}^{(0)}_{\beta\gamma,\sigma}(\omega)\overline{G}^{(0)}_{\gamma\beta,\sigma}(\omega)^* \;, 
\end{align}
\end{subequations}
where $\boldsymbol{\Gamma}_{\alpha}(\omega)=-{\rm Im}\boldsymbol{\Delta}_{\alpha}(\omega)$ and $\Gamma_{\alpha}(\omega)=\pi |V_{\alpha}|^2 \rho_{\alpha}(\omega)$. In the standard two-terminal setup, $\mathcal{T}(\omega)\equiv \mathcal{T}_{sd}(\omega)$.

We speculate that this result is in fact exact, and is derivable from the Kubo formula by utilizing Wick's theorem to factorize the two-particle susceptibility $K(t)$ into single-particle retarded Green's functions for the nanostructure and for the leads.\cite{note_EM_thesis}  We emphasize that this can only be done for non-interacting systems since the decoupling step relies here on Wick's theorem rather than a canonical transformation of the lead basis which is only possible for PC systems.

The generalization of Landauer-B\"{u}ttiker to multi-terminal transport with inequivalent leads is given by Eq.~\ref{eq:LBaux}, with $\mathcal{T}_{\gamma\beta}(\omega)$ given above in Eq.~\ref{eq:land_T_inequiv}.

%###########################

\section{Perturbative correction to non-equilibrium distribution for the AIM}\label{app:Ng}

The general form of the MW formula (Eqs.~\ref{eq:MW_I}, \ref{eq:MW_X}), applicable also for interacting non-PC systems, requires the lesser Green's function $\boldsymbol{G}_{\sigma}^{<}$. The latter is related to the lesser self-energy via the Keldysh equation  $\boldsymbol{G}^{<}_{\sigma}=\boldsymbol{G}_{\sigma}^{\phantom{<}}\boldsymbol{\Sigma}_{\sigma}^{<}\boldsymbol{G}_{\sigma}^*$.\cite{Keldysh1965,*LangrethLinearRespNEq1976} One may then express the lesser interaction self-energy in terms of its retarded counterpart through the nonequilibrium distribution function, 
$\boldsymbol{\Sigma}^{<}_{\sigma;{\rm int}}=-2i \boldsymbol{f}_{\sigma;{\rm int}}^{\rm NE} ~{\rm Im} \boldsymbol{\Sigma}_{\sigma}$. Of course, $\boldsymbol{f}_{\sigma;{\rm int}}^{\rm NE}$ is in general not known. 

The Ng Ansatz\cite{ng1996ac,*Dong2002} amounts to a simple approximation to this unknown function, $\boldsymbol{f}^{\rm NE}_{\sigma;{\rm int}} \to \boldsymbol{f}^{\rm NE}_0= [f_s\boldsymbol{\Gamma}^s + f_d\boldsymbol{\Gamma}^d] [\boldsymbol{\Gamma}^s+ \boldsymbol{\Gamma}^d + 2 \boldsymbol{{\rm I}}0^+]^{-1}$. The latter is the pseudo-equilibrium distribution\cite{jauho1994time} for a non-interacting system out of equilibrium. For the AIM Eq.~\ref{eq:aim}, the Ng Ansatz corresponds to 
 $f_{\sigma;{\rm int}}^{\rm NE} \to [f_s \Gamma_s + f_d \Gamma_d]/(\Gamma_s+\Gamma_d)$. Assuming a symmetric voltage bias split $\pm \tfrac{1}{2} \Delta V_b$ across source and drain leads, this becomes,
 \begin{equation}\label{eq:fNg}
 f_{\rm Ng}(\omega) = f_{\rm eq}(\omega) + \tfrac{1}{2}e \Delta V_b [-\partial_\omega f_{\rm eq}(\omega) ] \frac{\Gamma_s-\Gamma_d}{\Gamma_s+\Gamma_d} + \mathcal{O}(\Delta V_b^2) \;,
 \end{equation}
where we have expanded to linear-order in the bias, as required to determine the linear-response conductance via the MW formula.

Since the above is exact for $U_d=0$, a natural question pertains to perturbative corrections in $U_d$. To that end, we compute $f_{\sigma;{\rm int}}^{\rm NE}=i\Sigma^<_{\sigma;{\rm int}}/2{\rm Im}\Sigma_{\sigma}$ for the AIM up to second-order in the interaction $U_d$ using nonequilibrium Kadanoff-Keldysh perturbation theory.\cite{Keldysh1965,*LangrethLinearRespNEq1976} We then expand the result to linear order in the bias $\Delta V_b$ and compare to Eq.~\ref{eq:fNg}. After some tedious calculations we obtain,\cite{note_EM_thesis}
 \begin{equation}\label{eq:fcorr}
 \begin{split}
 f_{\sigma;{\rm int}}^{\rm NE}(\omega) = f_{\rm eq}(\omega) + \tfrac{1}{2}e \Delta V_b [-\partial_\omega f_{\rm eq}(\omega) ] \frac{\Gamma_s-\Gamma_d}{\Gamma_s+\Gamma_d}&\Big [ 1+ U_d^2 h_{\sigma}(\omega) \Big ] \\
 +& \mathcal{O}(\Delta V_b^2, U_d^3) \;,
 \end{split}
 \end{equation}
where the correction is controlled by the function,
\begin{equation}
h_{\sigma}(\omega) = \frac{\pi}{\Gamma_s+\Gamma_d}\int d\omega_1 \int d\omega_2~ \bar{f}(\omega,\omega_1,\omega_2)\times \bar{A}_{\sigma}(\omega,\omega_1,\omega_2)\;,
\end{equation}
with $\bar{f}(\omega,\omega_1,\omega_2)=f_{\rm eq}(\omega_1+\omega_2-\omega)[f_{\rm eq}(\omega_1)+f_{\rm eq}(\omega_2)-1]-f_{\rm eq}(\omega_1)f_{\rm eq}(\omega_2)$ and $\bar{A}_{\sigma}(\omega,\omega_1,\omega_2)=\mathcal{A}^0_{\sigma}(\omega_1)\mathcal{A}^0_{\bar{\sigma}}(\omega_2)\mathcal{A}^0_{\bar{\sigma}}(\omega_1+\omega_2-\omega)$ with the latter given in terms of the non-interacting impurity spectral function, $\mathcal{A}^0_{\sigma}(\omega)=\tfrac{1}{\pi\Gamma}\times[1+(\omega-\epsilon_{d\sigma})^2/\Gamma^2]^{-1}$.

 We therefore find a non-zero correction at order $U_d^2\Delta V_b^{\phantom{2}}$, which contributes to the linear-response conductance. Of course, this analysis is not really needed for the AIM itself, since the PC formulation Eq.~\ref{eq:MW_cond} here applies. However, we provide the above discussion to illustrate a general strategy for systematically improving upon the Ng Ansatz for non-PC systems, by including the effect of interactions perturbatively. 

\vfill

%%%%%%% give f a sigma label in main text!

%###########################
%###########################

%\bibliography{refs}

\begin{thebibliography}{157}
\bibitem [{foo({\natexlab{a}})}]{foot_rescale}%
  \BibitemOpen
  \href@noop {} {} ({\natexlab{a}}),\ \bibinfo {note} {in practice $H_N$ is
  rescaled at each step to ensure that its eigenvalues are $\mathcal{O}(1)$,
  but this detail is omitted here for clarity.}\BibitemShut {Stop}%
\bibitem [{foo({\natexlab{b}})}]{foot_N}%
  \BibitemOpen
  \href@noop {} {} ({\natexlab{b}}),\ \bibinfo {note} {we apply the usual NRG
  approximation to neglect contributions to observables at iteration $n$ from
  Wilson shells $m>n$.}\BibitemShut {Stop}%
\bibitem [{not({\natexlab{b}})}]{note_eo}%
  \BibitemOpen
  \href@noop {} {} ({\natexlab{b}}),\ \bibinfo {note} {this assumes that
  $j'_{ee}>j'_{oo}$, which is the conventional situation. However, it is also
  possible that $j'_{oo}>j'_{ee}$, in which case the roles of the even and odd
  channels in the following are reversed.}\BibitemShut {Stop}%
\end{thebibliography}

%###########################
%###########################

\end{document}